%% file: arxiv_2026.tex
\documentclass[10pt]{article} 
\usepackage[preprint]{tmlr}

\input{math_commands.tex}

\usepackage{adjustbox}
\usepackage{algorithm}
\usepackage{algpseudocode}
\usepackage{amsfonts}
\usepackage{amsmath}
\usepackage{amssymb}
\usepackage{amsthm}
\usepackage{array}
\usepackage{bm}
\usepackage{bbm}
\usepackage{dsfont}
\usepackage{booktabs}
\usepackage{calc}
\usepackage{calculator}
\usepackage{colortbl}
\usepackage{enumitem}
\usepackage{etoc}
\usepackage{float}
\usepackage{fp}
\usepackage{framed}
\usepackage{graphicx}
\usepackage{hhline}
\usepackage{hyperref}
\usepackage{latexsym}
\usepackage{listings}
\usepackage{makecell}
\usepackage{mathtools}
\usepackage{mdframed}
\usepackage{microtype}
\usepackage{multicol}
\usepackage{multirow}
\usepackage{nicefrac}
\usepackage{pifont}
\usepackage{pgf}
\usepackage{url}
\usepackage{utfsym}
\usepackage{setspace}
\usepackage{subcaption}
\usepackage{tabularx}
\usepackage{tcolorbox}
\usepackage{tikz}
\usepackage{titletoc}
\usepackage{tabularx}
\usepackage{wrapfig}
\usepackage{xcolor}
\usepackage{xspace}

\definecolor{refcolor}{HTML}{6A9CE8}
\hypersetup{colorlinks=true, citecolor=refcolor, linkcolor=refcolor, urlcolor=refcolor, filecolor=refcolor}

\input{sections/macro}

\makeatletter
\def\@maketitle{\vbox{\hsize\textwidth
  {\LARGE\bf\sffamily \@title\par}\vskip \aftertitskip
  \@startauthor \@author \@endauthor
  \vskip 0.06in minus 0.03in}}
\renewenvironment{abstract}%
  {\vskip 0.02in\centerline{\large\bf\sffamily Abstract}\vspace{0.3ex}\begin{quote}}%
  {\par\end{quote}\vskip 0.4ex}
\makeatother

\makeatletter
\newcommand{\blfootnote}[1]{%
  \begingroup
  \renewcommand{\thefootnote}{}%
  \long\def\@makefntext##1{\noindent ##1}%
  \footnote{#1}%
  \addtocounter{footnote}{-1}%
  \endgroup
}
\makeatother

\title{Enhancing Software Engineering Through Closed-Loop Memory Optimization}

\author{%
  \centering
  \resizebox{\textwidth}{!}{%
    Xuehang Guo$^{*\dagger1}$,\;
    Xingyao Wang$^{*\dagger3,4}$,\;
    Zora Zhiruo Wang$^{2}$,\;
    Qingyun Wang$^{1}$,\;
    Graham Neubig$^{2,3}$,\;
    Heng Ji$^{\dagger4}$%
  } \\[0.4em]
  \footnotesize
  $^{1}$William \& Mary \;
  $^{2}$Carnegie Mellon University \;
  $^{3}$OpenHands \;
  $^{4}$University of Illinois Urbana-Champaign%
  \\[0.2em]
  \footnotesize$^{*}$Equal contribution%
}

\begin{document}

\vspace*{-0.35in}

\maketitle

\blfootnote{$^{\dagger}$Correspondence to: Xuehang Guo \texttt{\textless xguo15@wm.edu\textgreater}, Xingyao Wang \texttt{\textless xingyao6@illinois.edu\textgreater}, Heng Ji \texttt{\textless hengji@illinois.edu\textgreater}.}

\begin{abstract}

Large language models (LLMs) have enabled powerful software engineering (SE) agents capable of navigating complex codebases and resolving real-world issues. However, these agents remain fundamentally episodic: they fail to retain, refine, and reuse experiences across tasks, repeatedly reconstructing context from scratch and reproducing similar mistakes. Even with memory support, they offer no remedy for the absence of a principled, task-agnostic \textit{memory utility}, making them difficult to evaluate rigorously or generalize across agents and settings. To tackle these limitations, we introduce \ours, a closed-loop framework for memory augmentation in SE agents. \ours grounds memory utility in \textit{validated downstream impact}, establishing utility as both a task-agnostic \textbf{evaluation benchmark} and an annotation-free \textbf{optimization signal}. Evaluating \ours across four benchmarks, three programming languages, and two SE agents, under complementary \textit{single-episode} and \textit{cross-episode} memory augmentation, results demonstrate the \textbf{effectiveness} of \ours, achieving absolute gains of up to $\uparrow5.25\%$ in success rate and $\uparrow4.63\%$ in resolve efficiency, while notably reducing deployment cost by $\geq10.33\%$. \ours further exhibits \textbf{generalizability} across SE agents, memory backbones, and optimization algorithms, together with \textbf{transferability} of \ours trained memory model to unseen programming languages and SE activities.
Our project page: \href{https://xhguo7.github.io/MemOp/}{https://xhguo7.github.io/MemOp/}.

\end{abstract}

\input{sections/1_intro}
\input{sections/2_method}
\input{sections/3_train}
\input{sections/4_exps}
\input{sections/5_results}

\input{sections/6_related}
\input{sections/7_conclusions}




\subsubsection*{Broader Impact Statement}

\ours advances memory-augmented AI software engineering, with the potential to meaningfully improve developer productivity and software quality (\S\ref{appendix:subsec:improved_efficiency}). By enabling SE agents to learn from past problem-solving experience and apply distilled knowledge to future tasks (\S\ref{sec:method}), \ours reduces the number of iterations required to resolve issues and increases the success rate of task resolution across \textit{single-episode} and \textit{cross-episode} settings (Tabs.~\ref{tab:exp:single_episode}-\ref{tab:exp:cross_episode}). These efficiency gains could meaningfully lower the barrier to deploying automated AI software engineering agents in real-world tasks, particularly for open-source projects with limited human contributor bandwidth.
Moreover, \ours operates exclusively on the publicly available benchmark (\S\ref{subsec:exp_setup}), and does not involve additional collection of private codebases, user data, or personally identifiable information, thereby excluding privacy and data security concerns.



\bibliography{tmlr}
\bibliographystyle{tmlr}


\newpage
\appendix
\input{sections/appendix}

\end{document}

%% file: math_commands.tex
\usepackage{amsmath,amsfonts,bm}

\def\eqref#1{equation~\ref{#1}}

\def\1{\bm{1}}

\DeclareMathAlphabet{\mathsfit}{\encodingdefault}{\sfdefault}{m}{sl}
\SetMathAlphabet{\mathsfit}{bold}{\encodingdefault}{\sfdefault}{bx}{n}

\DeclareMathOperator*{\argmax}{arg\,max}

%% file: sections/macro.tex
\definecolor{BASELINE_SE_AGENT}{HTML}{F39025}
\definecolor{PURPLE_BG}{HTML}{EAE2FA}
\definecolor{ourscolor}{HTML}{BC8FD6}
\definecolor{notunecolor}{HTML}{6D92EF}
\definecolor{reflectcolor}{HTML}{858585}

\definecolor{BASELINE_BG}{HTML}{E1E1E1}
\definecolor{FROZEN_BG}{HTML}{D4E3F9}
\definecolor{EVOLVE_BG}{HTML}{E7D3F9}

\definecolor{deltagreen}{HTML}{5CB24B}
\definecolor{deltared}{HTML}{ED7777}
\definecolor{deltaredbg}{HTML}{F4AAAA}
\colorlet{deltacellcolor}{white}

\newcommand{\ours}{\textcolor{ourscolor}{\textbf{MemOp}}\xspace}

\newcommand{\bcnum}[1]{%
  \ding{\numexpr 181 + #1\relax}%
}

\newcommand{\cmark}{\textcolor[HTML]{2E9B4F}{\ding{51}}}
\newcommand{\xmark}{\textcolor[HTML]{C43C3C}{\ding{55}}}

\algrenewcommand\algorithmiccomment[1]{\textcolor{ourscolor}{\hfill$\triangleright$ #1}}

\newcommand{\absdeltapct}[2]{%
  \FPeval{\result}{round(#2-#1:2)}%
  #2%
  \FPifpos\result%
    ~{\scriptsize\textbf{\textcolor{deltagreen}{↑\result}}}%
  \else%
    \FPeval{\absresult}{round(0-\result:2)}%
    ~{\scriptsize\textbf{\textcolor{deltared}{↓\absresult}}}%
  \fi%
}

\newcommand{\absdeltapctbf}[2]{%
  \FPeval{\result}{round(#2-#1:2)}%
  \textbf{#2}%
  \FPifpos\result%
    ~{\scriptsize\textbf{\textcolor{deltagreen}{↑\result}}}%
  \else%
    \FPeval{\absresult}{round(0-\result:2)}%
    ~{\scriptsize\textbf{\textcolor{deltared}{↓\absresult}}}%
  \fi%
}

\newcommand{\applydeltacellcolor}[1]{%
  \pgfmathsetmacro{\myintensity}{int(min(100, abs(#1) * 6))}%
  \ifdim#1pt>0pt%
    \edef\tmpcolor{deltagreen!\myintensity!white}%
  \else%
    \edef\tmpcolor{deltaredbg!\myintensity!white}%
  \fi%
  \expandafter\cellcolor\expandafter{\tmpcolor}%
}

\newcommand{\myul}[1]{\leavevmode\vtop{\vbox{\hbox{#1}}\hrule height 0.4pt}}

\newcommand{\absdeltaonlypct}[2]{%
  \pgfmathsetmacro{\deltares}{#2-#1}%
  \pgfmathtruncatemacro{\deltawhole}{int(abs(\deltares))}%
  \pgfmathtruncatemacro{\deltadec}{int(round((abs(\deltares)-\deltawhole)*100))}%
  \ifnum\deltadec<10\edef\deltadecstr{0\deltadec}\else\edef\deltadecstr{\deltadec}\fi%
  \ifdim\deltares pt>0pt%
    \pgfmathsetmacro{\deltaint}{int(min(100, abs(\deltares) * 6))}%
    \edef\deltacolor{deltagreen!\deltaint!white}%
    \expandafter\cellcolor\expandafter{\deltacolor}\small\rmfamily\textcolor{deltagreen}{↑\deltawhole.\deltadecstr}%
  \else%
    \pgfmathsetmacro{\deltaint}{int(min(100, abs(\deltares) * 4))}%
    \edef\deltacolor{deltared!\deltaint!white}%
    \expandafter\cellcolor\expandafter{\deltacolor}\small\rmfamily\textcolor{deltared}{↓\deltawhole.\deltadecstr}%
  \fi%
}

\newcommand{\absdeltaonlypctlight}[2]{%
  \pgfmathsetmacro{\deltares}{#2-#1}%
  \pgfmathtruncatemacro{\deltawhole}{int(abs(\deltares))}%
  \pgfmathtruncatemacro{\deltadec}{int(round((abs(\deltares)-\deltawhole)*100))}%
  \ifnum\deltadec<10\edef\deltadecstr{0\deltadec}\else\edef\deltadecstr{\deltadec}\fi%
  \ifdim\deltares pt>0pt%
    \pgfmathsetmacro{\deltaint}{int(min(100, abs(\deltares) * 3))}%
    \edef\deltacolor{deltagreen!\deltaint!white}%
    \expandafter\cellcolor\expandafter{\deltacolor}\small\rmfamily\textcolor{deltagreen}{↑\deltawhole.\deltadecstr}%
  \else%
    \pgfmathsetmacro{\deltaint}{int(min(100, abs(\deltares) * 2))}%
    \edef\deltacolor{deltared!\deltaint!white}%
    \expandafter\cellcolor\expandafter{\deltacolor}\small\rmfamily\textcolor{deltared}{↓\deltawhole.\deltadecstr}%
  \fi%
}

\newcommand{\absdeltaonlypctbf}[2]{%
  \pgfmathsetmacro{\deltares}{#2-#1}%
  \pgfmathtruncatemacro{\deltawhole}{int(abs(\deltares))}%
  \pgfmathtruncatemacro{\deltadec}{int(round((abs(\deltares)-\deltawhole)*100))}%
  \ifnum\deltadec<10\edef\deltadecstr{0\deltadec}\else\edef\deltadecstr{\deltadec}\fi%
  \ifdim\deltares pt>0pt%
    \pgfmathsetmacro{\deltaint}{int(min(100, abs(\deltares) * 6))}%
    \edef\deltacolor{deltagreen!\deltaint!white}%
    \begingroup\rmfamily\bfseries\small\expandafter\cellcolor\expandafter{\deltacolor}\textcolor{deltagreen}{↑\deltawhole.\deltadecstr}\endgroup%
  \else%
    \pgfmathsetmacro{\deltaint}{int(min(100, abs(\deltares) * 4))}%
    \edef\deltacolor{deltared!\deltaint!white}%
    \begingroup\rmfamily\bfseries\small\expandafter\cellcolor\expandafter{\deltacolor}\textcolor{deltared}{↓\deltawhole.\deltadecstr}\endgroup%
  \fi%
}

\newcommand{\absdeltaonlypctul}[2]{%
  \pgfmathsetmacro{\deltares}{#2-#1}%
  \pgfmathsetmacro{\deltaint}{int(min(100, abs(\deltares) * 4))}%
  \pgfmathtruncatemacro{\deltawhole}{int(abs(\deltares))}%
  \pgfmathtruncatemacro{\deltadec}{int(round((abs(\deltares)-\deltawhole)*100))}%
  \ifnum\deltadec<10\edef\deltadecstr{0\deltadec}\else\edef\deltadecstr{\deltadec}\fi%
  \ifdim\deltares pt>0pt%
    \edef\deltacolor{deltagreen!\deltaint!white}%
    \expandafter\cellcolor\expandafter{\deltacolor}\small\rmfamily\myul{\textcolor{deltagreen}{↑\deltawhole.\deltadecstr}}%
  \else%
    \edef\deltacolor{deltared!\deltaint!white}%
    \expandafter\cellcolor\expandafter{\deltacolor}\small\rmfamily\myul{\textcolor{deltared}{↓\deltawhole.\deltadecstr}}%
  \fi%
}

\newcommand{\absdeltaonlypctbful}[2]{%
  \pgfmathsetmacro{\deltares}{#2-#1}%
  \pgfmathsetmacro{\deltaint}{int(min(100, abs(\deltares) * 4))}%
  \pgfmathtruncatemacro{\deltawhole}{int(abs(\deltares))}%
  \pgfmathtruncatemacro{\deltadec}{int(round((abs(\deltares)-\deltawhole)*100))}%
  \ifnum\deltadec<10\edef\deltadecstr{0\deltadec}\else\edef\deltadecstr{\deltadec}\fi%
  \ifdim\deltares pt>0pt%
    \edef\deltacolor{deltagreen!\deltaint!white}%
    \begingroup\rmfamily\bfseries\small\expandafter\cellcolor\expandafter{\deltacolor}\myul{\textcolor{deltagreen}{↑\deltawhole.\deltadecstr}}\endgroup%
  \else%
    \edef\deltacolor{deltared!\deltaint!white}%
    \begingroup\rmfamily\bfseries\small\expandafter\cellcolor\expandafter{\deltacolor}\myul{\textcolor{deltared}{↓\deltawhole.\deltadecstr}}\endgroup%
  \fi%
}



%% file: sections/1_intro.tex
\input{figures/teaser}

\section{Introduction}
\label{sec:intro}

The emergence of large language models has catalyzed a paradigm shift in software engineering, enabling LLM agents capable of addressing complex real-world software engineering tasks~\citep{jin2025surveyllmsllmbasedagentssoftware,guo2025surveybenchmarkssolutions}. Through tool use, code execution, and multi-step reasoning, LLM-powered SE agents have demonstrated remarkable capabilities in repository-level code generation~\citep{Yang2024SWEagent,Zhang2024CodeAgent,Tao2024MAGIS}.
Recent advances, such as SWE-agent~\citep{Yang2024SWEagent}, OpenHands~\citep{wang2025openhands}, and Claude Code~\citep{ClaudeCode}, showcase the great potential of SE agents to autonomously navigate codebases, understand complex requirements, and implement solutions that span multiple files and functions within continuously evolving SE environments.

Despite these advances, SE agents remain fundamentally episodic: \textit{they fail to develop an adaptive and evolvable mental model across tasks}.
When solving repository-level issues, SE agents repeatedly reconstruct context from scratch, rediscover codebase structure, repeat ineffective implementation strategies, and overlook transferrable knowledge and lessons (\S\ref{appendix:sec:preliminary} \& Figs.~\ref{fig:preliminary:SE_agent_failure_analysis}-\ref{fig:preliminary:failure_example7:hallucination}).
This limitation is especially acute in software engineering, where repositories evolve continuously and many tasks share project-specific conventions, debugging patterns, and architectural constraints. These failures are not due to inherent capability limits, but stem from \textit{the absence of principled memory mechanisms that continuously distill experience into reusable knowledge}~\citep{wang2026humansmissing}.

However, simply augmenting agents with memory fails to resolve the problem. Existing memory-augmented approaches lack a principled and general definition of \textit{memory utility}: they can store, retrieve, or summarize past experience, but whether a given memory actually improves downstream performance, and whether that judgment transfers across different tasks and agents, remains fundamentally unclear.
As a result, memory design and optimization typically rely on task-specific heuristics~\citep{Wang2025ImprovingCLA,ong2025dialogue,yan2025generalagenticmemorydeep,fang2025lightmemlightweightefficientmemoryaugmented}, complex memory architectures~\citep{Chen2024CompressTIA}, and human expertise to determine what should be remembered and how memories should be refined~\citep{wang2026memgovernenhancingcodeagents,Chen2024CompressTIA} (\S\ref{sec:related_work}). Yet without a task-agnostic notion of \textit{memory utility}, it remains unclear (i) how to determine whether a memory is genuinely beneficial, (ii) which properties make memory effective, and (iii) how to leverage utility to guide memory optimization in a generalizable, annotation-free manner. These shortcomings make existing methods difficult to \textit{evaluate rigorously, optimize effectively, and transfer across agents, tasks, and settings}.

To address these limitations, we introduce \ours (\S\ref{sec:method}), a closed-loop memory optimization framework for SE agents (Fig.~\ref{fig:teaser}).
\ours grounds memory utility in \textit{validated downstream impact}: a memory is useful \textit{if and only if} it measurably improves SE agent performance on downstream tasks. As such, \ours turns memory from a hand-designed prompt artifact into an optimizable component: the system learns \textit{what to remember} by measuring \textit{what improves downstream performance}. \ours present two key innovations:
\begin{enumerate}[
  leftmargin=*,
  itemsep=-2pt,
  topsep=-2pt,
  label=\bcnum{\arabic*}
]
\item \textbf{Principled memory utility as dual harness:} \ours introduces a formal notion of \textit{memory utility} grounded in \textit{validated downstream impact} (\S\ref{subsec:memory_utility}): a memory is useful \textit{if and only if} it causally improves SE agent performance on downstream tasks. This outcome-grounded definition serves a dual role: (i) an \textit{evaluation benchmark} that enables rigorous, task-agnostic memory assessment (\S\ref{subsec:eval_metrics}), and (ii) an \textit{optimization signal} that directly drives memory optimization without costly annotation (\S\ref{subsec:rejection_sampling}). By anchoring utility to empirical SE performance rather than heuristic preferences (\S\ref{sec:related_work}), \ours transforms memory quality from an ill-defined property into a measurable, optimizable quantity.
\item \textbf{Closed-loop memory optimization:}
Building on this utility definition, \ours implements \textit{closed-loop training} that adaptively optimizes memory without external supervision (\S\ref{subsec:memop_methodology}). \textit{Trajectory-level reflection} distills candidate memories from completed SE trajectories, and \textit{performance-grounded validation} converts each candidate into training signal by measuring its causal impact on downstream outcomes (\S\ref{subsec:two_stage_sft_rl_training}).
\end{enumerate}

%
%

To systematically evaluate \textit{memory effectiveness}, we study \ours under two regimes (\S\ref{subsec:two_eval_regimes}):
\textit{single-episode memory generation}, where memory is distilled from one trajectory and immediately reused, and \textit{cross-episode memory evolution}, where memory is progressively refined as the agent accumulates experience across tasks.
Results shows \ours's
\textbf{(1) effectiveness}: \ours notably improves SE agents in both regimes, with $\Delta_{abs}$ up to $\uparrow5.25\%$ in $SR$ (\S\ref{subsec:exps:effectiveness}), at substantially reduced deployment cost ($\Delta_{\textit{rel}} \geq 10.33\%$, Tab.~\ref{tab:exp:computation_overhead}, \S\ref{appendix:subsec:computation_overhead});
\textbf{(2) generalizability}: \ours demonstrates consistent gains generalizable across different SE agents, $M_\theta$ backbones, and optimization algorithms (\S\ref{subsec:exps:generalizability});
\textbf{(3) transferability}: \ours trained $M_\theta$ improves all ten metrics in different programming languages and improves the native criteria of different benchmarks covering SE activities beyond issue resolution (\S\ref{subsec:exps:transferability}).

%% file: figures/teaser.tex






\begin{wrapfigure}{r}{0.6\linewidth}

\vspace{-36pt}

\centering
\includegraphics[width=1.0\linewidth]{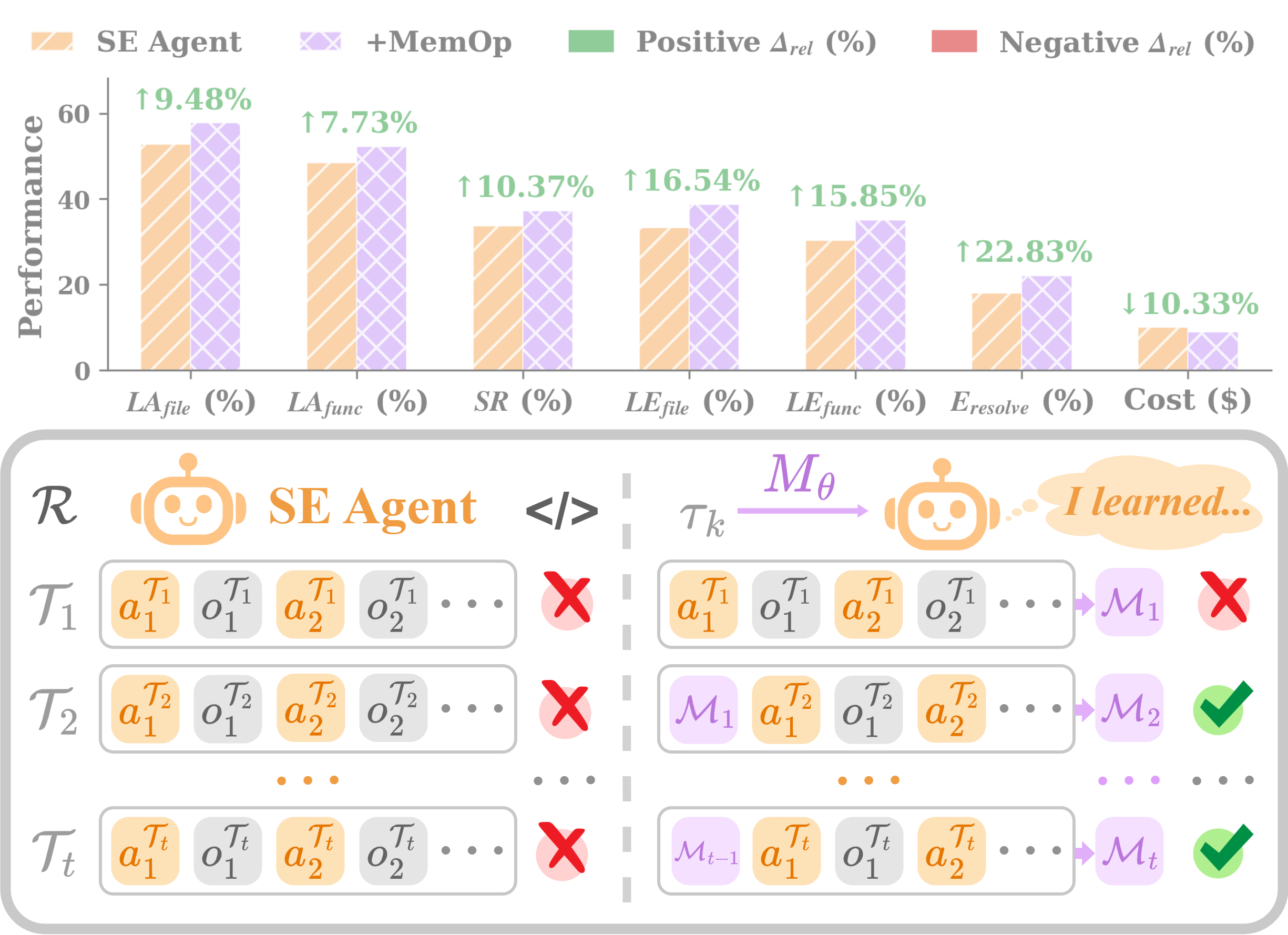}

\vspace{-3pt}

\caption{\textbf{Memory-Augmented Software Engineering.} Compared with \textcolor{BASELINE_SE_AGENT}{\textit{no-$M_\theta$} SE agent} (\textit{left}), \ours (\textit{right}) equips SE agents with adaptively distilled memories to better tackle dynamic real-world SE challenges.}


\label{fig:teaser}

\vspace{-18pt}

\end{wrapfigure}

%% file: sections/2_method.tex
\section{Memory Augmentation for Agentic Software Engineering}
\label{sec:method}

\subsection{Problem Formulation}
\label{subsec:problem_formulation}

We consider an SE agent operating on a repository $\mathcal{R}$ to solve the $k$-th task $\mathcal{T}_k$ $(k=1,2,3,\ldots)$.
Starting from the initial state $\mathcal{S}_{k-1}$, the SE agent interacts with $\mathcal{R}$ through a sequence of actions $\mathcal{A} = \{a_1, a_2, \ldots, a_n\}$ ($n \leq N_{\mathcal{A}}$, $N_{\mathcal{A}}$ is the maximum problem-solving iterations allowed for the SE agent), including code navigation, file reading, command execution, and function modification.
Upon task completion at state $\mathcal{S}_k$, this interaction produces a problem-solving trajectory $\tau_k = \{(a_1, o_1), (a_2, o_2), \ldots, (a_n, o_n)\}$, where $o_i$ represents the outputs and observations of $a_i$ ($1 \leq i \leq n$).

\textbf{Memory-Augmented SE Agent.}
Motivated by our preliminary studies that expose fundamental limitations in how agents process and retain problem-solving experience (§\ref{appendix:subsec:preliminary:SE_agent_failure}), we augment the SE agent with an evolving memory state $\mathcal{M}_k$, stored in \texttt{Memory.md} (\S\ref{appendix:subsec:preliminary:memory_structure}), that distills lessons learned to facilitate the SE agent on future tasks. In deployment, \ours operates in two phases:
\begin{enumerate}[
  leftmargin=*,
  itemsep=-2pt,
  topsep=-2pt,
  label=\bcnum{\arabic*}
]
    \item \textbf{Reflective Memory Evolution}: After completing task $\mathcal{T}_k$ on repository $\mathcal{R}$, a dedicated memory model $M_\theta$ reflects on problem-solving trajectory $\tau_k$ to evolve memory state: $\mathcal{M}_k = M_{\theta}(\mathcal{R}, \mathcal{S}_k)$, where $\mathcal{S}_k = \mathbf{S}(\mathcal{M}_{k-1}, \tau_k)$ incorporates prior memory and current trajectory, and $\mathbf{S}(\cdot)$ constructs the agent state from previous memory state and current trajectory.
    
    \item \textbf{Memory-Augmented Execution}: When solving $\mathcal{T}_{k+1}$, the agent is equipped with $\mathcal{M}_k$, which provides distilled insights guiding more effective navigation, reasoning, and problem-solving.
\end{enumerate}
This raises two fundamental challenges that \ours addresses: (1) \textit{how to measure whether a memory is useful}, and (2) \textit{how to optimize memory toward utility}.

\subsection{Memory Utility: If It Doesn't Help, It Doesn't Count}
\label{subsec:memory_utility}

\textbf{Grounding Utility in Downstream Impact.}
Memory-augmented agentic systems face a critical problem: \textit{there is no principled way to quantify whether a given memory is genuinely useful} (\S\ref{sec:related_work}).

\input{figures/eval_module}

\vspace{-4pt}
Without such measures, it is impossible to distinguish good memory from noise, or even to leverage quality signals to drive learning (\S\ref{sec:related_work}\&\ref{appendix:sec:related_work_extension}).
We tackle this with a concrete, outcome-grounded definition (Fig.~\ref{fig:eval_module}): \textit{a memory is useful if and only if it demonstrably improves SE agent performance on downstream tasks}. 


\textbf{Performance-Grounded Memory Utility.}
Given a candidate memory $\mathcal{M}_j$ generated from trajectory $\tau_k$ on task $\mathcal{T}_k$ in repository $\mathcal{R}$, we measure the performance change $\Delta_i(\mathcal{M}_j)$ for each metric $i$ ($i \in \{1, \ldots, N_Q\}$, where $N_Q = 10$ multi-dimensional metrics defined in \S\ref{subsec:eval_metrics}) by comparing memory-augmented problem-solving against no-memory baseline.

\textbf{Memory Acceptance Criterion.}
Based on empirical measurements, we establish a rigorous criterion for memory quality.
A candidate memory $\mathcal{M}_j$ is \textbf{\textit{accepted}} as high-quality if and only if it satisfies:
\begin{equation}
    \forall i: \Delta_i(\mathcal{M}_j) \geq 0 \quad \land \quad \exists i: \Delta_i(\mathcal{M}_j) > 0
\label{eq:memory_acceptance}
\end{equation}
Memory candidates that fail this criterion are \textbf{\textit{rejected}}.
This \textit{acceptance criteria} serves as a dual role in \ours: (1) an evaluation benchmark for rigorous, task-agnostic memory assessment (\S\ref{subsec:eval_metrics}), and (2) an optimization signal that drives memory learning without external annotation (\S\ref{subsec:rejection_sampling}).

\subsection{\ours For Closed-Loop Memory Optimization}
\label{subsec:memop_methodology}

\vspace{-3pt}

\textbf{From Experience to Memory.}
At the core of \ours is to learn a memory model $M_\theta: \mathcal{S}_k \rightarrow \mathcal{M}_k$ that distills agent experience into memory, mapping current agent state $\mathcal{S}_k=\mathbf{S}(\mathcal{M}_{k-1}, \tau_k)$ to a new memory state $\mathcal{M}_k$ that captures key patterns, observations, and actionable insights (\S\ref{appendix:sec:preliminary}):
%
%
\begin{equation}
    \mathcal{M}_k = M_{\theta}(\mathcal{R}, \mathcal{S}_k) = M_{\theta}(\mathcal{R}, (\mathcal{M}_{k-1}, \tau_k))
\label{eq:memory_generation}
\end{equation}
Each memory encodes lessons learned, codebase structures identified, reasoning patterns discovered, and task-relevant contextual information that the SE agent can leverage in future problem-solving. Our preliminary study on memory structure (\S\ref{appendix:subsec:preliminary:memory_structure}) and generation instructions (\S\ref{appendix:subsec:preliminary:memory_prompt}) informs how $\mathcal{M}_k$ is represented and how $M_\theta$ is guided to generate it effectively.

\textbf{Memory Optimization Objective.}
Our goal is to learn $M_{\theta}$ such that the SE agent augmented with memory $\mathcal{M}_k$ achieves superior performance on future tasks $T_{k'}$ ($k'>k$):
\begin{equation}
        M_{\theta}^* = \argmax_{M_{\theta}} \mathbb{E}_{\mathcal{T}_{k'}, \mathcal{R}} \left[ \mathbf{Q}(\mathcal{A}_{k'} | \mathcal{R}, \mathcal{S}_{k'}) \right]
\end{equation}
where $\mathbf{Q}(\cdot)$ is a composite memory utility function combining all $N_{Q}$ multi-dimensional metrics (\S\ref{subsec:eval_metrics}) across task performance and problem-solving efficiency, and $M_{\theta}^*$ is the optimal memory evolution function \ours aims to learn.
Crucially, $\mathbf{Q}$ is precisely the utility signal defined in \S\ref{subsec:memory_utility}, directly connecting \textit{how we measure} memory quality to \textit{what we optimize for}.
This alignment is the conceptual core of \ours: by anchoring the training objective to empirical SE outcomes rather than heuristic proxies, memory quality becomes a measurable, optimizable quantity.


%% file: figures/eval_module.tex
\begin{wrapfigure}{r}{0.69\linewidth}

\vspace{-6pt}

\centering
\includegraphics[width=1.0\linewidth]{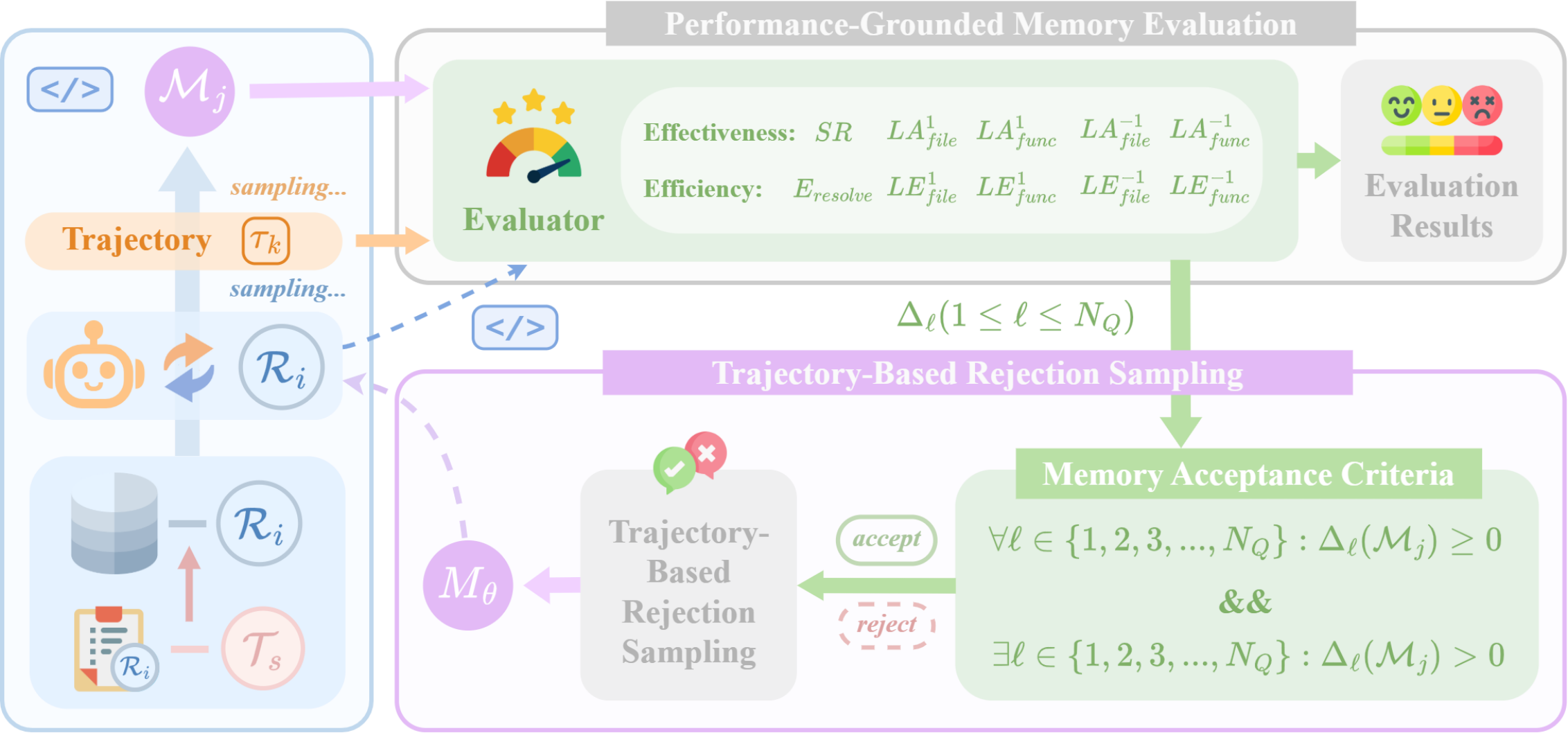}

\vspace{0pt}

\caption{\textbf{Memory Utility.} We tackle the fundamental challenge (\S\ref{subsec:memory_utility}) by proposing \textit{memory utility} with \textit{performance-grounded memory evaluation} and \textit{trajectory-level rejection sampling}.}
\label{fig:eval_module}

\vspace{-6pt}

\end{wrapfigure}

%% file: sections/3_train.tex
\section{Memory Optimization via Performance-Validated Finetuning}
\label{sec:training}

\input{figures/train}

Turning \textit{memory utility} into \textit{optimization signals} (Fig.~\ref{fig:two_stage_train}), we first define how \textit{memory utility} is measured (\S\ref{subsec:eval_metrics}), then use that signal to curate training data (\S\ref{subsec:rejection_sampling}), and finally train $M_\theta$ in a closed loop (\S\ref{subsec:two_stage_sft_rl_training}).

\subsection{Grounding Memory Utility in Downstream Performance}
\label{subsec:eval_metrics}

To ensure the evaluation quality and comprehensiveness during both memory training datasets curation (\S\ref{subsec:rejection_sampling}) and SE performance evaluation (\S\ref{sec:exps}), we employ a systematic suite of four metric families across $N_Q=10$ evaluation metrics, covering the effectiveness and efficiency of SE agent's problem-solving.

\textbf{Success Rate} ($SR$) (Eq.~\ref{eq:eval_metric:SR}) measures the proportion of tasks successfully resolved.
\begin{equation}
    SR = \frac{1}{|\mathcal{T}|} \sum_{\mathcal{T}_t \in \mathcal{T}} \mathds{1} (SR_t=1)
\label{eq:eval_metric:SR}
\end{equation}
where $\mathcal{T}$ is the set of all tasks, $\mathds{1}(\cdot)$ is the indicator function, and $SR_t \in \{0, 1\}$ indicates whether the $t$-th task is successfully resolved.

\textbf{Localization Accuracy} ($LA$) (Eq.~\ref{eq:eval_metric:LA}) evaluates the SE agent's ability to localize required targets at two levels, including: (a) \textit{file} ($LA_{\textit{file}}$): accurately identifying the target files, and (b) \textit{function} ($LA_{\textit{func}}$): accurately pinpointing the target functions; and two criteria, including: (a) \textit{first success} ($LA^{(1)}$): measures whether the agent successfully localizes at least one required target, and (b) \textit{full success} ($LA^{(-1)}$): measures whether the agent successfully localizes all required targets.
\begin{equation} LA^{(e)}_f = \frac{1}{|\mathcal{T}|} \sum_{\mathcal{T}_t \in \mathcal{T}} \mathds{1}(LA^{(e)}_{f,t} = 1) \label{eq:eval_metric:LA} \end{equation}
%
%
%
%
%
%
%
where $e \in \{1, -1\}$ denotes the success criterion (\textit{first} vs. \textit{full}), and $f \in \{\textit{file}, \textit{func}\}$ denotes the target granularity.
For a given task $\mathcal{T}_t$, let $\mathcal{F}^*_t$ and $\mathcal{G}^*_t$ denote the ground-truth target files and functions, respectively, and let $\mathcal{F}_{t,i}$ and $\mathcal{G}_{t,i}$ represent the files and functions accessed by the agent at step $i$.
\textit{First success} ($e=1$) measures whether the agent identifies \emph{any} target during problem-solving: the agent achieves first file localization if it accesses at least one target file across all steps, and similarly for functions.
\textit{Full success} ($e=-1$) measures whether the agent identifies \emph{all} targets: the agent achieves full file localization only if all target files are eventually accessed, and similarly for functions.

\textbf{Resolve Efficiency} ($\mathit{E_{\textit{resolve}}}$) (Eq.~\ref{eq:eval_metric:Eresolve}) measures the average proportion of SE agent's problem-solving iterations saved when resolving tasks.
\begin{equation}
    \mathit{E_{\textit{resolve}}} = \frac{1}{|\mathcal{T}|} \sum_{\mathcal{T}_t \in \mathcal{T}} \mathit{E_{\textit{resolve},t}}
\label{eq:eval_metric:Eresolve}
\end{equation}
where $\mathit{E_{\textit{resolve},t}} = 1 - \frac{n_t}{N_{\mathcal{A}}}$. Higher values indicate faster resolution, with $\mathit{E_{\textit{resolve},t}} = 0$ when the agent uses up all $N_{\mathcal{A}}$ iterations on task $\mathcal{T}_t$.

\textbf{Localization Efficiency} ($LE$) (Eq.~\ref{eq:eval_metric:LE}) measures the average efficiency in localizing required targets at two levels (\textit{file} \& \textit{function}) and two criteria (\textit{first} \& \textit{full}): (a) \textit{first success}: $LE^{(1)}$ evaluates SE agent efficiency in localizing the first required targets; and (b) \textit{full success}: $LE^{(-1)}$ evaluates SE agent efficiency in localizing all required targets.
\vspace{-5pt}
\begin{equation}
    LE^{(e)}_f = \frac{1}{|\mathcal{T}|} \sum_{\mathcal{T}_t \in \mathcal{T}} LE^{(e)}_{f,t}
\label{eq:eval_metric:LE}
\end{equation}
\vspace{-8pt}

where $e \in \{1, -1\}$ denotes the efficiency criterion (\textit{first} vs.\ \textit{full}), $f \in \{\textit{file}, \textit{func}\}$ denotes the target granularity, and $LE^{(e)}_{f,t} = 1 - \nicefrac{n^{(e)}_{f,t}}{N_{\mathcal{A}}}$ with $n^{(e)}_{f,t}$ as the number of steps to achieve localization success. For \textit{first success} ($e=1$), $n^{(1)}_{f,t}$ is the step at which the agent first identifies \textit{any} target file or function; for \textit{full success} ($e=-1$), $n^{(-1)}_{f,t}$ is the step by which \textit{all} targets are identified.

\textbf{Performance Difference} is measured by \textit{absolute} difference ($\Delta_{\textit{abs}}$) (Eq.~\ref{eq:eval_metric:deltaabs}) and \textit{relative} difference ($\Delta_{\textit{rel}}$) (Eq.~\ref{eq:eval_metric:deltarel}) in performance for each quality metric above (Eqs.~\ref{eq:eval_metric:SR}-\ref{eq:eval_metric:LE}).
\vspace{-3pt}
\begin{equation}
    \Delta_{\textit{abs},i} = (Q_{i,\text{\ours}} - Q_{i,\textit{baseline}}) \times 100\%
\label{eq:eval_metric:deltaabs}
\end{equation}
\begin{equation}
    \Delta_{\textit{rel},i} = (Q_{i,\text{\ours}} - Q_{i,\textit{baseline}}) / Q_{i,\textit{baseline}} \times 100\%
\label{eq:eval_metric:deltarel}
\end{equation}
where $Q_{i,\text{\ours}}$ and $Q_{i,\textit{baseline}}$ denote the $i$-th metric with \ours and baseline, respectively.

\input{tables/rejection_sampling_algorithm}

\subsection{Curating Performance-Validated Memory Utility}
\label{subsec:rejection_sampling}

The quality of memory instances directly determines their utility for SE agent problem-solving.
To construct high-quality training data, we propose \textit{trajectory-based rejection sampling}, a rigorous data curation method that applies our evaluation module (\S\ref{subsec:memory_utility}) to filter memory candidates, retaining only those empirically validated to improve SE agent performance.
%
%
%
%
%
%
%
%
%
%
%
Alg.~\ref{alg:rejection_sampling} formalizes this rejection sampling algorithm on $N_{\mathcal{R}}$ repositories randomly sampled from SWE-Bench-\textit{Verified} \citep{jimenez2024swebench}, with $N_{\mathcal{T}}$ SE tasks for each repository. The SE agent attempts to solve each task with $N_{\tau}$ rollouts, during which each trajectory is used to generate $N_{\mathcal{M}}$ memory candidates.
Our memory acceptance criteria (Eq.~\ref{eq:memory_acceptance}) then determines whether $\mathcal{M}_p$ is categorized to the accepted set $\mathcal{D}_{\text{\textit{accept}}}$ or rejected set $\mathcal{D}_{\text{\textit{reject}}}$. 
As such, this rejection sampling yields two datasets for our two-stage training:
\begin{enumerate}[
  leftmargin=*,
  itemsep=-2pt,
  topsep=-2pt,
  label=\bcnum{\arabic*}
]
    \item \textbf{Stage I Dataset $\mathcal{D}_{\text{\textit{SFT}}}$}: The Stage I dataset, comprising accepted memory samples filtered from $\mathcal{D}_{\text{\textit{accept}}}$ to train the memory model to generate performance-enhancing memories.
    \item \textbf{Stage II Dataset $\mathcal{D}_{\text{\textit{RL}}}$}: The Stage II dataset, where the ground-truth memory of each sample is drawn from $\mathcal{D}_{\text{\textit{accept}}}$, and contrastive candidates are drawn from the filtered subset of $\mathcal{D}_{\text{\textit{reject}}}$ corresponding to the same SE task and repository.
\end{enumerate}

\subsection{Training Memory Model from Validated Reflections}
\label{subsec:two_stage_sft_rl_training}


\textbf{Stage I: Supervised Finetuning for Memory Generation.}
The memory model $M_{\theta}$ is first trained through supervised finetuning (SFT) on $\mathcal{D}_{\text{\textit{SFT}}}$ (\S\ref{subsec:rejection_sampling}) to acquire fundamental memory generation abilities from SE agents' trajectories.
Given a trajectory-memory pair $(\tau, \mathcal{M}) \in \mathcal{D}_{\text{\textit{SFT}}}$, we optimize:
\begin{equation}
    \mathcal{J}_{\text{StageI}} = -\mathbb{E}_{(\tau, \mathcal{M}) \sim \mathcal{D}_{\text{\textit{SFT}}}} \left[ \log P_{\theta}(\mathcal{M} | \tau, \mathcal{T}, \mathcal{R}) \right]
\end{equation}
This stage teaches $M_{\theta}$ the critical characteristics of effective memories, establishing the foundation for high-quality memory generation and evolution.

\textbf{Stage II: Reinforcement Learning with Preference Optimization.}
As Stage I equips $M_\theta$ with foundational memory generation abilities, Stage II further optimizes $M_\theta$ toward generating high-quality memory that improves SE agent performance via RL on $\mathcal{D}_{\text{\textit{RL}}}$ (\S\ref{subsec:rejection_sampling}).
$M_\theta$ is trained with the reward (Eq.~\ref{eq:rl_reward}) grounded in memory utility (\S\ref{subsec:memory_utility}), directly optimizing memory generation toward validated downstream SE impact.
%
%
Concretely, for each sample in $\mathcal{D}_{\text{\textit{RL}}}$, $M_\theta$ generates $c$ memory candidates $\{\mathcal{M}_1, \mathcal{M}_2, \ldots, \mathcal{M}_c\}$ from the same task $\mathcal{T}^\mathcal{R}$ on repository $\mathcal{R}$, where ground-truth memory is drawn from $\mathcal{D}_{\text{\textit{accept}}}$ and the remaining $(c-1)$ candidates are drawn from $\mathcal{D}_{\text{\textit{reject}}}$.
The reward function directly measures the SE performance improvement induced by each generated memory:
\vspace{-6pt}
\begin{equation}
    r(\mathcal{M}_i) = \frac{1}{N_Q} \sum_{\ell=1}^{N_Q} \Delta_\ell(\mathcal{M}_i)
    \label{eq:rl_reward}
\end{equation}
where $\Delta_\ell(\mathcal{M}_i)$ is the performance change on metric $\ell$ when the SE agent is augmented with $\mathcal{M}_i$ compared to the no-memory baseline (Eq.~\ref{eq:eval_metric:deltaabs}). This grounds the RL signal directly in downstream SE outcomes, where $M_\theta$ is rewarded for generating memory that measurably improves agent performance, and penalized for generating memory that is redundant or harmful.
Accordingly, Stage II optimizes $M_\theta$ to maximize the expected reward advantage relative to the group baseline:
\begin{equation}
    \mathcal{J}_{\text{StageII}} = -\mathbb{E}_{(\mathcal{T}, \mathcal{R}, \mathcal{M}_i) \sim \mathcal{D}_{\text{\textit{RL}}}} \left[ A_i \log P_{\theta}(\mathcal{M}_i | \mathcal{T}, \mathcal{R}) \right]
    \label{eq:rl_objective}
\end{equation}
%
%
%
%
%
%
%
where $A_i = r(\mathcal{M}_i) - \bar{r}$ is the advantage of $\mathcal{M}_i$, and $\bar{r} = \frac{1}{c} \sum_{j=1}^{c} r(\mathcal{M}_j)$ is the group baseline reward computed over a rollout batch of $c$ generated memory candidates.
%
Through Stage II optimization, $M_\theta$ learns to generate memory that shares properties with validated high-quality memories while avoiding patterns common in rejected candidates.

\input{tables/exp_main_single_episode}

\subsection{Evaluation Regimes for Effective Reflection and Adaptive Evolution}
\label{subsec:two_eval_regimes}

A useful memory framework should be \textit{effective across downstream tasks} and \textit{adaptive in memory evolution}. Therefore, we evaluate \ours in two modes:

\textbf{Single-Episode Memory Generation.}
\ours generates a memory $\mathcal{M}_k$ from each completed SE trajectory 
$\tau_k$, which is immediately reused to augment the SE agent.
This regime evaluates the most fundamental ability of \ours: \textit{whether memory distilled from a single trajectory can effectively improve downstream SE performance}.
It validates the quality of \ours's memory generation in isolation, 
independent of cross-task transfer or cross-episode generalization.

\textbf{Cross-Episode Memory Evolution.}
\ours evolves memory progressively across a sequence of tasks.
When SE agent performs the ($k$+1)-th task, it is equipped with the latest $\mathcal{M}_k$ generated at $\mathcal{S}_k=\mathbf{S}(\mathcal{M}_{k-1}, \tau_k)$.
This regime validates the adaptability and robustness of \ours: \textit{whether memory can remain coherent, non-redundant, and increasingly useful as the SE agent encounters a growing diversity of tasks}.

%% file: figures/train.tex
\begin{figure}[!t]

\centering
\includegraphics[width=1.0\linewidth]{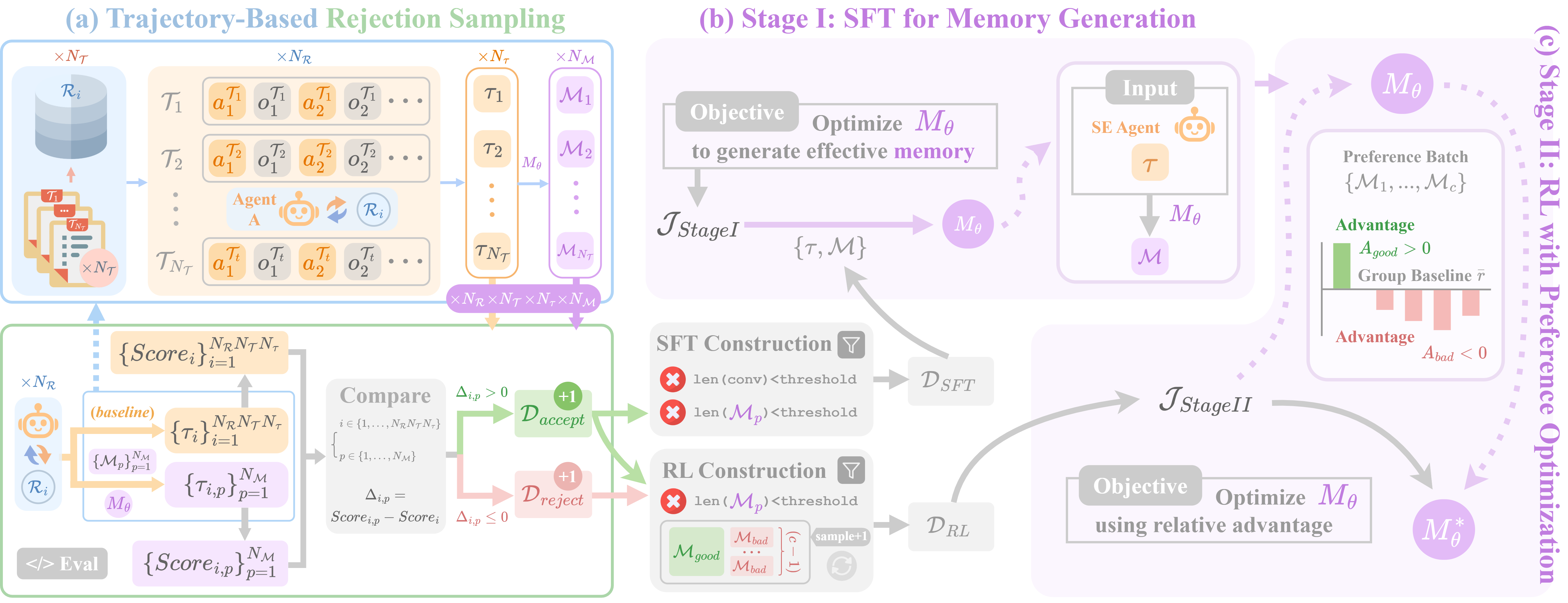}

\caption{\textbf{Memory Model Finetuning.} By preparing training datasets (Tab.~\ref{tab:training_data_overview}) through \textit{trajectory-based rejection sampling} (\S\ref{subsec:rejection_sampling}), $M_\theta$ is finetuned through two-stage training via SFT and RL (\S\ref{subsec:two_stage_sft_rl_training}).}
\label{fig:two_stage_train}

\end{figure}

%% file: tables/rejection_sampling_algorithm.tex
\begin{algorithm}[t]
\caption{Trajectory-Based Rejection Sampling}
\label{alg:rejection_sampling}
\begin{algorithmic}[1]

\Statex \textbf{Input:} Repositories $\{\mathcal{R}_i\}_{i=1}^{N_{\mathcal{R}}}$, tasks per repo $N_{\mathcal{T}}$, trajectory rollouts per task $N_{\tau}$, memory candidates per trajectory $N_{\mathcal{M}}$
\Statex \textbf{Output:} Accepted dataset $\mathcal{D}_{\textit{accept}}$, rejected dataset $\mathcal{D}_{\textit{reject}}$
\Statex \textbf{Models:} Memory model $M_{\theta}$, SE agent \textsc{Agent}
\Statex \textbf{Metrics:} $N_Q{=}10$ multi-dimensional metrics ($SR$, $LA$, $E_{\textit{resolve}}$, $LE$) defined in \S\ref{subsec:eval_metrics}

\vspace{4pt}

\Function{SampleMemory}{$\mathcal{R}, \mathcal{T}, \tau$} \Comment{Memory sampling function definition (\S\ref{subsec:rejection_sampling})}
    \State $\text{\textsc{Score}}_{\textit{base}} \gets \mathbf{Q}(\textsc{Agent}(\mathcal{R}, \mathcal{T}, \emptyset))$ \Comment{Baseline utility (\S\ref{subsec:memop_methodology})}
    \For{$p = 1$ \textbf{to} $N_{\mathcal{M}}$}
        \State $\mathcal{M}_p \gets M_{\theta}(\mathcal{R}, \tau)$ \Comment{Generate candidate (Eq.~\ref{eq:memory_generation})}
        \State $\text{\textsc{Score}}_{\textit{p}} \gets \mathbf{Q}(\textsc{Agent}(\mathcal{R}, \mathcal{T}, \mathcal{M}_p))$ \Comment{Candidate memory utility (\S\ref{subsec:memop_methodology})}
        \State $\Delta(\mathcal{M}_p) \gets \text{\textsc{Score}}_{\textit{p}} - \text{\textsc{Score}}_{\textit{base}}$ \Comment{Utility gain (Eq.~\ref{eq:eval_metric:deltaabs})}
        \If{$(\forall \ell \in \{1,\ldots,N_Q\}: \Delta_{\ell}(\mathcal{M}_p) \geq 0)\ \land\ (\exists \ell: \Delta_{\ell}(\mathcal{M}_p) > 0)$} \Comment{Criterion (Eq.~\ref{eq:memory_acceptance})}
            \State $\mathcal{D}_{\textit{accept}} \gets \mathcal{D}_{\textit{accept}} \cup \{(\mathcal{R}, \tau, \mathcal{M}_p)\}$
        \Else
            \State $\mathcal{D}_{\textit{reject}} \gets \mathcal{D}_{\textit{reject}} \cup \{(\mathcal{R}, \tau, \mathcal{M}_p)\}$
        \EndIf
    \EndFor
\EndFunction

\vspace{4pt}

\State $\mathcal{D}_{\textit{accept}} \gets \emptyset$, $\mathcal{D}_{\textit{reject}} \gets \emptyset$
\For{$i = 1$ \textbf{to} $N_{\mathcal{R}}$} \Comment{Iterate over $N_{\mathcal{R}}$ repositories}
    \For{$j = 1$ \textbf{to} $N_{\mathcal{T}}$} \Comment{Iterate over $N_{\mathcal{T}}$ tasks}
        \For{$t = 1$ \textbf{to} $N_{\tau}$} \Comment{Iterate over $N_\tau$ trajectories}
            \State \Call{SampleMemory}{$\mathcal{R}_i,\ \mathcal{T}_{ij},\ \tau_{ijt}$} \Comment{Memory sampling}
        \EndFor
    \EndFor
\EndFor
\State \Return $\mathcal{D}_{\textit{accept}}$, $\mathcal{D}_{\textit{reject}}$ \Comment{Dataset curation (\S\ref{subsec:two_stage_sft_rl_training})}

\end{algorithmic}
\end{algorithm}

%% file: tables/exp_main_single_episode.tex

\begin{table}[t]

\small
\centering
\renewcommand{\arraystretch}{1.1}
\setlength{\tabcolsep}{0pt}

\caption{\textbf{Single-Episode Memory Augmentation.} Single-episode accuracy and efficiency of SE agents (\S\ref{subsec:eval_metrics}). We compare \textcolor{gray}{\textbf{no-$M_\theta$ baselines}} against three memory conditions: (1) a \textcolor{reflectcolor}{\textbf{reflection baseline}} (\texttt{Claude-4-Sonnet}), which applies the same memory format and generation instruction (Fig.~\ref{fig:two_prompt_for_single_and_cross_episode}) without utility filtering or training; (2) \textcolor{notunecolor}{\textbf{non-finetuned $M_\theta$ (NFT)}}; and (3) \textcolor{ourscolor}{\textbf{\ours-finetuned $M_\theta$ (FT)}}. Here, \textbf{\textit{absolute performance differences}} (Eq.~\ref{eq:eval_metric:deltaabs}) are shown as \textit{positive} \textcolor{deltagreen}{$\uparrow \Delta_{\textit{abs}}\%$} or \textit{negative} \textcolor{deltared}{$\downarrow \Delta_{\textit{abs}}\%$}. ``-T" indicates \textit{thinking} mode.}
\label{tab:exp:single_episode}

\begin{tabularx}{\textwidth}{c*{10}{>{\centering\arraybackslash}X}}

\toprule

\multirow{3}{*}{$\mathbf{\textit{M}}_{\theta}$} & \multicolumn{5}{c}{\textbf{Accuracy (\%)}} & \multicolumn{5}{c}{\textbf{Efficiency (\%)}} \\
\cmidrule(lr){2-6} \cmidrule(lr){7-11}
& $\bm{\textbf{\textit{LA}}_{\textit{file}}^{(1)}}$ & $\bm{\textbf{\textit{LA}}_{\textit{func}}^{(1)}}$ & $\bm{\textbf{\textit{LA}}_{\textit{file}}^{(-1)}}$ & $\bm{\textbf{\textit{LA}}_{\textit{func}}^{(-1)}}$ & \textbf{\textit{SR}}
& $\bm{\textbf{\textit{LE}}_{\textit{file}}^{(1)}}$ & $\bm{\textbf{\textit{LE}}_{\textit{func}}^{(1)}}$ & $\bm{\textbf{\textit{LE}}_{\textit{file}}^{(-1)}}$ & $\bm{\textbf{\textit{LE}}_{\textit{func}}^{(-1)}}$ & $\bm{E_{\textit{resolve}}}$ \\

\midrule


\multicolumn{11}{c}{\cellcolor{FROZEN_BG}\textbf{\textit{SE Agent}: Devstral-Small-2507}} \\
\addlinespace[3pt]

\textcolor{gray}{\textbf{(Baseline)}}      & \textcolor{gray}{69.75} & \textcolor{gray}{62.50} & \textcolor{gray}{64.89} & \textcolor{gray}{51.10} & \textcolor{gray}{39.25} & \textcolor{gray}{32.43} & \textcolor{gray}{27.51} & \textcolor{gray}{27.23} & \textcolor{gray}{15.34} & \textcolor{gray}{10.01} \\

\arrayrulecolor{gray!50}\cmidrule(l){1-11}\arrayrulecolor{black}

\textcolor{reflectcolor}{\textbf{Claude-4-Sonnet}} & \absdeltaonlypct{69.75}{71.25} & \absdeltaonlypct{62.50}{64.00} & \absdeltaonlypct{64.89}{67.26} & \absdeltaonlypct{51.10}{52.11} & \absdeltaonlypct{39.25}{41.25} & \absdeltaonlypct{32.43}{35.26} & \absdeltaonlypct{27.51}{30.64} & \absdeltaonlypct{27.23}{30.98} & \absdeltaonlypct{15.34}{19.17} & \absdeltaonlypct{10.01}{13.75} \\

\arrayrulecolor{gray!50}\cmidrule(l){1-11}\arrayrulecolor{black}

\textcolor{notunecolor}{\textbf{DeepSeek-1.5B}} & \absdeltaonlypct{69.75}{66.25} & \absdeltaonlypct{62.50}{58.75} & \absdeltaonlypct{64.89}{62.67} & \absdeltaonlypct{51.10}{47.84} & \absdeltaonlypct{39.25}{34.75} & \absdeltaonlypct{32.43}{28.64} & \absdeltaonlypct{27.51}{24.22} & \absdeltaonlypct{27.23}{25.67} & \absdeltaonlypct{15.34}{13.62} & \absdeltaonlypct{10.01}{8.38} \\

\textcolor{notunecolor}{\textbf{DeepSeek-7B}} & \absdeltaonlypct{69.75}{68.00} & \absdeltaonlypct{62.50}{61.00} & \absdeltaonlypct{64.89}{62.83} & \absdeltaonlypct{51.10}{48.45} & \absdeltaonlypct{39.25}{36.00} & \absdeltaonlypct{32.43}{29.61} & \absdeltaonlypct{27.51}{24.98} & \absdeltaonlypct{27.23}{26.54} & \absdeltaonlypct{15.34}{14.36} & \absdeltaonlypct{10.01}{9.59} \\

\textcolor{notunecolor}{\textbf{Qwen2.5-3B}} & \absdeltaonlypct{69.75}{67.75} & \absdeltaonlypct{62.50}{60.50} & \absdeltaonlypct{64.89}{62.87} & \absdeltaonlypct{51.10}{48.60} & \absdeltaonlypct{39.25}{35.00} & \absdeltaonlypct{32.43}{29.72} & \absdeltaonlypct{27.51}{25.64} & \absdeltaonlypct{27.23}{25.17} & \absdeltaonlypct{15.34}{14.27} & \absdeltaonlypct{10.01}{8.98} \\

\textcolor{notunecolor}{\textbf{Qwen2.5-7B}} & \absdeltaonlypct{69.75}{69.25} & \absdeltaonlypct{62.50}{61.50} & \absdeltaonlypct{64.89}{63.92} & \absdeltaonlypct{51.10}{50.13} & \absdeltaonlypct{39.25}{37.50} & \absdeltaonlypct{32.43}{31.36} & \absdeltaonlypct{27.51}{26.67} & \absdeltaonlypct{27.23}{26.61} & \absdeltaonlypct{15.34}{14.81} & \absdeltaonlypct{10.01}{9.33} \\

\textcolor{notunecolor}{\textbf{Qwen3-4B}} & \absdeltaonlypct{69.75}{67.75} & \absdeltaonlypct{62.50}{61.25} & \absdeltaonlypct{64.89}{63.08} & \absdeltaonlypct{51.10}{48.33} & \absdeltaonlypct{39.25}{35.25} & \absdeltaonlypct{32.43}{29.86} & \absdeltaonlypct{27.51}{25.17} & \absdeltaonlypct{27.23}{24.82} & \absdeltaonlypct{15.34}{14.14} & \absdeltaonlypct{10.01}{8.77} \\

\textcolor{notunecolor}{\textbf{Qwen3-4B-T}} & \absdeltaonlypct{69.75}{68.25} & \absdeltaonlypct{62.50}{62.00} & \absdeltaonlypct{64.89}{63.76} & \absdeltaonlypct{51.10}{49.64} & \absdeltaonlypct{39.25}{36.00} & \absdeltaonlypct{32.43}{30.91} & \absdeltaonlypct{27.51}{26.95} & \absdeltaonlypct{27.23}{26.63} & \absdeltaonlypct{15.34}{14.75} & \absdeltaonlypct{10.01}{9.40} \\

\arrayrulecolor{gray!50}\cmidrule(l){1-11}\arrayrulecolor{black}

\textcolor{ourscolor}{\textbf{DeepSeek-1.5B (FT)}} & \absdeltaonlypct{69.75}{70.50} & \absdeltaonlypct{62.50}{63.00} & \absdeltaonlypct{64.89}{66.22} & \absdeltaonlypct{51.10}{51.73} & \absdeltaonlypct{39.25}{40.00} & \absdeltaonlypct{32.43}{33.62} & \absdeltaonlypct{27.51}{28.16} & \absdeltaonlypct{27.23}{28.57} & \absdeltaonlypct{15.34}{16.73} & \absdeltaonlypct{10.01}{12.21} \\

\textcolor{ourscolor}{\textbf{DeepSeek-7B (FT)}} & \absdeltaonlypct{69.75}{72.00} & \absdeltaonlypctbf{62.50}{65.50} & \absdeltaonlypct{64.89}{67.67} & \absdeltaonlypct{51.10}{54.11} & \absdeltaonlypct{39.25}{44.00} & \absdeltaonlypct{32.43}{33.78} & \absdeltaonlypct{27.51}{30.21} & \absdeltaonlypct{27.23}{29.74} & \absdeltaonlypct{15.34}{19.47} & \absdeltaonlypct{10.01}{13.82} \\

\textcolor{ourscolor}{\textbf{Qwen2.5-3B (FT)}} & \absdeltaonlypct{69.75}{70.75} & \absdeltaonlypct{62.50}{63.25} & \absdeltaonlypct{64.89}{65.92} & \absdeltaonlypct{51.10}{51.81} & \absdeltaonlypct{39.25}{40.00} & \absdeltaonlypct{32.43}{33.19} & \absdeltaonlypct{27.51}{28.14} & \absdeltaonlypct{27.23}{28.27} & \absdeltaonlypct{15.34}{16.23} & \absdeltaonlypct{10.01}{11.93} \\

\textcolor{ourscolor}{\textbf{Qwen2.5-7B (FT)}} & \absdeltaonlypct{69.75}{72.00} & \absdeltaonlypct{62.50}{64.50} & \absdeltaonlypct{64.89}{67.24} & \absdeltaonlypctbf{51.10}{55.74} & \absdeltaonlypctbf{39.25}{44.50} & \absdeltaonlypct{32.43}{34.17} & \absdeltaonlypct{27.51}{30.50} & \absdeltaonlypct{27.23}{29.44} & \absdeltaonlypct{15.34}{19.35} & \absdeltaonlypct{10.01}{12.68} \\

\textcolor{ourscolor}{\textbf{Qwen3-4B (FT)}} & \absdeltaonlypct{69.75}{71.75} & \absdeltaonlypct{62.50}{63.75} & \absdeltaonlypct{64.89}{67.34} & \absdeltaonlypct{51.10}{52.47} & \absdeltaonlypct{39.25}{41.00} & \absdeltaonlypct{32.43}{34.69} & \absdeltaonlypct{27.51}{28.67} & \absdeltaonlypct{27.23}{29.72} & \absdeltaonlypct{15.34}{18.80} & \absdeltaonlypct{10.01}{12.23} \\

\textcolor{ourscolor}{\textbf{Qwen3-4B-T (FT)}} & \absdeltaonlypctbf{69.75}{72.50} & \absdeltaonlypct{62.50}{64.25} & \absdeltaonlypctbf{64.89}{68.37} & \absdeltaonlypct{51.10}{52.55} & \absdeltaonlypct{39.25}{41.75} & \absdeltaonlypctbf{32.43}{35.60} & \absdeltaonlypctbf{27.51}{30.98} & \absdeltaonlypctbf{27.23}{31.14} & \absdeltaonlypctbf{15.34}{20.12} & \absdeltaonlypctbf{10.01}{14.64} \\[1pt]


\multicolumn{11}{c}{\cellcolor{FROZEN_BG}\textbf{\textit{SE Agent}: Qwen3-Coder-30B-A3B}} \\
\addlinespace[3pt]

\textcolor{gray}{\textbf{(Baseline)}}      & \textcolor{gray}{52.75} & \textcolor{gray}{48.50} & \textcolor{gray}{49.06} & \textcolor{gray}{40.42} & \textcolor{gray}{33.75} & \textcolor{gray}{33.25} & \textcolor{gray}{30.29} & \textcolor{gray}{28.69} & \textcolor{gray}{21.06} & \textcolor{gray}{17.96} \\

\arrayrulecolor{gray!50}\cmidrule(l){1-11}\arrayrulecolor{black}

\textcolor{reflectcolor}{\textbf{Claude-4-Sonnet}} & \absdeltaonlypct{52.75}{56.75} & \absdeltaonlypct{48.50}{51.25} & \absdeltaonlypct{49.06}{53.25} & \absdeltaonlypct{40.42}{43.12} & \absdeltaonlypct{33.75}{36.00} & \absdeltaonlypct{33.25}{38.27} & \absdeltaonlypct{30.29}{34.43} & \absdeltaonlypct{28.69}{33.76} & \absdeltaonlypct{21.06}{25.42} & \absdeltaonlypct{17.96}{21.74} \\

\arrayrulecolor{gray!50}\cmidrule(l){1-11}\arrayrulecolor{black}

\textcolor{notunecolor}{\textbf{DeepSeek-1.5B}} & \absdeltaonlypct{52.75}{48.00} & \absdeltaonlypct{48.50}{44.50} & \absdeltaonlypct{49.06}{44.75} & \absdeltaonlypct{40.42}{38.82} & \absdeltaonlypct{33.75}{31.00} & \absdeltaonlypct{33.25}{30.78} & \absdeltaonlypct{30.29}{28.15} & \absdeltaonlypct{28.69}{26.23} & \absdeltaonlypct{21.06}{19.14} & \absdeltaonlypct{17.96}{15.87} \\

\textcolor{notunecolor}{\textbf{DeepSeek-7B}} & \absdeltaonlypct{52.75}{51.00} & \absdeltaonlypct{48.50}{47.00} & \absdeltaonlypct{49.06}{47.78} & \absdeltaonlypct{40.42}{38.37} & \absdeltaonlypct{33.75}{32.00} & \absdeltaonlypct{33.25}{31.31} & \absdeltaonlypct{30.29}{28.84} & \absdeltaonlypct{28.69}{27.62} & \absdeltaonlypct{21.06}{19.53} & \absdeltaonlypct{17.96}{15.72} \\

\textcolor{notunecolor}{\textbf{Qwen2.5-3B}} & \absdeltaonlypct{52.75}{48.50} & \absdeltaonlypct{48.50}{44.50} & \absdeltaonlypct{49.06}{44.82} & \absdeltaonlypct{40.42}{38.02} & \absdeltaonlypct{33.75}{31.00} & \absdeltaonlypct{33.25}{30.34} & \absdeltaonlypct{30.29}{27.21} & \absdeltaonlypct{28.69}{26.76} & \absdeltaonlypct{21.06}{18.58} & \absdeltaonlypct{17.96}{15.68} \\

\textcolor{notunecolor}{\textbf{Qwen2.5-7B}} & \absdeltaonlypct{52.75}{50.00} & \absdeltaonlypct{48.50}{47.25} & \absdeltaonlypct{49.06}{47.33} & \absdeltaonlypct{40.42}{39.57} & \absdeltaonlypct{33.75}{33.00} & \absdeltaonlypct{33.25}{32.31} & \absdeltaonlypct{30.29}{28.94} & \absdeltaonlypct{28.69}{27.72} & \absdeltaonlypct{21.06}{20.03} & \absdeltaonlypct{17.96}{15.92} \\

\textcolor{notunecolor}{\textbf{Qwen3-4B}} & \absdeltaonlypct{52.75}{48.25} & \absdeltaonlypct{48.50}{44.75} & \absdeltaonlypct{49.06}{44.73} & \absdeltaonlypct{40.42}{37.92} & \absdeltaonlypct{33.75}{32.00} & \absdeltaonlypct{33.25}{30.27} & \absdeltaonlypct{30.29}{27.42} & \absdeltaonlypct{28.69}{26.14} & \absdeltaonlypct{21.06}{18.77} & \absdeltaonlypct{17.96}{15.49} \\

\textcolor{notunecolor}{\textbf{Qwen3-4B-T}} & \absdeltaonlypct{52.75}{49.25} & \absdeltaonlypct{48.50}{45.50} & \absdeltaonlypct{49.06}{45.87} & \absdeltaonlypct{40.42}{39.77} & \absdeltaonlypct{33.75}{32.75} & \absdeltaonlypct{33.25}{31.64} & \absdeltaonlypct{30.29}{28.61} & \absdeltaonlypct{28.69}{27.52} & \absdeltaonlypct{21.06}{19.85} & \absdeltaonlypct{17.96}{16.11} \\

\arrayrulecolor{gray!50}\cmidrule(l){1-11}\arrayrulecolor{black}

\textcolor{ourscolor}{\textbf{DeepSeek-1.5B (FT)}} & \absdeltaonlypct{52.75}{54.00} & \absdeltaonlypct{48.50}{51.00} & \absdeltaonlypct{49.06}{51.33} & \absdeltaonlypct{40.42}{42.54} & \absdeltaonlypct{33.75}{35.00} & \absdeltaonlypct{33.25}{32.98} & \absdeltaonlypct{30.29}{31.65} & \absdeltaonlypct{28.69}{30.52} & \absdeltaonlypct{21.06}{22.78} & \absdeltaonlypct{17.96}{19.64} \\

\textcolor{ourscolor}{\textbf{DeepSeek-7B (FT)}} & \absdeltaonlypct{52.75}{54.75} & \absdeltaonlypct{48.50}{51.50} & \absdeltaonlypct{49.06}{52.68} & \absdeltaonlypct{40.42}{43.23} & \absdeltaonlypct{33.75}{36.25} & \absdeltaonlypct{33.25}{36.32} & \absdeltaonlypct{30.29}{34.17} & \absdeltaonlypct{28.69}{33.73} & \absdeltaonlypct{21.06}{23.11} & \absdeltaonlypct{17.96}{19.38} \\

\textcolor{ourscolor}{\textbf{Qwen2.5-3B (FT)}} & \absdeltaonlypct{52.75}{53.50} & \absdeltaonlypct{48.50}{49.00} & \absdeltaonlypct{49.06}{50.00} & \absdeltaonlypct{40.42}{41.04} & \absdeltaonlypct{33.75}{34.25} & \absdeltaonlypct{33.25}{33.44} & \absdeltaonlypct{30.29}{31.98} & \absdeltaonlypct{28.69}{29.51} & \absdeltaonlypct{21.06}{21.88} & \absdeltaonlypct{17.96}{18.65} \\

\textcolor{ourscolor}{\textbf{Qwen2.5-7B (FT)}} & \absdeltaonlypct{52.75}{55.25} & \absdeltaonlypctbf{48.50}{52.00} & \absdeltaonlypct{49.06}{52.92} & \absdeltaonlypctbf{40.42}{43.98} & \absdeltaonlypctbf{33.75}{37.25} & \absdeltaonlypct{33.25}{37.18} & \absdeltaonlypct{30.29}{34.66} & \absdeltaonlypct{28.69}{33.84} & \absdeltaonlypct{21.06}{24.88} & \absdeltaonlypct{17.96}{21.57} \\

\textcolor{ourscolor}{\textbf{Qwen3-4B (FT)}} & \absdeltaonlypct{52.75}{54.75} & \absdeltaonlypct{48.50}{49.75} & \absdeltaonlypct{49.06}{51.43} & \absdeltaonlypct{40.42}{41.88} & \absdeltaonlypct{33.75}{34.75} & \absdeltaonlypct{33.25}{36.28} & \absdeltaonlypct{30.29}{32.71} & \absdeltaonlypct{28.69}{31.98} & \absdeltaonlypct{21.06}{23.15} & \absdeltaonlypct{17.96}{19.37} \\

\textcolor{ourscolor}{\textbf{Qwen3-4B-T (FT)}} & \absdeltaonlypctbf{52.75}{57.00} & \absdeltaonlypct{48.50}{51.75} & \absdeltaonlypctbf{49.06}{54.50} & \absdeltaonlypct{40.42}{43.63} & \absdeltaonlypct{33.75}{36.00} & \absdeltaonlypctbf{33.25}{38.75} & \absdeltaonlypctbf{30.29}{35.09} & \absdeltaonlypctbf{28.69}{34.62} & \absdeltaonlypctbf{21.06}{26.18} & \absdeltaonlypctbf{17.96}{22.06} \\

\bottomrule
\end{tabularx}

\end{table}

%% file: sections/4_exps.tex
\section{Experiments}
\label{sec:exps}

\subsection{Setup}
\label{subsec:exp_setup}

\textbf{Model.}
For SE agents, we evaluate two LLMs, including \texttt{Devstral-Small-2507} \citep{devstral2025} and \texttt{Qwen3-Coder-30B-A3B} \citep{qwen3technicalreport}.
During training dataset construction (\S\ref{subsec:rejection_sampling}), we employ \texttt{Claude-4-Sonnet} \citep{Claude4Sonnet} as $M_\theta$ to generate memory candidates.
For \ours, we finetune six LLMs for memory reflection, including \texttt{Qwen2.5} (\texttt{3B} \& \texttt{7B})~\citep{qwen25technicalreport}, \texttt{DeepSeek-R1} (\texttt{1.5B} \& \texttt{7B})~\citep{deepseekr12025}, together with \texttt{Qwen3-4B} and \texttt{Qwen3-4B-Thinking}~\citep{qwen3technicalreport}.



%
%
%

\textbf{Data.}
We leverage SWE-Bench-\textit{Verified} \citep{jimenez2024swebench} for training and evaluation.
(1) For training dataset construction (\S\ref{subsec:rejection_sampling}), we randomly sample $N_{\mathcal{T}}=10$ tasks from each of $N_{\mathcal{R}}=10$ repositories, yielding $N_{\mathcal{R}} \times N_{\mathcal{T}} = 100$ training tasks. We employ two LLMs (\texttt{Devstral-Small-2505}~\citep{devstral2025} and \texttt{Qwen3-Coder-480B-A35B}~\cite{qwen3technicalreport}) to power the SE agents on these $100$ tasks, where we generate $N_{\tau}=4$ trajectories per task and $N_{\mathcal{M}}=4$ memory candidates per trajectory (Alg.~\ref{alg:rejection_sampling}), contributing to $100 \times 2 \times 4 \times 4 = 3,200$ memory candidates. Through quality filtering (\S\ref{appendix:sec:dataset_construction}), we summarize our datasets in Tab.~\ref{tab:training_data_overview}.
(2) During evaluation (\S\ref{subsec:two_eval_regimes}), we assess \ours in both \textit{single-episode} and \textit{cross-episode} settings. To avoid evaluation circularity (\S\ref{appendix:sec:dataset_construction}), we randomly sample $100$ evaluation instances (Fig.~\ref{fig:eval_dataset_distribution}) non-overlapped with the $100$ training instances (\S\ref{subsec:rejection_sampling}). We compute \textit{avg@4} across all metrics for evaluation accuracy.
(3) To evaluate \ours's transferability to new programming languages and SE activities (\S\ref{subsec:exps:transferability}), we further evaluate on Multi-SWE-Bench~\citep{zan2025multiswebench} (C and Java), Commit0-Lite~\citep{zhao2024commit0}, and TestGenEval~\citep{jain2025testgeneval} ($100$ randomly sampled tasks), with each scored under its own native evaluation criteria.

%% file: sections/5_results.tex
\input{figures/exp_ablation_sft_rl}

\subsection{Effectiveness of \ours: Self-Evolving Memory Augmentation for SE Agent}
\label{subsec:exps:effectiveness}

\input{tables/exp_main_cross_episode}

\paragraph{\ours consistently improves SE agents in \textit{single-episode} software engineering.}
Compared to \textit{\textcolor{gray}{no-$M_\theta$ baselines}} (Tab.~\ref{tab:exp:single_episode}), \ours-augmented SE agents consistently outperform across ten metrics (\S\ref{subsec:eval_metrics}), achieving absolute gains of up to $\uparrow 5.25\%$ in $SR$ and $\uparrow 4.63\%$ in $E_{\textit{resolve}}$. \texttt{Claude-4-Sonnet} serves as a strong \textcolor{reflectcolor}{\textit{reflection baseline}}: a frontier model distilling lessons-learned memory under the same format and generation instruction as \ours, but without rejection sampling, utility modeling, or training. Most \ours-FT $M_\theta$ variants surpass it by up to $\Delta_{abs}=\uparrow3.25\%$ in $SR$, while being served locally at $1.5$B--$7$B scale instead of \$$53.89$ per $100$ instances (Tab.~\ref{tab:exp:computation_overhead}). In contrast, all six \textcolor{notunecolor}{NFT-$M_\theta$} variants degrade SE performance, indicating that, without \ours, base models alone are insufficient for effective SE memory augmentation. Collectively, \ours exhibits significant improvements on single-episode SE, underscoring the effectiveness of \textit{memory reflection} in enhancing intra-episode reasoning and problem-solving.

\paragraph{\ours generalizes effectively to \textit{cross-episode} software engineering.}
Tab.~\ref{tab:exp:cross_episode} shows consistent performance gains over \textit{\textcolor{gray}{baseline no-$M_\theta$}} SE agents, with absolute improvements of up to $\uparrow3.00\%$ in $SR$ and $\uparrow3.17\%$ in $LA$. Cross-episode reflective memory evolution is inherently challenging, as $M_\theta$ needs to determine which information to retain and share across episodes versus what to update. As a result, \textcolor{notunecolor}{NFT-$M_\theta$} exhibits degraded performance compared to their single-episode counterparts (Tab.~\ref{tab:exp:single_episode}). In contrast, \ours demonstrates strong adaptability in cross-episode reflective memory evolution, allowing $M_\theta$ to accumulate repository-level shared knowledge base while updating critical information. As shown in Tab.~\ref{tab:exp:cross_episode}, \ours-powered $M_\theta$ variants outperform both \textit{\textcolor{gray}{no-$M_\theta$ baselines}} ($\Delta_{abs} \geq 1.00\%$ in $SR$) and the \textcolor{notunecolor}{NFT-$M_\theta$} of the same backbone ($\Delta_{abs} \geq 5.50\%$ in $SR$) across evolution episodes, most of which surpass the \textcolor{reflectcolor}{\textit{reflection baseline}} \texttt{Claude-4-Sonnet} by up to $\Delta_{abs}=\uparrow1.75\%$ in $SR$. \texttt{Qwen3-4B} also showcases robust reflective memory evolution ability, presenting $\uparrow1.50\%$ higher than its single-episode $SR$. These observations highlight the effectiveness of \ours in progressively refining and evolving knowledge across episodes to enable more informed problem-solving and improved long-horizon software engineering.

\paragraph{\ours resolves more tasks while spending less per task.}
Because \ours both raises $SR$ and shortens problem-solving trajectories, the informative quantity is cost per resolved instance rather than raw cost. As shown in Tab.~\ref{tab:exp:deployment_cost}, deployment cost per resolved instance falls from \$$0.299$ to \$$0.246$ on \texttt{Devstral-Small-2507} ($\Delta_{rel}=\downarrow17.77\%$) and from \$$0.295$ to \$$0.248$ on \texttt{Qwen3-Coder-30B-A3B} ($\Delta_{rel}=\downarrow15.94\%$). Both exceed the corresponding reductions in raw API cost ($\Delta_{rel}=\downarrow12.53\%$ and $\downarrow10.33\%$), as the accuracy gain and the cost saving compound. \ours therefore improves accuracy and cost together, rather than purchasing accuracy with additional iterations.

\paragraph{Construction and training are one-time costs that amortize across settings.}
Tab.~\ref{tab:exp:computation_overhead} (\S\ref{appendix:subsec:computation_overhead}) itemizes the construction API cost for both SE agents, the memory-generation cost, and the GPU allocation for Stage~I and Stage~II. These are incurred \textit{once} and yield a reusable $M_\theta$, whereas the deployment saving above recurs on every task. Critically, the resulting $M_\theta$ is neither per-task, per-repository, nor per-agent: each \ours-FT $M_\theta$ can serve different SE agents (Tabs.~\ref{tab:exp:single_episode}-\ref{tab:exp:cross_episode}), and \ours is not only generalizable across varying backbones and algorithms (\S\ref{subsec:exps:effectiveness}), but also transferable across different benchmarks and programming languages (\S\ref{subsec:exps:transferability}). Therefore, the one-time cost amortizes over all of them. At deployment, $M_\theta$ is a $1.5$B-$7$B model served locally alongside the SE agent, and the SE agent's own cost decreases at the same time. Consequently, the recurring cost of \ours stays below the \textit{\textcolor{gray}{no-$M_\theta$ baseline}} from the first task onward.

\subsection{Generalizability of \ours: Different SE Agents, Backbones, and Algorithms}
\label{subsec:exps:generalizability}

\paragraph{\ours is generalizable across different SE agents and $M_\theta$ backbones.}
As shown in Tabs.~\ref{tab:exp:single_episode}-\ref{tab:exp:cross_episode}, for both \texttt{Devstral-Small-2507} and \texttt{Qwen3-Coder-30B-A3B}, \ours-augmented SE agents consistently outperform \textit{\textcolor{gray}{no-$M_\theta$ baselines}} and six \textcolor{notunecolor}{NFT-$M_\theta$} across \textit{single-episode} and \textit{cross-episode} scenarios. Against the \textcolor{notunecolor}{NFT-$M_\theta$} of the \textit{same backbone}, \textcolor{ourscolor}{\ours-FT $M_\theta$} achieves $\Delta_{\textit{abs}}$ of up to $\uparrow$8.00\% in $SR$ and $\uparrow$5.24\% in $E_{\textit{resolve}}$ for \texttt{Devstral-Small-2507}, and $\uparrow$6.75\% in $SR$ and $\uparrow$6.56\% in $E_{\textit{resolve}}$ for \texttt{Qwen3-} \texttt{Coder-30B-A3B}.
Meanwhile, deployment cost is notably reduced by $\Delta_{\textit{rel}}=$12.53\% for \texttt{Devstral-Small-2507} and $\Delta_{\textit{rel}}=$10.33\% for \texttt{Qwen3-Coder-30B-A3B} (Tab.~\ref{tab:exp:computation_overhead}).

\paragraph{All training stages contribute to performance improvements.}
As shown in Fig.~\ref{fig:exp_ablation_sft_rl}, both training stages (\S\ref{subsec:two_stage_sft_rl_training})
independently contribute to SE performance gains, where Stage I equips $M_\theta$ with foundational generation capabilities, while Stage II further refines memory quality via comparative 
performance signals. Notably, while Stage I or II alone independently improves SE performance, Stage I+II yields superior gains, demonstrating their complementary roles in memory optimization.

\input{tables/exp_deployment_cost}
\input{figures/exp_ablation_rl_algorithm}

\paragraph{\ours is generalizable across repositories.}
Compared with \textit{\textcolor{gray}{no-$M_\theta$ baselines}}, Fig.~\ref{fig:repo_wise_delta} highlights the generalizability of \ours to different SE repositories. 
Specifically, by evaluating repository-wise performance across 9 repositories and 100 instances with and without \ours, respectively, results demonstrate consistent performance gains across ten metrics, underscoring the generalizability and robustness of \ours across diverse repository structures and contexts.

\paragraph{\ours is also useful when applying different RL algorithms.}
As shown in Fig.~\ref{fig:exp_ablation_rl_algorithm}, by adapting three RL algorithms, including GRPO~\citep{2024grpo}, DAPO~\citep{2025dapo}, and GSPO~\citep{2025gspo}, \ours yields consistent gains over both \textit{\textcolor{gray}{no-$M_\theta$ baselines}} and \textcolor{notunecolor}{NFT-$M_\theta$}, with $\Delta_{abs}$ of up to $\uparrow3.50\%$ on SR and $\uparrow4.03\%$ on $LA$. These demonstrate that \ours is algorithm-agnostic and can be seamlessly integrated with various algorithms for improved reflective memory modeling.

\subsection{Transferability of \ours: New Languages and SE Activities}
\label{subsec:exps:transferability}

Memory content is repository-specific by design: repository layout, build and test commands, and project conventions are useful precisely because they are local (\S\ref{appendix:subsec:qualitative_analysis}). What generalizes (\S\ref{subsec:exps:generalizability}) and transfers (\S\ref{subsec:exps:transferability}) is therefore not the memory itself but $M_\theta$, the ability to distill a useful memory from a trajectory. To isolate this, we apply \ours-FT $M_\theta$ (\texttt{Qwen3-4B-T}), with neither retraining nor prompt re-tuning, to benchmarks, programming languages, and SE activities absent from the $100$ training tasks (\S\ref{subsec:rejection_sampling}).
Different from \S\ref{subsec:exps:generalizability}, which varies how \ours is built, here we keep \ours-FT $M_\theta$ unchanged and apply it directly to new programming languages and SE activities.
%

\input{tables/exp_transfer_cross_language}
\input{tables/exp_transfer_language}

\paragraph{\ours transfers to unseen programming languages.}
As shown in Tab.~\ref{tab:exp:transfer_language}, \ours-FT $M_\theta$ improves $SR$ on Multi-SWE-Bench~\citep{zan2025multiswebench} in both C ($\Delta_{abs}=\uparrow2.50\%$) and Java ($\Delta_{abs}=\uparrow4.04\%$), neither of which appears in training. The Java gain exceeds the in-distribution Python gain on SWE-Bench-\textit{Verified} ($\Delta_{abs}=\uparrow2.50\%$), indicating that what $M_\theta$ acquires is a language-agnostic reflection mechanism rather than memorized Python repository content.

\paragraph{\ours transfers to SE activities beyond issue resolution.}
Tab.~\ref{tab:exp:transfer_activity} evaluates two benchmarks that score entirely different SE activities: Commit0-Lite~\citep{zhao2024commit0} scores from-scratch library implementation, and TestGenEval~\citep{jain2025testgeneval} scores unit-test generation and completion. \ours improves the native metrics of both benchmarks, including the resolve rate of Commit0-Lite ($\Delta_{abs}=\uparrow3.12\%$) and the all Pass@1 of TestGenEval ($\Delta_{abs}=\uparrow3.50\%$). Notably, each benchmark is scored under \textit{its own} criteria that $M_\theta$ is never optimized against, \textit{e.g.}, TestGenEval's code coverage ($\Delta_{abs}=\uparrow8.38\%$) lying outside our acceptance criteria (Eq.~\ref{eq:memory_acceptance}) entirely. This demonstrates that the knowledge \ours distills is broadly useful for software engineering, benefiting library implementation and unit-test generation as effectively as issue resolution, and improving criteria well beyond those it directly optimizes toward.

\paragraph{Gains hold across all ten metrics in every language.}
Tab.~\ref{tab:exp:transfer_cross_language} applies the same \ours-FT $M_\theta$ to a second SE agent powered by \texttt{Qwen3-Coder-30B-A3B} and evaluates on our full metric suite (\S\ref{subsec:eval_metrics}) for all three languages. Scores across all thirty language-metrics are improved, ranging from $\Delta_{abs}=\uparrow2.25\%$ to $\Delta_{abs}=\uparrow5.93\%$, with no degradation on any metric in any language. This uniform improvement across both accuracy and efficiency shows that \ours-FT $M_\theta$ transfers as a complete capability rather than a language-specific effect, guiding the SE agent to the right targets and reaching them in fewer iterations regardless of which language the repository is written in.

\input{tables/exp_transfer_activity}

%% file: figures/exp_ablation_sft_rl.tex
\begin{figure}[t]

\vspace{12pt}

\centering
\includegraphics[width=1.0\linewidth]{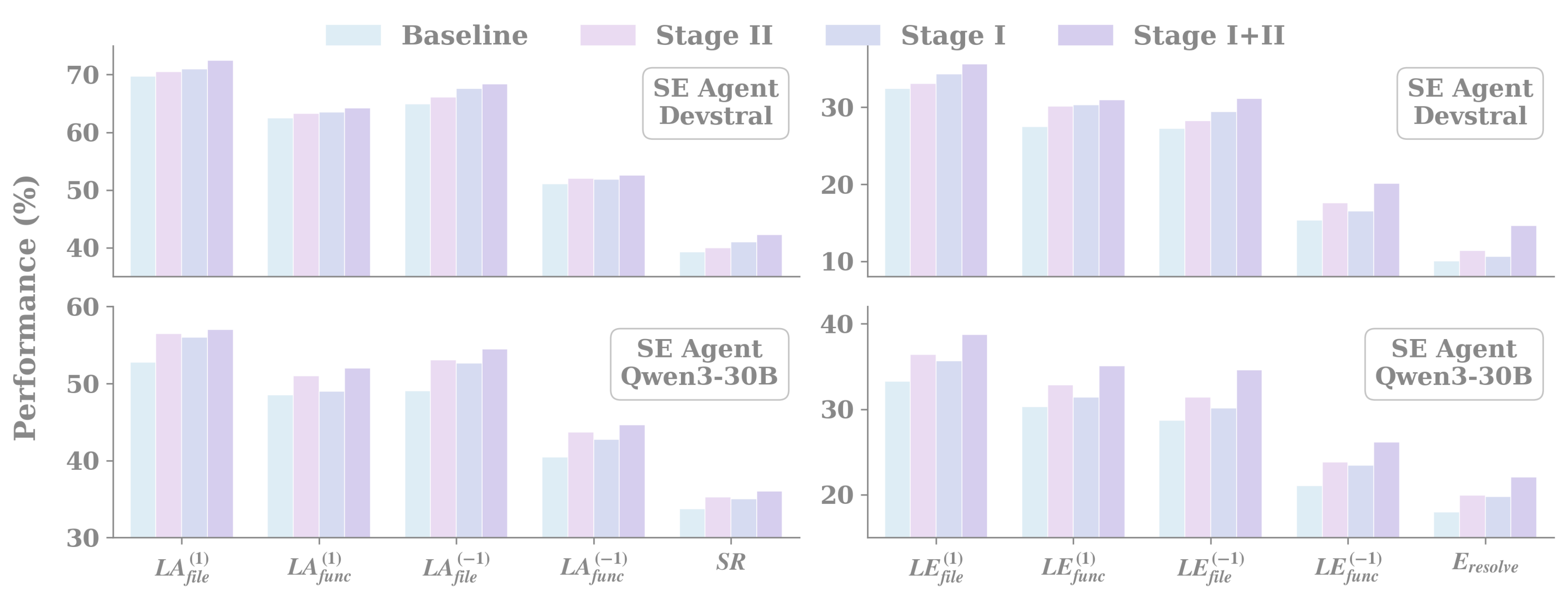}

\caption{\textbf{Effects of Two-Stage Finetuning on $M_\theta$.} We compare SE agent performance across our ten metrics (\S\ref{subsec:eval_metrics}) with $M_\theta$ trained through Stage~I (SFT) alone, Stage~II (RL) alone, and Stage~I+II (\S\ref{subsec:two_stage_sft_rl_training}), against the \textit{\textcolor{gray}{no-$M_\theta$ baseline}}, on SE agents powered by \texttt{Devstral-Small-2507} (\textit{top}) and \texttt{Qwen3-Coder-30B-A3B} (\textit{bottom}), respectively. Each stage independently improves over the baseline, while Stage~I+II attains the largest gains, indicating that the two stages supply complementary optimization signals.}
\label{fig:exp_ablation_sft_rl}

\vspace{18pt}

\end{figure}

%% file: tables/exp_main_cross_episode.tex

\renewcommand{\absdeltaonlypct}[2]{%
  \pgfmathsetmacro{\deltares}{#2-#1}%
  \pgfmathtruncatemacro{\deltawhole}{int(abs(\deltares))}%
  \pgfmathtruncatemacro{\deltadec}{int(round((abs(\deltares)-\deltawhole)*100))}%
  \ifnum\deltadec<10\edef\deltadecstr{0\deltadec}\else\edef\deltadecstr{\deltadec}\fi%
  \ifdim\deltares pt>0pt%
    \pgfmathsetmacro{\deltaint}{int(min(100, abs(\deltares) * 10))}%
    \edef\deltacolor{deltagreen!\deltaint!white}%
    \expandafter\cellcolor\expandafter{\deltacolor}\small\rmfamily\textcolor{deltagreen}{↑\deltawhole.\deltadecstr}%
  \else%
    \pgfmathsetmacro{\deltaint}{int(min(100, abs(\deltares) * 4))}%
    \edef\deltacolor{deltared!\deltaint!white}%
    \expandafter\cellcolor\expandafter{\deltacolor}\small\rmfamily\textcolor{deltared}{↓\deltawhole.\deltadecstr}%
  \fi%
}

\renewcommand{\absdeltaonlypctbf}[2]{%
  \pgfmathsetmacro{\deltares}{#2-#1}%
  \pgfmathtruncatemacro{\deltawhole}{int(abs(\deltares))}%
  \pgfmathtruncatemacro{\deltadec}{int(round((abs(\deltares)-\deltawhole)*100))}%
  \ifnum\deltadec<10\edef\deltadecstr{0\deltadec}\else\edef\deltadecstr{\deltadec}\fi%
  \ifdim\deltares pt>0pt%
    \pgfmathsetmacro{\deltaint}{int(min(100, abs(\deltares) * 10))}%
    \edef\deltacolor{deltagreen!\deltaint!white}%
    \begingroup\rmfamily\bfseries\small\expandafter\cellcolor\expandafter{\deltacolor}\textcolor{deltagreen}{↑\deltawhole.\deltadecstr}\endgroup%
  \else%
    \pgfmathsetmacro{\deltaint}{int(min(100, abs(\deltares) * 4))}%
    \edef\deltacolor{deltared!\deltaint!white}%
    \begingroup\rmfamily\bfseries\small\expandafter\cellcolor\expandafter{\deltacolor}\textcolor{deltared}{↓\deltawhole.\deltadecstr}\endgroup%
  \fi%
}

\begin{table}[t]

\small
\centering
\setlength{\tabcolsep}{0pt}
\renewcommand{\arraystretch}{1.1}

\caption{\textbf{Cross-Episode Memory Augmentation.} Cross-episode accuracy and efficiency of SE agents (\S\ref{subsec:eval_metrics}). We compare \textcolor{gray}{\textbf{no-$M_\theta$ baselines}} against three memory conditions: (1) a \textcolor{reflectcolor}{\textbf{reflection baseline}} (\texttt{Claude-4-Sonnet}), which applies the same memory format and generation instruction (Fig.~\ref{fig:two_prompt_for_single_and_cross_episode}) without utility filtering or training; (2) \textcolor{notunecolor}{\textbf{non-finetuned $M_\theta$ (NFT)}}; and (3) \textcolor{ourscolor}{\textbf{\ours-finetuned $M_\theta$ (FT)}}. Here, \textbf{\textit{absolute performance differences}} (Eq.~\ref{eq:eval_metric:deltaabs}) are shown as \textcolor{deltagreen}{$\uparrow \Delta_{\textit{abs}}\%$} or \textcolor{deltared}{$\downarrow \Delta_{\textit{abs}}\%$}.}
\label{tab:exp:cross_episode}

\begin{tabularx}{\textwidth}{c*{10}{>{\centering\arraybackslash}X}}

\toprule

\multirow{3}{*}{$\mathbf{\textit{M}}_{\theta}$} & \multicolumn{5}{c}{\textbf{Accuracy (\%)}} & \multicolumn{5}{c}{\textbf{Efficiency (\%)}} \\
\cmidrule(lr){2-6} \cmidrule(lr){7-11}
& $\bm{\textbf{\textit{LA}}_{\textit{file}}^{(1)}}$ & $\bm{\textbf{\textit{LA}}_{\textit{func}}^{(1)}}$ & $\bm{\textbf{\textit{LA}}_{\textit{file}}^{(-1)}}$ & $\bm{\textbf{\textit{LA}}_{\textit{func}}^{(-1)}}$ & \textbf{\textit{SR}}
& $\bm{\textbf{\textit{LE}}_{\textit{file}}^{(1)}}$ & $\bm{\textbf{\textit{LE}}_{\textit{func}}^{(1)}}$ & $\bm{\textbf{\textit{LE}}_{\textit{file}}^{(-1)}}$ & $\bm{\textbf{\textit{LE}}_{\textit{func}}^{(-1)}}$ & $\bm{E_{\textit{resolve}}}$ \\

\midrule


\multicolumn{11}{c}{\cellcolor{FROZEN_BG}\textbf{\textit{SE Agent}: Devstral-Small-2507}} \\
\addlinespace[3pt]

\textcolor{gray}{\textbf{(Baseline)}}      & \textcolor{gray}{69.75} & \textcolor{gray}{62.50} & \textcolor{gray}{64.89} & \textcolor{gray}{51.10} & \textcolor{gray}{39.25} & \textcolor{gray}{32.43} & \textcolor{gray}{27.51} & \textcolor{gray}{27.23} & \textcolor{gray}{15.34} & \textcolor{gray}{10.01} \\

\arrayrulecolor{gray!50}\cmidrule(l){1-11}\arrayrulecolor{black}

\textcolor{reflectcolor}{\textbf{Claude-4-Sonnet}} & \absdeltaonlypct{69.75}{70.50} & \absdeltaonlypct{62.50}{63.75} & \absdeltaonlypct{64.89}{66.36} & \absdeltaonlypct{51.10}{51.78} & \absdeltaonlypct{39.25}{40.50} & \absdeltaonlypct{32.43}{34.78} & \absdeltaonlypct{27.51}{29.85} & \absdeltaonlypct{27.23}{29.64} & \absdeltaonlypct{15.34}{17.41} & \absdeltaonlypct{10.01}{12.64} \\

\arrayrulecolor{gray!50}\cmidrule(l){1-11}\arrayrulecolor{black}

\textcolor{notunecolor}{\textbf{DeepSeek-7B}} & \absdeltaonlypct{69.75}{63.75} & \absdeltaonlypct{62.50}{55.25} & \absdeltaonlypct{64.89}{60.35} & \absdeltaonlypct{51.10}{44.73} & \absdeltaonlypct{39.25}{34.00} & \absdeltaonlypct{32.43}{24.36} & \absdeltaonlypct{27.51}{23.43} & \absdeltaonlypct{27.23}{21.76} & \absdeltaonlypct{15.34}{11.75} & \absdeltaonlypct{10.01}{5.52} \\

\textcolor{notunecolor}{\textbf{Qwen2.5-7B}} & \absdeltaonlypct{69.75}{66.50} & \absdeltaonlypct{62.50}{57.25} & \absdeltaonlypct{64.89}{62.37} & \absdeltaonlypct{51.10}{47.72} & \absdeltaonlypct{39.25}{35.75} & \absdeltaonlypct{32.43}{28.42} & \absdeltaonlypct{27.51}{25.24} & \absdeltaonlypct{27.23}{26.62} & \absdeltaonlypct{15.34}{13.11} & \absdeltaonlypct{10.01}{8.36} \\

\textcolor{notunecolor}{\textbf{Qwen3-4B}} & \absdeltaonlypct{69.75}{64.50} & \absdeltaonlypct{62.50}{55.25} & \absdeltaonlypct{64.89}{60.84} & \absdeltaonlypct{51.10}{44.78} & \absdeltaonlypct{39.25}{34.50} & \absdeltaonlypct{32.43}{27.65} & \absdeltaonlypct{27.51}{24.87} & \absdeltaonlypct{27.23}{24.37} & \absdeltaonlypct{15.34}{13.38} & \absdeltaonlypct{10.01}{8.06} \\

\arrayrulecolor{gray!50}\cmidrule(l){1-11}\arrayrulecolor{black}

\textcolor{ourscolor}{\textbf{DeepSeek-7B (FT)}} & \absdeltaonlypct{69.75}{70.75} & \absdeltaonlypct{62.50}{62.75} & \absdeltaonlypct{64.89}{65.37} & \absdeltaonlypct{51.10}{51.52} & \absdeltaonlypct{39.25}{40.25} & \absdeltaonlypct{32.43}{33.07} & \absdeltaonlypct{27.51}{27.98} & \absdeltaonlypct{27.23}{27.88} & \absdeltaonlypct{15.34}{15.73} & \absdeltaonlypct{10.01}{10.69} \\

\textcolor{ourscolor}{\textbf{Qwen2.5-7B (FT)}} & \absdeltaonlypct{69.75}{71.00} & \absdeltaonlypct{62.50}{63.50} & \absdeltaonlypct{64.89}{66.08} & \absdeltaonlypct{51.10}{51.96} & \absdeltaonlypctbf{39.25}{42.00} & \absdeltaonlypctbf{32.43}{34.79} & \absdeltaonlypctbf{27.51}{30.25} & \absdeltaonlypct{27.23}{28.75} & \absdeltaonlypct{15.34}{17.77} & \absdeltaonlypctbf{10.01}{12.74} \\

\textcolor{ourscolor}{\textbf{Qwen3-4B (FT)}} & \absdeltaonlypctbf{69.75}{71.50} & \absdeltaonlypctbf{62.50}{63.75} & \absdeltaonlypctbf{64.89}{66.58} & \absdeltaonlypctbf{51.10}{51.88} & \absdeltaonlypct{39.25}{41.00} & \absdeltaonlypct{32.43}{34.71} & \absdeltaonlypct{27.51}{29.32} & \absdeltaonlypctbf{27.23}{29.58} & \absdeltaonlypctbf{15.34}{17.93} & \absdeltaonlypct{10.01}{11.34} \\[1pt]


\multicolumn{11}{c}{\cellcolor{FROZEN_BG}\textbf{\textit{SE Agent}: Qwen3-Coder-30B-A3B}} \\
\addlinespace[3pt]

\textcolor{gray}{\textbf{(Baseline)}}      & \textcolor{gray}{52.75} & \textcolor{gray}{48.50} & \textcolor{gray}{49.06} & \textcolor{gray}{40.42} & \textcolor{gray}{33.75} & \textcolor{gray}{33.25} & \textcolor{gray}{30.29} & \textcolor{gray}{28.69} & \textcolor{gray}{21.06} & \textcolor{gray}{17.96} \\

\arrayrulecolor{gray!50}\cmidrule(l){1-11}\arrayrulecolor{black}

\textcolor{reflectcolor}{\textbf{Claude-4-Sonnet}} & \absdeltaonlypct{52.75}{54.50} & \absdeltaonlypct{48.50}{50.00} & \absdeltaonlypct{49.06}{50.76} & \absdeltaonlypct{40.42}{41.86} & \absdeltaonlypct{33.75}{35.00} & \absdeltaonlypct{33.25}{34.62} & \absdeltaonlypct{30.29}{31.14} & \absdeltaonlypct{28.69}{30.61} & \absdeltaonlypct{21.06}{22.63} & \absdeltaonlypct{17.96}{18.66} \\

\arrayrulecolor{gray!50}\cmidrule(l){1-11}\arrayrulecolor{black}

\textcolor{notunecolor}{\textbf{DeepSeek-7B}} & \absdeltaonlypct{52.75}{46.50} & \absdeltaonlypct{48.50}{40.75} & \absdeltaonlypct{49.06}{42.64} & \absdeltaonlypct{40.42}{33.26} & \absdeltaonlypct{33.75}{29.25} & \absdeltaonlypct{33.25}{26.64} & \absdeltaonlypct{30.29}{23.66} & \absdeltaonlypct{28.69}{22.67} & \absdeltaonlypct{21.06}{15.27} & \absdeltaonlypct{17.96}{12.78} \\

\textcolor{notunecolor}{\textbf{Qwen2.5-7B}} & \absdeltaonlypct{52.75}{48.50} & \absdeltaonlypct{48.50}{44.75} & \absdeltaonlypct{49.06}{45.32} & \absdeltaonlypct{40.42}{36.89} & \absdeltaonlypct{33.75}{31.25} & \absdeltaonlypct{33.25}{28.78} & \absdeltaonlypct{30.29}{25.18} & \absdeltaonlypct{28.69}{25.37} & \absdeltaonlypct{21.06}{17.77} & \absdeltaonlypct{17.96}{14.11} \\

\textcolor{notunecolor}{\textbf{Qwen3-4B}} & \absdeltaonlypct{52.75}{46.75} & \absdeltaonlypct{48.50}{41.00} & \absdeltaonlypct{49.06}{42.29} & \absdeltaonlypct{40.42}{33.64} & \absdeltaonlypct{33.75}{29.50} & \absdeltaonlypct{33.25}{27.43} & \absdeltaonlypct{30.29}{24.16} & \absdeltaonlypct{28.69}{23.71} & \absdeltaonlypct{21.06}{16.21} & \absdeltaonlypct{17.96}{13.62} \\

\arrayrulecolor{gray!50}\cmidrule(l){1-11}\arrayrulecolor{black}

\textcolor{ourscolor}{\textbf{DeepSeek-7B (FT)}} & \absdeltaonlypct{52.75}{53.25} & \absdeltaonlypct{48.50}{49.75} & \absdeltaonlypct{49.06}{50.86} & \absdeltaonlypct{40.42}{41.32} & \absdeltaonlypct{33.75}{35.50} & \absdeltaonlypct{33.25}{33.63} & \absdeltaonlypct{30.29}{30.88} & \absdeltaonlypct{28.69}{29.24} & \absdeltaonlypct{21.06}{22.27} & \absdeltaonlypct{17.96}{18.34} \\

\textcolor{ourscolor}{\textbf{Qwen2.5-7B (FT)}} & \absdeltaonlypct{52.75}{54.00} & \absdeltaonlypct{48.50}{50.25} & \absdeltaonlypct{49.06}{51.32} & \absdeltaonlypct{40.42}{42.30} & \absdeltaonlypctbf{33.75}{36.75} & \absdeltaonlypct{33.25}{34.46} & \absdeltaonlypct{30.29}{31.37} & \absdeltaonlypctbf{28.69}{30.86} & \absdeltaonlypct{21.06}{22.98} & \absdeltaonlypct{17.96}{19.29} \\

\textcolor{ourscolor}{\textbf{Qwen3-4B (FT)}} & \absdeltaonlypctbf{52.75}{55.50} & \absdeltaonlypctbf{48.50}{50.75} & \absdeltaonlypctbf{49.06}{52.23} & \absdeltaonlypctbf{40.42}{42.87} & \absdeltaonlypct{33.75}{36.25} & \absdeltaonlypctbf{33.25}{35.33} & \absdeltaonlypctbf{30.29}{32.18} & \absdeltaonlypct{28.69}{30.84} & \absdeltaonlypctbf{21.06}{23.32} & \absdeltaonlypctbf{17.96}{20.18} \\

\bottomrule
\end{tabularx}

\end{table}

%% file: tables/exp_deployment_cost.tex

\begin{table}[t]

\small
\centering
\renewcommand{\arraystretch}{1.2}

\caption{\textbf{Deployment Cost per Resolved Instance.} Deployment cost is the recurring cost of solving tasks, separate from the one-time construction and training accounted in Tab.~\ref{tab:exp:computation_overhead} (\S\ref{appendix:subsec:computation_overhead}). We compare the \textit{\textcolor{gray}{no-$M_\theta$ baseline}} against \ours-FT $M_\theta$ (\texttt{Qwen3-4B-T}), which is served locally alongside the SE agent. Cost per resolved instance divides API cost by the number of instances resolved. Lower is better for both cost columns, where \textcolor{deltagreen}{$\downarrow$} marks the \textit{\textbf{relative}} reduction $\Delta_{\textit{rel}}$ (Eq.~\ref{eq:eval_metric:deltarel}) against the \textit{\textcolor{gray}{no-$M_\theta$ baseline}}.}
\label{tab:exp:deployment_cost}

\begin{tabular*}{\textwidth}{@{\extracolsep{\fill}}llccc@{}}

\toprule

\textbf{SE Agent} & \textbf{Condition} & \makecell{\textbf{API Cost}\\\textbf{(\$ / 100 Inst.)}} & \makecell{$\bm{SR}$\\\textbf{(\%, avg@4)}} & \makecell{\textbf{Cost / Resolved}\\\textbf{(\$)}} \\

\midrule

\multirow{2}{*}{\textbf{Devstral-Small-2507}}
  & \textcolor{gray}{no memory} & \textcolor{gray}{11.73} & \textcolor{gray}{39.25} & \textcolor{gray}{0.299} \\
  & \textcolor{ourscolor}{+ \ours} & \textbf{10.26}~{\scriptsize\textbf{\textcolor{deltagreen}{$\downarrow$12.53\%}}} & \textbf{41.75} & \textbf{0.246}~{\scriptsize\textbf{\textcolor{deltagreen}{$\downarrow$17.77\%}}} \\

\arrayrulecolor{gray!50}\cmidrule(l){1-5}\arrayrulecolor{black}

\multirow{2}{*}{\textbf{Qwen3-Coder-30B-A3B}}
  & \textcolor{gray}{no memory} & \textcolor{gray}{9.97} & \textcolor{gray}{33.75} & \textcolor{gray}{0.295} \\
  & \textcolor{ourscolor}{+ \ours} & \textbf{8.94}~{\scriptsize\textbf{\textcolor{deltagreen}{$\downarrow$10.33\%}}} & \textbf{36.00} & \textbf{0.248}~{\scriptsize\textbf{\textcolor{deltagreen}{$\downarrow$15.94\%}}} \\

\bottomrule

\end{tabular*}

\end{table}

%% file: figures/exp_ablation_rl_algorithm.tex
\begin{figure}[!t]

\vspace{6pt}

\centering
\includegraphics[width=1.0\linewidth]{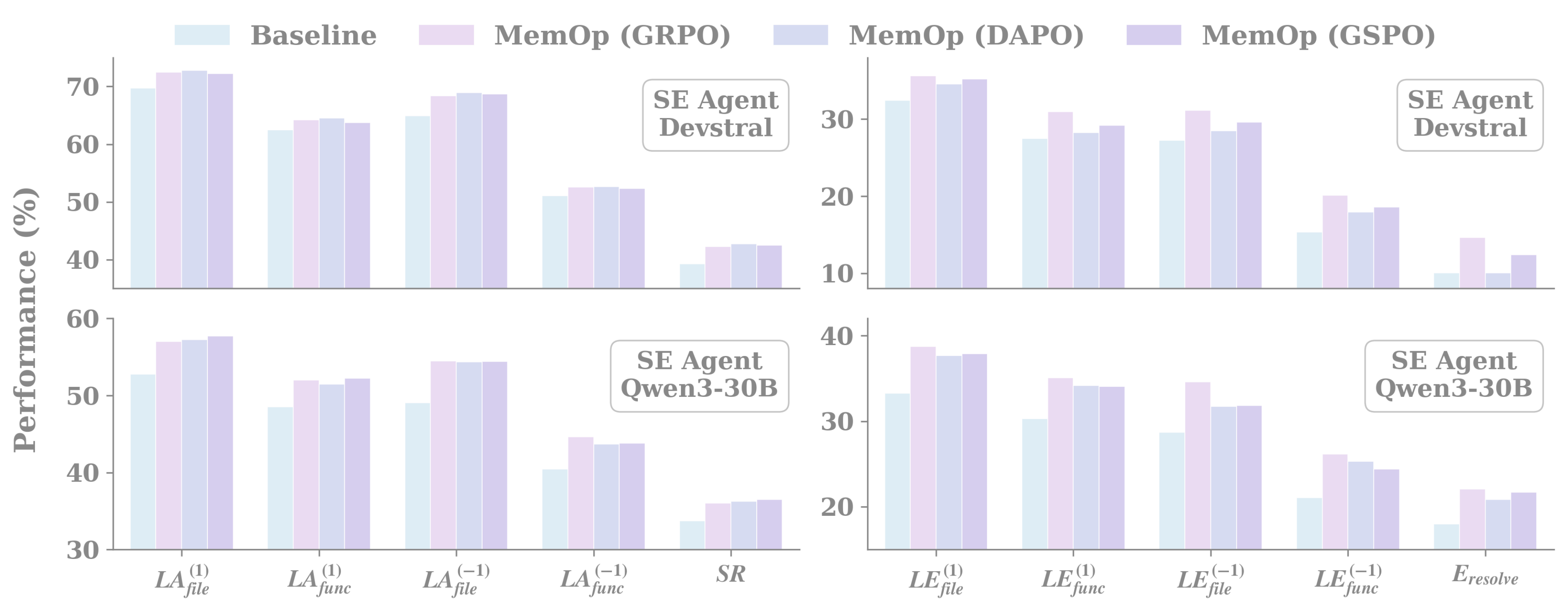}

\caption{\textbf{\ours Generalizes across RL Algorithms.} We instantiate Stage~II (\S\ref{subsec:two_stage_sft_rl_training}) with GRPO~\citep{2024grpo}, DAPO~\citep{2025dapo}, and GSPO~\citep{2025gspo}, respectively. Each is evaluated across our ten metrics (\S\ref{subsec:eval_metrics}) on SE agents powered by \texttt{Devstral-Small-2507} (\textit{top}) and \texttt{Qwen3-Coder-30B-A3B} (\textit{bottom}). All three algorithms yield consistent gains over the \textit{\textcolor{gray}{no-$M_\theta$ baseline}}, showing that the utility-grounded reward of \ours (Eq.~\ref{eq:rl_reward}) is not tied to a particular policy-optimization algorithm.}
\label{fig:exp_ablation_rl_algorithm}

\vspace{0pt}

\end{figure}

%% file: tables/exp_transfer_cross_language.tex

\begin{table}[t]

\small
\centering
\renewcommand{\arraystretch}{1.1}
\setlength{\tabcolsep}{0pt}

\caption{\textbf{Cross-Language Memory Augmentation across Ten Metrics.} We apply the same \ours-FT $M_\theta$ (\texttt{Qwen3-4B-T}) to the SE agent powered by \texttt{Qwen3-Coder-30B-A3B}, which improves every metric of our suite (\S\ref{subsec:eval_metrics}) in all three languages. This table summarizes \textbf{\textit{absolute performance differences}} (Eq.~\ref{eq:eval_metric:deltaabs}) against the \textit{\textcolor{gray}{no-$M_\theta$ baseline}} of the same language, shown as \textit{positive} \textcolor{deltagreen}{$\uparrow \Delta_{\textit{abs}}\%$} or \textit{negative} \textcolor{deltared}{$\downarrow \Delta_{\textit{abs}}\%$}. Python is evaluated on SWE-Bench-\textit{Verified}, C and Java are evaluated on Multi-SWE-Bench (\S\ref{subsec:rejection_sampling}).}
\label{tab:exp:transfer_cross_language}

\begin{tabularx}{\textwidth}{c*{10}{>{\centering\arraybackslash}X}}

\toprule

\multirow{3}{*}{\textbf{Language}} & \multicolumn{5}{c}{\textbf{Accuracy (\%)}} & \multicolumn{5}{c}{\textbf{Efficiency (\%)}} \\
\cmidrule(lr){2-6} \cmidrule(lr){7-11}
& $\bm{\textbf{\textit{LA}}_{\textit{file}}^{(1)}}$ & $\bm{\textbf{\textit{LA}}_{\textit{func}}^{(1)}}$ & $\bm{\textbf{\textit{LA}}_{\textit{file}}^{(-1)}}$ & $\bm{\textbf{\textit{LA}}_{\textit{func}}^{(-1)}}$ & \textbf{\textit{SR}}
& $\bm{\textbf{\textit{LE}}_{\textit{file}}^{(1)}}$ & $\bm{\textbf{\textit{LE}}_{\textit{func}}^{(1)}}$ & $\bm{\textbf{\textit{LE}}_{\textit{file}}^{(-1)}}$ & $\bm{\textbf{\textit{LE}}_{\textit{func}}^{(-1)}}$ & $\bm{E_{\textit{resolve}}}$ \\

\midrule

\textbf{Python} & \absdeltaonlypctlight{0}{4.25} & \absdeltaonlypctlight{0}{3.25} & {\bfseries\absdeltaonlypctlight{0}{5.44}} & \absdeltaonlypctlight{0}{3.21} & \absdeltaonlypctlight{0}{2.25} & {\bfseries\absdeltaonlypctlight{0}{5.50}} & {\bfseries\absdeltaonlypctlight{0}{4.80}} & {\bfseries\absdeltaonlypctlight{0}{5.93}} & {\bfseries\absdeltaonlypctlight{0}{5.12}} & {\bfseries\absdeltaonlypctlight{0}{4.10}} \\

\textbf{C} & \absdeltaonlypctlight{0}{3.93} & \absdeltaonlypctlight{0}{3.00} & \absdeltaonlypctlight{0}{3.63} & \absdeltaonlypctlight{0}{2.95} & \absdeltaonlypctlight{0}{2.85} & \absdeltaonlypctlight{0}{4.42} & \absdeltaonlypctlight{0}{3.14} & \absdeltaonlypctlight{0}{4.18} & \absdeltaonlypctlight{0}{2.87} & \absdeltaonlypctlight{0}{3.50} \\

\textbf{Java} & {\bfseries\absdeltaonlypctlight{0}{5.15}} & {\bfseries\absdeltaonlypctlight{0}{3.83}} & \absdeltaonlypctlight{0}{4.41} & {\bfseries\absdeltaonlypctlight{0}{3.67}} & {\bfseries\absdeltaonlypctlight{0}{4.55}} & \absdeltaonlypctlight{0}{5.13} & \absdeltaonlypctlight{0}{3.84} & \absdeltaonlypctlight{0}{4.85} & \absdeltaonlypctlight{0}{3.36} & \absdeltaonlypctlight{0}{4.01} \\

\bottomrule

\end{tabularx}

\vspace{12pt}

\end{table}

%% file: tables/exp_transfer_language.tex

\begin{wraptable}{r}{0.5\textwidth}

\vspace{-12pt}

\centering
\small
\renewcommand{\arraystretch}{1.15}

\caption{\textbf{Transferability to New Programming Languages.} We apply \ours-FT $M_\theta$ (\texttt{Qwen3-4B-T}) to the SE agent powered by \texttt{Devstral-Small-2507}, evaluated on programming languages absent from training. $SR$ (Eq.~\ref{eq:eval_metric:SR}) follows each benchmark's own resolution criterion, with SWE-Bench-\textit{Verified} (Python) as the in-distribution reference and Multi-SWE-Bench supplying C and Java. $N_{\mathcal{R}}$ and $|\mathcal{T}|$ denote the number of repositories and evaluation instances, and \textcolor{deltagreen}{$\uparrow \Delta_{\textit{abs}}\%$} the absolute gain (Eq.~\ref{eq:eval_metric:deltaabs}).}
\label{tab:exp:transfer_language}

\begin{tabular*}{\linewidth}{@{\extracolsep{\fill}}lcccc@{}}
\toprule
\multirow{2}{*}{\textbf{Language}} & \multirow{2}{*}{$\bm{N_{\mathcal{R}}}$} & \multirow{2}{*}{$\bm{|\mathcal{T}|}$} & \multicolumn{2}{c}{$\bm{SR}$ \textbf{(\%)}} \\
\cmidrule(lr){4-5}
& & & \textbf{Baseline} & \textbf{+\ours} \\
\midrule
Python & 9 & 100 & \textcolor{gray}{\textbf{39.25}} & \absdeltapctbf{39.25}{41.75} \\
C & 3 & 30 & \textcolor{gray}{\textbf{18.33}} & \absdeltapctbf{18.33}{20.83} \\
Java & 9 & 99 & \textcolor{gray}{\textbf{39.90}} & \absdeltapctbf{39.90}{43.94} \\
\bottomrule
\end{tabular*}

\vspace{0pt}

\end{wraptable}

%% file: tables/exp_transfer_activity.tex

\begin{wraptable}{r}{0.5\textwidth}

\vspace{-18pt}

\centering
\small
\renewcommand{\arraystretch}{1.15}

\caption{\textbf{Transferability to New Software Engineering Activities.} We apply \ours-FT $M_\theta$ (\texttt{Qwen3-4B-T}) to the SE agent powered by \texttt{Devstral-Small-2507}, evaluated on library implementation (Commit0-Lite) and unit-test generation (TestGenEval) covering different SE activities beyond issue resolution (Tab.~\ref{tab:exp:activity_coverage}, \S\ref{appendix:subsec:activity_coverage}). Each is scored under its own native metrics (\S\ref{subsec:eval_metrics}). \textcolor{deltagreen}{$\uparrow \Delta_{\textit{abs}}\%$} denotes the absolute gain (Eq.~\ref{eq:eval_metric:deltaabs}).}
\label{tab:exp:transfer_activity}

\resizebox{\linewidth}{!}{%
\begin{tabular}{llcc}
\toprule
\textbf{Benchmark} & \textbf{Native Metric} & \textbf{Baseline} & \textbf{+\ours} \\
\midrule
\multirow{4}{*}{Commit0-Lite}
  & $SR$ (resolve rate) & \textcolor{gray}{\textbf{4.69}}  & \absdeltapctbf{4.69}{7.81} \\
  & Pass rate (macro)   & \textcolor{gray}{\textbf{26.32}} & \absdeltapctbf{26.32}{30.02} \\
  & Pass rate (micro)   & \textcolor{gray}{\textbf{6.37}}  & \absdeltapctbf{6.37}{7.08} \\
  & Tests passed        & \textcolor{gray}{\textbf{41.44}} & \absdeltapctbf{41.44}{45.81} \\
\arrayrulecolor{gray!50}\cmidrule(l){1-4}\arrayrulecolor{black}
\multirow{3}{*}{TestGenEval}
  & $SR$ (all Pass@1)   & \textcolor{gray}{\textbf{63.75}} & \absdeltapctbf{63.75}{67.25} \\
  & Any Pass@1          & \textcolor{gray}{\textbf{88.25}} & \absdeltapctbf{88.25}{90.50} \\
  & Coverage            & \textcolor{gray}{\textbf{50.24}} & \absdeltapctbf{50.24}{58.62} \\
\bottomrule
\end{tabular}%
}

\vspace{-10pt}

\end{wraptable}

%% file: sections/6_related.tex
\section{Related Work}
\label{sec:related_work}

\vspace{-6pt}

\textbf{Memory-Augmented LLM Agents.}
As LLM agents become more capable~\citep{wang2025openhands,Yang2024SWEagent}, their real-world deployment demands complex reasoning~\citep{Liu2023LostInContexts, Kuratov2024BABILongTTA,li2024loogle,hosseini2025efficientlongcontexts} and long-context understanding~\citep{Chen2024CompressTIA}, underscoring the need for principled memory augmentation. Existing approaches range from human-inspired memory organization~\citep{fang2025lightmemlightweightefficientmemoryaugmented} to task-specific designs for lifelong dialogue agents~\citep{ong2025dialogue} and sophisticated workflows~\citep{Chen2024CompressTIA,yan2025generalagenticmemorydeep}, yet sharing critical limitations (\S\ref{appendix:subsec:related_work_extension:broader_use_of_memory}): (1) \textit{build upon complex architectural pipelines} that are difficult to adapt and generalize, (2) \textit{lack a principled, task-agnostic definition of memory utility} that can support rigorous evaluation and memory optimization, and (3) \textit{rely on task-specific designs or costly annotations} that impede transfer across agents and problem settings.

\textbf{Memory for SE Agents.}
LLM agents deployed in SE tasks interact with complex environments, identifying problematic code lines, executing commands, and implementing revisions across multiple files and repository levels \citep{wang2025openhands,Yang2024SWEagent,Claude4Sonnet}.
Each action expands the reasoning chain and accumulates environmental observations, overwhelming agents with redundant contextual information while driving up computational cost (\S\ref{sec:intro}).
Prior SE memory systems differ most in \textit{what defines a good memory}: human curation~\citep{wang2026memgovernenhancingcodeagents}, a structural prior over subtasks~\citep{shen2026structurallyalignedsubtasklevelmemory} or repositories~\citep{Wang2025ImprovingCLA}, an LLM judgment of past experience~\citep{mu2026experepair,chen2026sweexpexperience}, or no filter at all~\citep{lindenbauer2025knowledge2noise}. Each supplies that definition in advance rather than measuring it, so none can determine whether retained information is genuinely beneficial, and unfiltered memory is observed to actively degrade SE agents~\citep{lindenbauer2025knowledge2noise,chen2026sweexpexperience}. \ours instead defines memory quality as measured causal effect on downstream outcomes, which lets one signal serve as both evaluation benchmark and annotation-free optimization signal. We compare against these systems empirically in Tab.~\ref{tab:exp:memory_system_comparison} and organize them along five axes in Tab.~\ref{tab:related_work_taxonomy} (\S\ref{appendix:subsec:related_work_extension:memory4SE}).

%% file: sections/7_conclusions.tex
\section{Conclusion}
\label{sec:conclusion}

In this work, we propose \ours, a closed-loop framework that equips SE agents with memory augmentation through a principled notion of \textit{memory utility} (\S\ref{sec:method}-\ref{sec:training}).
By grounding memory utility in validated downstream impact and closing the training loop via \textit{trajectory-level reflection} and \textit{outcome-grounded validation}, \ours enables SE agents to continuously retain, refine, and reuse experience across episodes via self-supervision without external annotations or reward models (\S\ref{sec:training}).
Experiments across \textit{single-episode} and \textit{cross-episode} settings demonstrate the \textbf{effectiveness} of \ours, with consistent gains across multi-dimensional metrics at reduced deployment cost (\S\ref{subsec:exps:effectiveness}), its \textbf{generalizability} across SE agents, memory backbones, and optimization algorithms(\S\ref{subsec:exps:generalizability}), and its \textbf{transferability} to unseen programming languages and SE activities (\S\ref{subsec:exps:transferability}). We hope our work motivates broader exploration of \textit{memory utility} as a first-class design principle for building adaptive, generalizable, and continually-improving agents.


%% file: sections/appendix.tex
\clearpage
\appendix

\section{Preliminary: Memory-Augmented AI Software Engineering}
\label{appendix:sec:preliminary}

\subsection{Preliminary Exploration on SE Agent Failures}
\label{appendix:subsec:preliminary:SE_agent_failure}

\input{figures/preliminary1_SE_agent_failure_analysis}

\textbf{\textit{How do SE agents fail in software engineering tasks?}}
Understanding the nature of SE agent failures is critical for improving their problem-solving. Motivated by this, we perform a preliminary analysis on key failure patterns in SE agents, categorizing common causes to inform potential directions for enhancement.
Our preliminary study manually examines the complete problem-solving trajectories of $50$ software engineering failures ($SR=0.00\%$) of the SE agent powered by \texttt{GPT-4o-mini}, identifying seven key causes of failures in AI agent software engineering.
For each category, we present a representative case from our study. As shown in Figs.~\ref{fig:preliminary:failure_example1:repo_structure}-\ref{fig:preliminary:failure_example7:hallucination}, these cases illustrate different failure patterns and their impact on intermediate progress:
\textit{repository structure} (Fig.~\ref{fig:preliminary:failure_example1:repo_structure}), \textit{repetition} (Fig.~\ref{fig:preliminary:failure_example2:repetition}), \textit{reasoning} (Fig.~\ref{fig:preliminary:failure_example3:reasoning}),
\textit{coding} (Fig.~\ref{fig:preliminary:failure_example4:coding}), \textit{execution} (Fig.~\ref{fig:preliminary:failure_example5:execution}), \textit{inconsistency} (Fig.~\ref{fig:preliminary:failure_example6:inconsistency}), and \textit{hallucination} (Fig.~\ref{fig:preliminary:failure_example7:hallucination}).

As shown in Fig.~\ref{fig:preliminary:SE_agent_failure_analysis}, AI agent's misinterpretation and incorrect understanding of repository structure constitute the primary causes of problem-solving failures. As the number of actions and episodes increases, repetition in long-context reasoning emerges as the second major contributor, followed by reasoning errors as well as syntax and execution errors. These failures expose critical limitations of SE agents in comprehending, distilling, and memorizing key structures and patterns of the codebase, underscoring the significance of memory augmentation and optimization in complex software engineering tasks that demand long-context reasoning and problem-solving.

\input{tables/exp_preliminary_memory_structure}
\input{figures/preliminary2_SE_agent_failure_example_repo_structure}
\input{figures/preliminary3_memory_generation_prompt}
\input{figures/exp_repo_wise_delta}
\newpage

\subsection{Preliminary Exploration on Memory Structure}
\label{appendix:subsec:preliminary:memory_structure}

\textbf{\textit{Which memory structure can most effectively benefit SE agents?}}
The answer to this critical question informs the effective design of memory representation, shaping how memory is constructed, formatted, and evolved within memory-augmented frameworks.
To investigate this question, we conduct a preliminary study on memory structures, examining the impact of nine different memory structure settings on the performance of memory-augmented SE agents:

\begin{enumerate}[
  leftmargin=*,
  itemsep=1pt,
  topsep=2pt,
  label=\bcnum{\arabic*}
]
    \item \textbf{String:} Represent memory as a simple textual string, showing information in a linear format.
    \item \textbf{Dictionary:} Represent memory hierarchically using a JSON dictionary, where keys categorize high-level information and values store details. 
    \item \textbf{List:} Represent memory as an ordered JSON list, showing elements in sequence.
    \item \textbf{Tree:} Represents memory in a tree-like structure, displaying parent-child relationships between concepts or items.
    \item \textbf{Graph:} Represents memory as a graph-like structure, capturing relationships and connections between multiple elements.
    \item \textbf{Python:} Represents memory using Python code, enabling structured representation and potential implementation.
    \item \textbf{Given:} Represents memory using a fixed hierarchical structure with predefined fields and categories.
    \item \textbf{Discretionary:} Represents memory in a flexible format determined by the model, allowing adaptation to different types of information or structural arrangements.
    \item \textbf{\textcolor{ourscolor}{\textbf{\textit{Ours}}}:} Represents memory in a markdown-based structured format \texttt{Memory.md}, integrating key knowledge and information into hierarchical organization.
\end{enumerate}

Concretely, we employ \texttt{GPT-4o-mini} to power the SE agent, and \texttt{GPT-4o} as $M_\theta$ to learn from past experiences and generate memory instances. Both baseline setting (\textit{without} memory) and memory-augmented setting (\textit{with} memory) SE leverage the same test set that comprises $30$ instances randomly sampled from SWE-Bench-\textit{Verified} \citep{jimenez2024swebench}.
The SE agent is allowed a maximum of $20$ interaction turns to complete each task.
To isolate the fundamental effects of each memory structure, all experiments are performed under the \textit{single-episode} mode (\S\ref{subsec:exp_setup}).
As the $20$-\textit{turn} limit results in $SR=0\%$ across all settings, we evaluate SE performance using \textit{first} localization accuracy $LA_f^{(1)}$ (Eq.~\ref{eq:eval_metric:LA}) and \textit{first} localization efficiency $LE_f^{(1)}$ (Eq.~\ref{eq:eval_metric:LE}).

Table~\ref{tab:exp:preliminary:memory_structure} summarizes the evaluation results across all settings. Compared with the no-memory baseline, eight of the nine memory structures improve SE agent's performance on all four localization accuracy and efficiency metrics. The sole exception is \textit{List}, which degrades $LA_{\textit{file}}^{(1)}$ (Tab.~\ref{tab:exp:preliminary:memory_structure}).
Surprisingly, allowing the model to decide memory structures autonomously results in relatively modest improvements ($\Delta_{abs} \leq 16.66\%$), indicating limited adaptive memory reflection and evolution abilities in LLMs.
On the other hand, highly structured formats, such as \textit{dictionaries} ($\Delta_{abs} \geq 14.85\%$) and \textit{Python code} ($\Delta_{abs} \geq 16.76\%$), lead to substantial gains, while the memory structure of \ours delivers the most pronounced improvement ($\Delta_{abs} \geq 17.65\%$), especially in localization accuracy ($\Delta_{abs} \geq 20.10\%$), which is an essential prerequisite for successful SE problem-solving. This finding suggests a fundamental design principle: \textit{Memory structures that are easier for the SE agent to understand are also more effective.}

\input{figures/preliminary3_memory_instruction_result}

\subsection{Preliminary Exploration on Memory Generation Instructions}
\label{appendix:subsec:preliminary:memory_prompt}

Following the memory principle (\S\ref{appendix:subsec:preliminary:memory_structure}), generating effective memory faces a key challenge: \textbf{\textit{How to instruct $\bm{M_\theta}$ to perform memory generation and evolution more effectively?}}

Both memory generation and memory evolution critically depend on high-quality reflection and distillation. In long-horizon software engineering (SE) trajectories characterized by diverse actions, execution outputs, environmental feedback, intermediate reasoning, etc., it is critical for $M_\theta$ to continuously identify salient structures, extract reusable knowledge, capture useful patterns, and distill key insights. However, without explicit guidance, $M_\theta$ struggles to distinguish essential information from incidental details (Figs.~\ref{fig:preliminary3:memory_generation_prompt}-\ref{fig:preliminary3:memory_generation_prompt_results}), leading to suboptimal memory construction. Therefore, the central problem is how to guide $M_\theta$ to conduct reliable and effective learning and reflection over complex, multi-step software engineering processes.

To investigate this question, we conduct a preliminary study on the behaviors of $M_\theta$ under different memory generation instructions. Specifically, we examine three types of prompts with increasing levels of guidance:

\begin{enumerate}[
  leftmargin=*,
  itemsep=1pt,
  topsep=2pt,
  label=\bcnum{\arabic*}
]
    \item \textbf{Concise Instruction Without Targeted Requirements} is designed to assess and expose the intrinsic reasoning and reflection capabilities of LLMs in the absence of external guidance, allowing us to uncover their inherent limitations and typical failure modes in memory generation.
    \item \textbf{High-Level Instruction} delineates the categories of information that should be prioritized for reflection and retention, while explicitly indicating non-essential content to be excluded. This formulation introduces principled constraints on the scope of reflection and distillation without prescribing a rigid structure, thereby encouraging structured yet flexible abstraction.
    \item \textbf{Fine-Grained Instruction} imposes explicit structural and content-level requirements on memory generation, supplemented with illustrative examples. This instruction provides detailed operational specifications intended to standardize the abstraction process and reduce ambiguity in what constitutes high-quality memory.
\end{enumerate}

Fig.~\ref{fig:preliminary3:memory_generation_prompt} presents these three versions of memory generation instructions: (1) general and concise, (2) high-level instruction, and (3) fine-grained instruction. Results in Fig.~\ref{fig:preliminary3:memory_generation_prompt_results} reveal clear differences in memory quality across these settings.
Understandably, instruction (1) yields the lowest memory quality. In the absence of explicit criteria or structural constraints, $M_\theta$ lacks a principled mechanism for reliably identifying what is worth memorizing and what should be reflected on to distill reusable knowledge. As a result, the generated memory tends to be overly generic, fragmented, or populated with low-value operational details and transient context. This uncovers the inherent weaknesses of LLMs in unguided memory construction, especially their limited ability to accurately identify important information and to effectively reflect upon and distill useful insights.
Based on the failure analysis of Instruction (1) (Fig.~\ref{fig:preliminary3:memory_generation_prompt} (1)), we define Instructions (2) and (3), which are able to provide more effective guidance, enabling $M_\theta$ to better differentiate between critical insights and peripheral information. By explicitly outlining what to focus on and what to disregard, they guide $M_\theta$ toward more task-relevant reflection and knowledge organization.
Among all three variants, instruction (2) exhibits the optimal guidance. This result reveals that high-level guidance strikes an effective balance: It offers sufficient direction to mitigate the intrinsic weaknesses observed in setting (1), while still preserving flexibility for $M_\theta$ to generalize and distill across diverse software engineering repositories and scenarios. In comparison, instruction (3), although more detailed, can inadvertently over-constrain the reflection process or bias the model toward rigid compliance with specified formats and examples, potentially limiting broader abstraction.
These findings suggest that identifying the intrinsic deficiencies of LLMs in unguided reflection is essential for designing effective memory generation instructions. Simply relying on intrinsic reasoning or over-specified details is insufficient, whereas carefully designed high-level guidance can substantially enhance memory generation and evolution to benefit long-horizon agentic SE.


\section{Related Work: Extended Discussion}
\label{appendix:sec:related_work_extension}

We extend our discussion in \S\ref{sec:related_work} with additional related work that has emerged concurrently with \ours, further contextualizing our contributions.

\subsection{Memory for SE Agents}
\label{appendix:subsec:related_work_extension:memory4SE}

\input{tables/related_work_taxonomy}

Recent work has begun to explore memory augmentation specifically in the context of software engineering agents. Tab.~\ref{tab:related_work_taxonomy} organizes these systems along five axes, and we discuss each in turn.

\textbf{Benchmarks without an optimization mechanism.} SWE-Bench-CL~\citep{joshi2025swebenchcl} introduces a continual learning benchmark that organizes GitHub issues into chronologically ordered sequences to evaluate knowledge accumulation and transfer across tasks. However, it proposes no mechanism for \textit{how} to optimize memory, leaving the core methodological challenge open.

\textbf{Retrieval-based memory without a quality filter.} CTIM-Rover~\citep{lindenbauer2025knowledge2noise} proposes a Mixture-of-Experts inspired approach for SE agents, but finds that unfiltered memory retrieval introduces noise that actively degrades performance. Our reproduction (Tab.~\ref{tab:exp:memory_system_comparison}) recovers the same effect under our own harness, which we read as independent corroboration rather than as a weak baseline: without a principled utility signal to separate beneficial memories from noise, memory augmentation can actively harm an SE agent.

\textbf{Memory quality defined by human curation or structural priors.} MemGovern~\citep{wang2026memgovernenhancingcodeagents} learns from human-governed experience, so its quality signal is ultimately human curation. Subtask-level memory~\citep{shen2026structurallyalignedsubtasklevelmemory} addresses retrieval granularity by aligning memory storage and retrieval with the SE agent's functional decomposition, a structural prior rather than a measured outcome. Repository Memory~\citep{Wang2025ImprovingCLA} optimizes a single subtask and is correspondingly most helpful for localization. In each case the definition of a good memory is supplied in advance, by a human or by a design assumption, rather than measured.

\textbf{Memory quality judged by an LLM or against gold patches.} ExpeRepair~\citep{mu2026experepair} maintains dual episodic and semantic memory, labeling episodic entries by LLM-judged reproduction success and curating semantic memory through LLM selection. SWE-Exp~\citep{chen2026sweexpexperience} extracts experience by comparison against gold patches and reranks it with an LLM. Both differ from \ours in three further respects. First, both are full agent stacks, a two-agent repair pipeline gated on successful reproduction-test generation and a dual Instructor/Assistant pair over MCTS, respectively, so neither attaches to a different SE agent without adopting its entire control flow; a head-to-head comparison would therefore replace our SE agent, sandbox, and harness, and would no longer isolate the memory mechanism that Tab.~\ref{tab:exp:memory_system_comparison} controls. Second, both rely on large models throughout, whereas \ours obtains its gains from a $1.5$B--$7$B open model served locally. Third, both identify as open problems precisely what our utility signal supplies: ExpeRepair states that it does not optimize localization for lack of an automated oracle, which is the role of $LA$ (Eq.~\ref{eq:eval_metric:LA}) and $LE$ (Eq.~\ref{eq:eval_metric:LE}); and SWE-Exp states that its agent lacks a mechanism to assess the applicability of prior experience, which is the role of our acceptance criterion (Eq.~\ref{eq:memory_acceptance}). SWE-Exp additionally observes that performance declines as more experiences are retrieved, again corroborating the premise motivating \ours.

\textbf{\ours defines memory quality by measured downstream effect.} Across all of the above, what varies most is \textit{what defines a good memory}: human curation, a structural prior, an LLM judgment, or nothing at all. \ours defines it as measured causal effect on downstream outcomes (Eq.~\ref{eq:memory_acceptance}), which lets one signal serve as both a task-agnostic evaluation benchmark and an annotation-free optimization signal. No LLM judges memory quality, and ground truth enters only to compute $\Delta$, never as input to $M_\theta$. The result is a memory model that attaches to a fixed SE agent through a single file, requires no architectural change, and transfers across SE agents, benchmarks, languages, and SE activities (\S\ref{subsec:exps:transferability}).

\subsection{Broader Use of Memory}
\label{appendix:subsec:related_work_extension:broader_use_of_memory}

Memory augmentation for general LLM agents has been explored along several directions.
One line of work augments agents with retrieval-based memory, where past experiences or contextual knowledge are stored and retrieved to support downstream generation~\citep{qian2025memoragboostinglongcontext,chen2025comrag,wang2024mragreinforcinglargelanguage}. Across these approaches, a shared limitation persists: memory quality is assessed through proxy signals, such as retrieval similarity, coverage, or structural coherence, without empirical validation. Lacking such a principled utility signal, these systems fail to reliably distinguish memory that improves agent behaviors from memory that is redundant or harmful.
A different line of work evolves memory through heuristic update rules, such as recency, novelty, similarity, or psychologically inspired forgetting schedules~\citep{he2024camelotlargelanguagemodels,zhong2023memorybankenhancinglargelanguage,salama2025meminsightautonomousmemoryaugmentation}. While these rules offer intuitive appeal, they are designed without downstream grounding, leaving memory evolution disconnected from the outcomes it is meant to support.
Another direction organizes memory through hierarchical~\citep{sun2026hmem,hu2025hiagent} or graph-based~\citep{atri2025lifelong,kashmira2025tobugraph} structures, which improve interpretability but cannot adapt to the diversity of tasks encountered in practice.
To address these limitations, \ours defines memory utility as demonstrated causal improvement grounded in downstream performance (\S\ref{sec:method}), deriving a principled, annotation-free optimization signal (\S\ref{sec:training}) that is transferable across agents and generalizable across tasks and settings (\S\ref{sec:exps}).

\section{Limitations}
\label{appendix:sec:limitations}

In this work, we introduce \ours (\S\ref{sec:method}), a closed-loop framework with performance-grounded memory utility serving a dual role (\S\ref{sec:intro}-\ref{sec:training}), enabling principled evaluation and annotation-free optimization of memory-augmented SE agents across \textit{single-episode} and \textit{cross-episode} settings (\S\ref{subsec:two_eval_regimes}).

We further note several limitations of the current work, each pointing to a direction for future extension.
First, \ours is evaluated across four benchmarks, three programming languages, and eleven SE activity categories (\S\ref{subsec:exps:transferability}, \S\ref{appendix:subsec:activity_coverage}). Extending this evaluation to further programming languages and SE activities is a natural next step.
Second, memory utility is measured from agent rollouts, which are inherently stochastic, and we therefore evaluate every task under $N_{\tau}=4$ trajectories to stabilize each estimate (Alg.~\ref{alg:rejection_sampling}). Further improving the calibration of this utility signal offers additional headroom for \ours.
Third, as we focus on memory optimization, SE performance also depends on the intrinsic problem-solving abilities and behaviors of the SE agent itself (\S\ref{appendix:sec:in_depth_analysis}), and the size and curation efficiency of the training dataset depend on the rollout configuration (\S\ref{appendix:sec:implementation_details}), which we leave to future exploration.
In future work, we aim to extend \ours to broader benchmarks, programming languages, and SE activities, to design new inference-time selection mechanisms over generated memory candidates (\S\ref{appendix:subsec:matched_budget}), and to explore tighter integration between memory optimization and SE agent training to further improve problem-solving performance.




\section{Dataset Construction for \ours Optimization}
\label{appendix:sec:dataset_construction}

Based on our \textit{trajectory-based rejection sampling} (Alg.~\ref{alg:rejection_sampling}), we construct our memory optimization dataset via \textit{performance-validated memory supervision} (\S\ref{subsec:rejection_sampling}). We begin by randomly sampling $N_{\mathcal{T}}=10$ tasks from each of $N_{\mathcal{R}}=10$ repositories, yielding $N_{\mathcal{R}} \times N_{\mathcal{T}}=100$ training tasks in total. Then, we deploy SE agents powered by \texttt{Devstral-Small-2505} and \texttt{Qwen3-Coder-480B-A35B-Instruct}, respectively, with each executing these $100$ tasks. During \textit{trajectory-based rejection sampling} (Alg.~\ref{alg:rejection_sampling}), we set $N_{\tau}=4$ and $N_{\mathcal{M}}=4$, contributing to $100 \times 2 \times 4 \times 4 = 3{,}200$ memory candidates. After performance filtering, the remaining samples form our curated memory optimization dataset for two-stage \ours finetuning (Tab.~\ref{tab:training_data_overview}).

\input{figures/eval_dataset_distribution}

During experiments, to avoid evaluation circularity, we randomly sample $100$ evaluation instances that have no overlap with the $100$ instances used to construct our training dataset. In \textit{cross-episode} evaluation, all instances of each repository are evaluated according to their temporal order to simulate real-world codebase evolution. As shown in Fig.~\ref{fig:eval_dataset_distribution}, the 100 evaluation instances are distributed evenly across repositories to support reliable \textit{cross-episode} evaluation. We assign \texttt{django} twice as many instances as it constitutes a primary portion of SWE-bench, while there is no 10\textit{th} repository that reaches 10.00\% of the total instances as others. This prevents any smaller repository from being overrepresented, thereby maintaining a balanced evaluation distribution.


\newpage
\input{figures/memory_prompts}
\newpage

\section{Implementation Details}
\label{appendix:sec:implementation_details}

\input{figures/memory}

\subsection{Memory Generation Instructions}
\label{appendix:subsec:memory_prompts}

Supported by our preliminary studies (\S\ref{appendix:sec:preliminary}), the memory generation instruction prompts we use for \textit{single-episode} and \textit{cross-episode} memory generation are shown in Fig.~\ref{fig:two_prompt_for_single_and_cross_episode}.

\input{tables/training_data}

\subsection{Experiment Configuration}
\label{appendix:subsec:experiment_configuration}

Our experiments involve various configuration settings in different stages. We summarize the key configurations during trajectory rollout, memory model finetuning, and performance evaluation in Tab.~\ref{tab:exp:configuration} to provide a clear and reproducible overview of our experimental setup.

\input{tables/exp_configuration}

\subsection{Computation Overhead}
\label{appendix:subsec:computation_overhead}

During dataset construction, we leverage LLM APIs to power both SE agents and $M_\theta$. For model finetuning, the memory model is optimized through our two-stage training paradigm (\S\ref{subsec:two_stage_sft_rl_training}) consisting of supervised finetuning and reinforcement learning. During evaluation, we use LLM APIs to support SE agents, while employing our finetuned memory models to instantiate $M_\theta$. For computational resources, we use NVIDIA H100 (80GB) GPUs to serve and train LLMs. To provide a clear accounting of computation resources, we summarize the resource overhead for dataset construction, model finetuning, and performance evaluation in Tab.~\ref{tab:exp:computation_overhead}, and compare the computational time and efficiency across \textit{single-episode} and \textit{cross-episode} memory augmentation in Fig.~\ref{fig:improved_SE_efficiency}.

\input{tables/exp_computation_overhead}

\section{A Deeper Dive Beneath Results}
\label{appendix:sec:in_depth_analysis}

We extend our experiments (\S\ref{sec:exps}) to in-depth analysis (\S\ref{appendix:sec:in_depth_analysis}), covering \textit{evaluation coverage across SE activities} (\S\ref{appendix:subsec:activity_coverage}), repository-wise generalizability (\S\ref{appendix:subsec:exp_ablation_repo_wise_generalizability}), \textit{memory evolution levels} (\S\ref{appendix:subsec:exp_evolve_granularity}), \textit{configuration effects of rollout batch size $c$ in $\mathcal{D}_{\textit{RL}}$} (\S\ref{appendix:subsec:exp_ablation_rl_options}), and \textit{qualitative analysis} (\S\ref{appendix:subsec:qualitative_analysis}).


\subsection{Evaluation Coverage across SE Activities}
\label{appendix:subsec:activity_coverage}

\input{tables/exp_activity_coverage}

Repository-level issue resolution is not a single activity but a bundle of them, so characterizing what our evaluation covers requires looking beneath the benchmark names. We therefore label every evaluation instance across our four benchmarks (\S\ref{subsec:exp_setup}) by the SE activity it involves, using \texttt{Claude-4.5-Sonnet} with multi-label assignment.

As shown in Tab.~\ref{tab:exp:activity_coverage}, our evaluation spans eleven activity categories. Eight of them carry at least $25$ labeled instances, led by program comprehension ($345$), bug fixing ($151$), testing ($120$), error handling and input validation ($116$), and documentation change ($94$). SWE-Bench-\textit{Verified} alone covers ten of the eleven categories, five of them with at least ten instances, which indicates that even the single-benchmark evaluation of our submission already reached beyond bug fixing. Commit0-Lite contributes differently by construction: implementing a library from scratch entails feature implementation, testing, API design, and comprehension in every instance, which is why its column is dense despite comprising only $16$ instances.

Program comprehension is labeled on every instance of every benchmark, which is expected rather than incidental: an SE agent cannot act on a repository it has not understood. This is also the activity our metric suite measures most directly, through localization accuracy (Eq.~\ref{eq:eval_metric:LA}) and localization efficiency (Eq.~\ref{eq:eval_metric:LE}), and it is where memory contributes most (\S\ref{subsec:exps:transferability}). Taken together, the coverage above establishes that the gains of \ours are measured over a broad range of SE activities rather than over issue resolution alone.

\subsection{Repository-Wise Generalizability of \ours}
\label{appendix:subsec:exp_ablation_repo_wise_generalizability}

Extending our discussion in \S\ref{subsec:exps:generalizability}, Fig.~\ref{fig:repo_wise_delta} demonstrates consistent performance enhancement of \ours-augmented SE agents, with gains up to $\Delta_{abs}=\uparrow17.50\%$ in $SR$ and $\Delta_{abs}=\uparrow15.00\%$ in $LA$. Only four of the $90$ repository-metric comparisons show a decrease in localization accuracy or efficiency, further demonstrating the robustness of \ours when generalizing to different repository contexts.

\subsection{Why Episode-Level over Action-Level Memory Evolution?}
\label{appendix:subsec:exp_evolve_granularity}

\input{figures/exp_evolve_granularity}

Determining the appropriate granularity of memory evolution is critical for balancing SE efficiency and memory optimization effectiveness. On this account, a key question is (Fig.~\ref{fig:ours_memory_method}): \textit{Should memory evolve at the action level or the episode level?}

To investigate this fundamental question, we employ \texttt{Qwen3-Coder-30B-A3B} to power the SE agent, with \ours using \texttt{Qwen3-4B-T} as $M_\theta$ backbone. We evaluate memory-augmented software engineering on the same test set under both \textit{action-level} and \textit{episode-level} memory evolution, respectively. 

As shown in Fig.~\ref{fig:exp_evolution_granularity}, \ours with \textit{episode-level} evolution yields reduced overhead in cross-episode settings ($\Delta_{rel}=\downarrow2.23\%$). In contrast, \textit{action-level} memory evolution severely degrades efficiency, requiring up to $\times11.82$ more time on average to solve SE tasks. More importantly, this inefficiency is accompanied by a substantial drop in success rate ($\Delta_{abs}=\downarrow12.50\%$), as compared to the improved success rate ($\Delta_{abs}=\uparrow2.75\%$) of \ours in cross-episode memory evolution. These results highlight that \textit{episode-level} memory evolution provides a more favorable optimization for both efficiency and effectiveness, making it a better design choice for memory-augmented SE agents.

\subsection{Effects of Preference Batch Size on Memory Optimization}
\label{appendix:subsec:exp_ablation_rl_options}

To investigate the effect of rollout batch size $c$ in $\mathcal{D}_{\textit{RL}}$ (\S\ref{subsec:rejection_sampling}), we compare SE agent performance with $M_\theta$ using the same backbone LLM (\texttt{Qwen3-4B-Thinking}), finetuned on $\mathcal{D}_{\textit{RL}}$ with $c=2$ and $c=4$, respectively.

Results in Fig.~\ref{fig:exp_ablation_rl_option} demonstrate consistent gains from \textit{Stage II} finetuning across different values of $c$, consistent with the effectiveness of Stage II observed in Fig.~\ref{fig:exp_ablation_sft_rl} (\S\ref{subsec:exps:generalizability}). Notably, $c=4$ presents more robust and comprehensive $M_\theta$ optimization effects across ten metrics (e.g., $c=2$ with $\Delta_{abs} \leq 0.75\%$ in $SR$, as compared to $c=4$ with $\Delta_{abs} \geq 2.25\%$ in $SR$).

\input{figures/exp_ablation_rl_option}



%


\subsection{\ours Improves Accuracy and Efficiency Together}
\label{appendix:subsec:compare_with_other_memory_methods}

\input{tables/exp_memory_system_comparison}

We compare against existing SE memory systems that attach to a fixed SE agent without replacing the agent stack and provide a usable public implementation, integrating each into our own harness so that all systems share the same SE agent, evaluation set, and metric suite (Tab.~\ref{tab:exp:memory_system_comparison}). For reproduction fidelity, MemGovern~\citep{wang2026memgovernenhancingcodeagents} is designed as a plug-in to the SWE-Agent scaffold~\citep{Yang2024SWEagent}, and we port its released memory cards and retrieval store into our harness so that every system is compared under an identical SE agent.
As shown in Tab.~\ref{tab:exp:memory_system_comparison}, efficiency separates the systems more sharply than accuracy. MemGovern~\citep{wang2026memgovernenhancingcodeagents} gains $\Delta_{abs}=\uparrow1.25\%$ in $SR$ at a cost of $\Delta_{abs}=\downarrow1.14\%$ in $E_{\textit{resolve}}$, consuming more iterations than an agent carrying no memory at all, and CTIM-Rover~\citep{lindenbauer2025knowledge2noise} declines on both axes. Notably, this reproduces rather than contradicts the finding of CTIM-Rover, whose own analysis attributes the degradation to unfiltered retrieval introducing noise, offering independent corroboration of the premise motivating \ours: \textit{memory without a validated notion of utility can actively harm an SE agent}.
\ours is the only system improving both axes at once ($\Delta_{abs}=\uparrow2.50\%$ in $SR$ and $\uparrow4.63\%$ in $E_{\textit{resolve}}$), showing that grounding memory in measured downstream utility enhances accuracy without paying for it in additional problem-solving iterations.

Systems without a public implementation, together with those whose memory mechanism cannot be detached from their own agent stack, are compared conceptually along five axes in Tab.~\ref{tab:related_work_taxonomy} (\S\ref{appendix:subsec:related_work_extension:memory4SE}).

\subsection{Deployed versus Matched-Budget Memory Gain}
\label{appendix:subsec:matched_budget}

\input{tables/exp_matched_budget}

The gains in Tabs.~\ref{tab:exp:single_episode}-\ref{tab:exp:cross_episode} are obtained by deploying a \textit{single} memory candidate. Since \ours generates $N_{\mathcal{M}}=4$ candidates per instance (Alg.~\ref{alg:rejection_sampling}) and selects none of them at deployment, the deployed number is the expected outcome of one unselected candidate rather than the contribution the memory mechanism is capable of.

To separate the two, we additionally compute \textit{max@4}, scoring the \textit{\textcolor{gray}{no-$M_\theta$ baseline}} as the best of four independent rollouts so that the selection budget is identical on both sides. As shown in Tab.~\ref{tab:exp:matched_budget}, the memory-attributable gain grows from $\Delta_{abs}=\uparrow2.50\%$ to $\uparrow5.50\%$ on \texttt{Devstral-Small-2507} and from $\uparrow2.25\%$ to $\uparrow6.00\%$ on \texttt{Qwen3-Coder-30B-A3B}, that is $2.20\times$ and $2.67\times$ the deployed gain. Because both sides receive four attempts, this isolates the contribution of memory from the fact that a stochastic agent benefits from repeated attempts regardless. Notably, the two SE agents yield closely comparable matched-budget gains despite differing by $5.50\%$ in baseline $SR$, indicating that the effect is a property of the memory mechanism rather than of a particular SE agent.


\subsection{\ours for Improved SE Efficiency}
\label{appendix:subsec:improved_efficiency}

Extending our study in \S\ref{appendix:subsec:exp_evolve_granularity}, we employ \texttt{Qwen3-Coder-30B-A3B} to power the SE agent, with \ours using \texttt{Qwen3-4B-T} as $M_\theta$ (FT) backbone.
Fig.~\ref{fig:improved_SE_efficiency} compares the average computational time used per task, $SR$, and $E_{\textit{resolve}}$ among no-$M_\theta$ baseline, single-episode \ours, and cross-episode \ours. Across both settings, \ours showcases consistently improved task success rate and problem-solving efficiency while reducing computational cost, demonstrating that \ours meaningfully enhances SE agent performance without sacrificing efficiency.

\input{figures/exp_improved_efficiency}

\subsection{\ours for More Robust Software Engineering}
\label{appendix:subsec:error_bar_robustness}

To systematically compare the augmentation effects of \ours, we use \texttt{Qwen3-Coder-30B} to power the SE agent with \texttt{Qwen3-4B-T} as the memory backbone of \ours-FT $M_\theta$, and compare their mean performance and variance across all ten evaluation metrics (\S\ref{subsec:eval_metrics}). As shown in Fig.~\ref{fig:error_bar_robustness}, \ours consistently achieves higher mean performance across all accuracy and efficiency metrics, while exhibiting notably tighter variance bands, particularly on localization accuracy (Fig.~\ref{fig:error_bar_robustness}, \textit{left}), as compared to the no-$M_\theta$ baseline that shows substantially wider performance spread across rollouts. This suggests that \ours not only improves average SE agent performance but also stabilizes problem-solving behaviors, reducing sensitivity to the stochasticity inherent in long-horizon agentic software engineering trajectories.

\input{figures/exp_error_bar}

\subsection{Qualitative Analysis on Memory Augmentation Success \& Failure}
\label{appendix:subsec:qualitative_analysis}

To better understand the effectiveness of $M_\theta$ in memory generation and evolution, we conduct case studies to qualitatively examine memory examples generated by \ours-FT $M_\theta$, analyzing their strengths in successful memory augmentation as well as their limitations in cases where SE agents fail, in hopes of meaningfully informing future research.

\paragraph{Effective Memory Augmentation Examples ($\Delta_{abs}>0$ \textit{across all metrics}).}

Fig.~\ref{fig:qualitative_analysis:success_examples} shows a few examples of high-quality memory generation, where generated memories effectively enable the SE agent to successfully resolve the task with improved problem-solving efficiency. As shown in these examples, good memories are able to maintain a high level of abstraction, enhancing their applicability and generalizability for more adaptive problem-solving across different SE scenarios. Moreover, their effective distillation of repository structure, key patterns, and workflows allows the SE agent to quickly familiarize itself with the codebase and start problem-solving attempts with the support of best practices and informative insights.

\paragraph{Ineffective Memory Augmentation Examples ($\Delta_{abs} \leq 0$ in $SR$).}
In contrast to effective memory augmentation (Fig.~\ref{fig:qualitative_analysis:success_examples}), ineffective memory reflection and distillation can misdirect the SE agent toward unproductive exploration and ultimately lead to task failure. Examples shown in Fig.~\ref{fig:qualitative_analysis:failure_examples} unveil several common issues: over-specific details, insufficient reflective synthesis, overly task-bound patterns, misdirected focus on suboptimal information, as well as repetitive and redundant content. Compared to successful examples (Fig.~\ref{fig:qualitative_analysis:success_examples}), effective memories emphasize codebase-level abstractions rather than task-specific details, underscoring the significance of distilling key structures, generalizable patterns, and actionable insights for memory-augmented software engineering.



\newpage

\input{figures/qualitative_analysis_success}
\input{figures/qualitative_analysis_failure}

%% file: figures/preliminary1_SE_agent_failure_analysis.tex
\begin{wrapfigure}{r}{0.5\linewidth}

\vspace{0pt}

\centering
\includegraphics[width=1.0\linewidth]{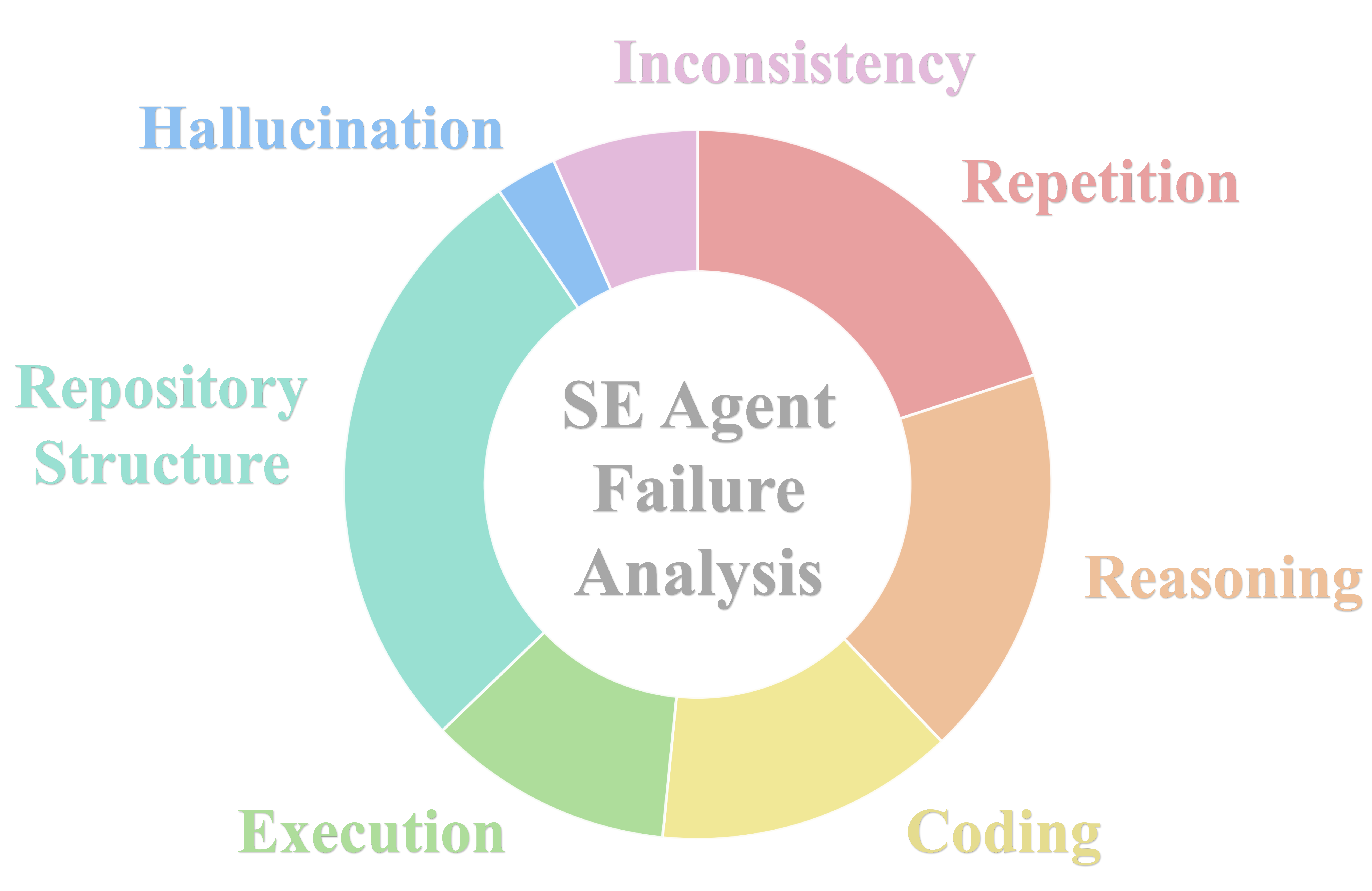}

\vspace{3pt}

\caption{\textbf{Preliminary Analysis on SE Agent Failures.} Through manual analysis, we identify seven failure patterns in SE agent problem-solving.}
\label{fig:preliminary:SE_agent_failure_analysis}

\vspace{12pt}

\end{wrapfigure}

%% file: tables/exp_preliminary_memory_structure.tex
\begin{table}[H]

\centering
\renewcommand{\arraystretch}{1.2}

\caption{\textbf{Preliminary Exploration of Memory Structures.} To investigate which memory structure can most effectively benefit SE agents, we perform a preliminary evaluation of nine memory structures compared to the no-memory baseline (highlighted in \colorbox{BASELINE_BG}{\textit{gray}}). The \textbf{\textit{absolute performance differences}} (Eq.~\ref{eq:eval_metric:deltaabs}) are shown as \textit{positive} \textcolor{deltagreen}{$\uparrow \Delta_{\textit{abs}}\%$} or \textit{negative} \textcolor{deltared}{$\downarrow \Delta_{\textit{abs}}\%$}.}
\label{tab:exp:preliminary:memory_structure}

\begin{tabularx}{\textwidth}{>{\bfseries}c>{\bfseries}cXXXXX}

\toprule

\textbf{Memory} & \textbf{Structure} & $\bm{SR}$ & $\bm{LA_{file}^{(1)}}$ & $\bm{LA_{func}^{(1)}}$ & $\bm{LE_{file}^{(1)}}$ & $\bm{LE_{func}^{(1)}}$ \\

\midrule

\rowcolor{BASELINE_BG}
\xmark & -- & 0.00 & 56.57 & 16.67 & 35.50 & 11.00 \\

\cmark & String & 0.00 & \absdeltapct{56.57}{63.33} & \absdeltapct{16.67}{43.33} & \absdeltapct{35.50}{41.85} & \absdeltapctbf{11.00}{31.65} \\
\cmark & Dictionary & 0.00 & \absdeltapctbf{56.57}{76.67} & \absdeltapct{16.67}{46.67} & \absdeltapct{35.50}{50.35} & \absdeltapct{11.00}{28.85} \\
\cmark & List & 0.00 & \absdeltapct{56.57}{53.33} & \absdeltapct{16.67}{26.67} & \absdeltapct{35.50}{42.00} & \absdeltapct{11.00}{20.35} \\
\cmark & Tree & 0.00 & \absdeltapct{56.57}{60.00} & \absdeltapct{16.67}{30.00} & \absdeltapct{35.50}{43.00} & \absdeltapct{11.00}{21.50} \\
\cmark & Graph & 0.00 & \absdeltapct{56.57}{66.67} & \absdeltapct{16.67}{46.67} & \absdeltapct{35.50}{46.50} & \absdeltapct{11.00}{28.35} \\
\cmark & Python & 0.00 & \absdeltapct{56.57}{73.33} & \absdeltapct{16.67}{40.00} & \absdeltapctbf{35.50}{56.35} & \absdeltapct{11.00}{33.00} \\
\cmark & Given & 0.00 & \absdeltapct{56.57}{60.00} & \absdeltapct{16.67}{36.67} & \absdeltapct{35.50}{42.65} & \absdeltapct{11.00}{23.50} \\
\cmark & Discretionary & 0.00 & \absdeltapct{56.57}{63.33} & \absdeltapct{16.67}{33.33} & \absdeltapct{35.50}{45.15} & \absdeltapct{11.00}{26.35} \\
\cmark & \textcolor{ourscolor}{\textit{Ours}} & 0.00 & \absdeltapctbf{56.57}{76.67} & \absdeltapctbf{16.67}{50.00} & \absdeltapct{35.50}{53.15} & \absdeltapct{11.00}{29.65} \\

\bottomrule

\end{tabularx}

\end{table}

%% file: figures/preliminary2_SE_agent_failure_example_repo_structure.tex
\begin{figure}[H]

\vspace{6pt}

\centering
\includegraphics[width=1.0\linewidth]{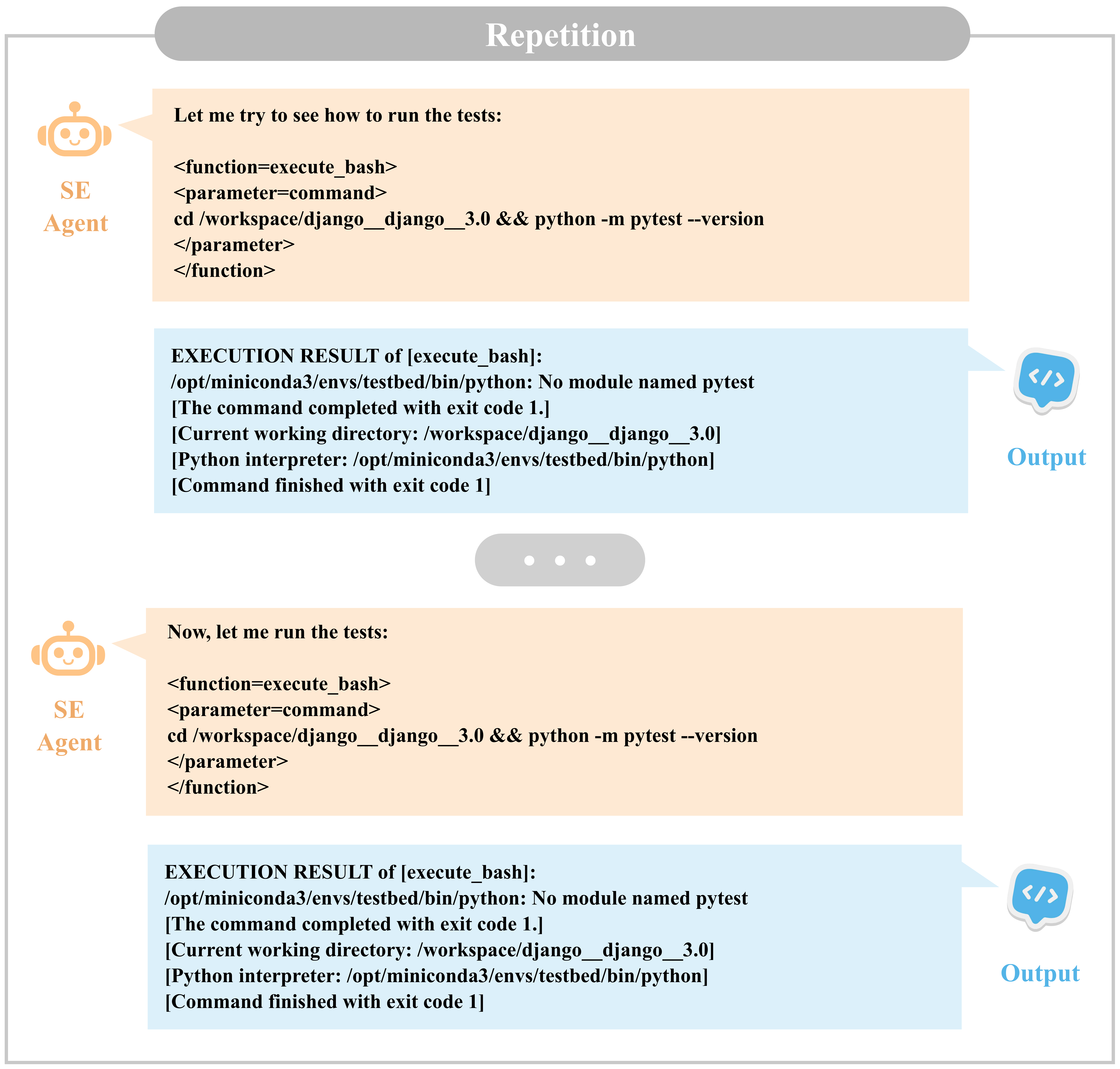}

\caption{\textbf{Failure Case of Repetition.} The SE agent repeats the same error as in earlier attempts.}
\label{fig:preliminary:failure_example2:repetition}

\vspace{-3pt}

\end{figure}

\begin{figure}[H]

\centering
\includegraphics[width=1.0\linewidth]{assets/preliminary_SE_agent_failure_example3_reasoning.pdf}

\caption{\textbf{Failure Case of Reasoning Error.} The SE agent fails due to its incorrect reasoning across multiple attempts.}
\label{fig:preliminary:failure_example3:reasoning}

\end{figure}

\begin{figure}[H]

\vspace{-12pt}

\centering
\includegraphics[width=1.0\linewidth]{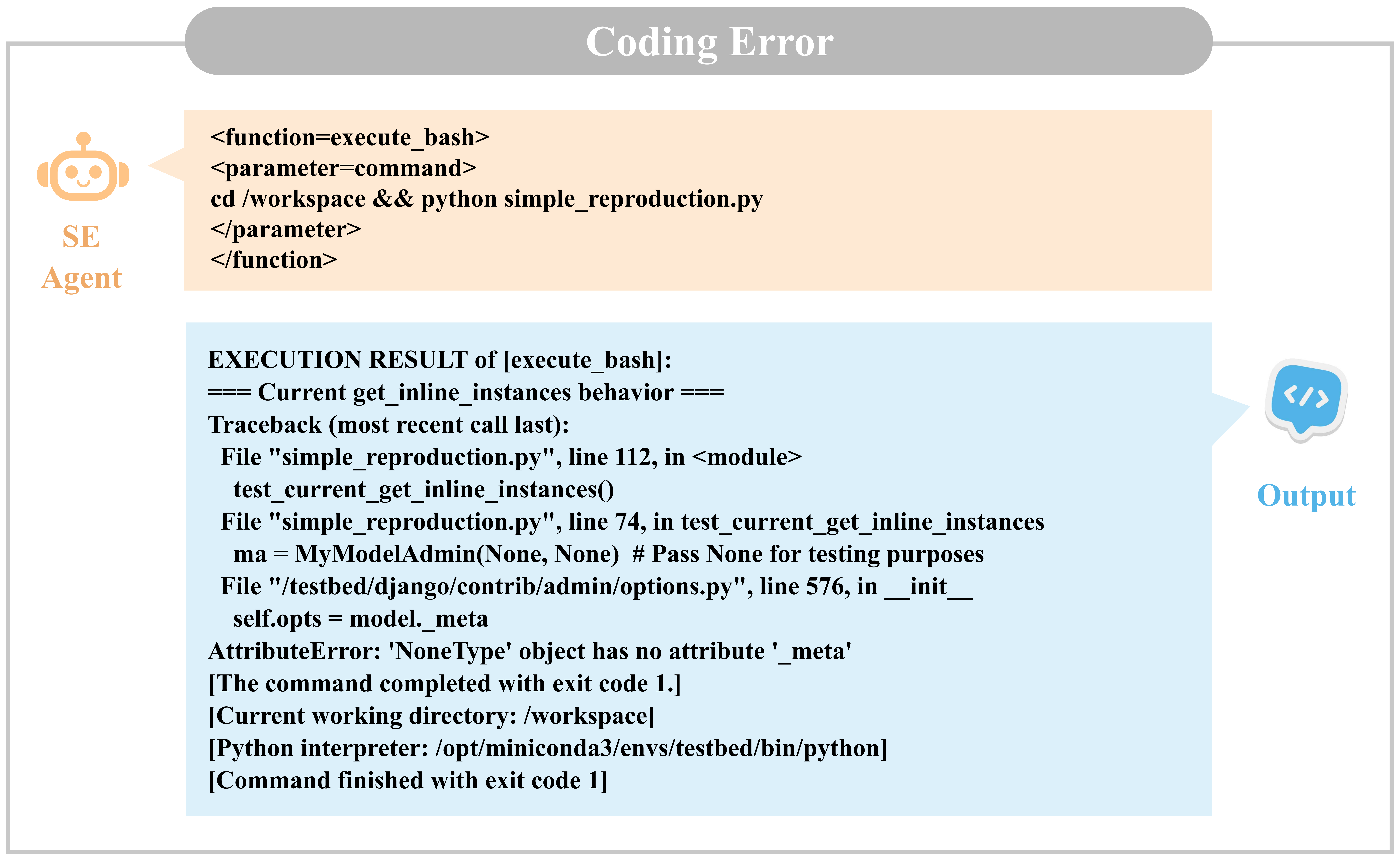}

\caption{\textbf{Failure Case of Coding Error.} The SE agent fails due to coding errors, such as \texttt{SyntaxError}, \texttt{NameError}, \texttt{AttributeError}, etc.}
\label{fig:preliminary:failure_example4:coding}

\vspace{-3pt}

\end{figure}

\begin{figure}[H]

\vspace{-18pt}

\centering
\includegraphics[width=1.0\linewidth]{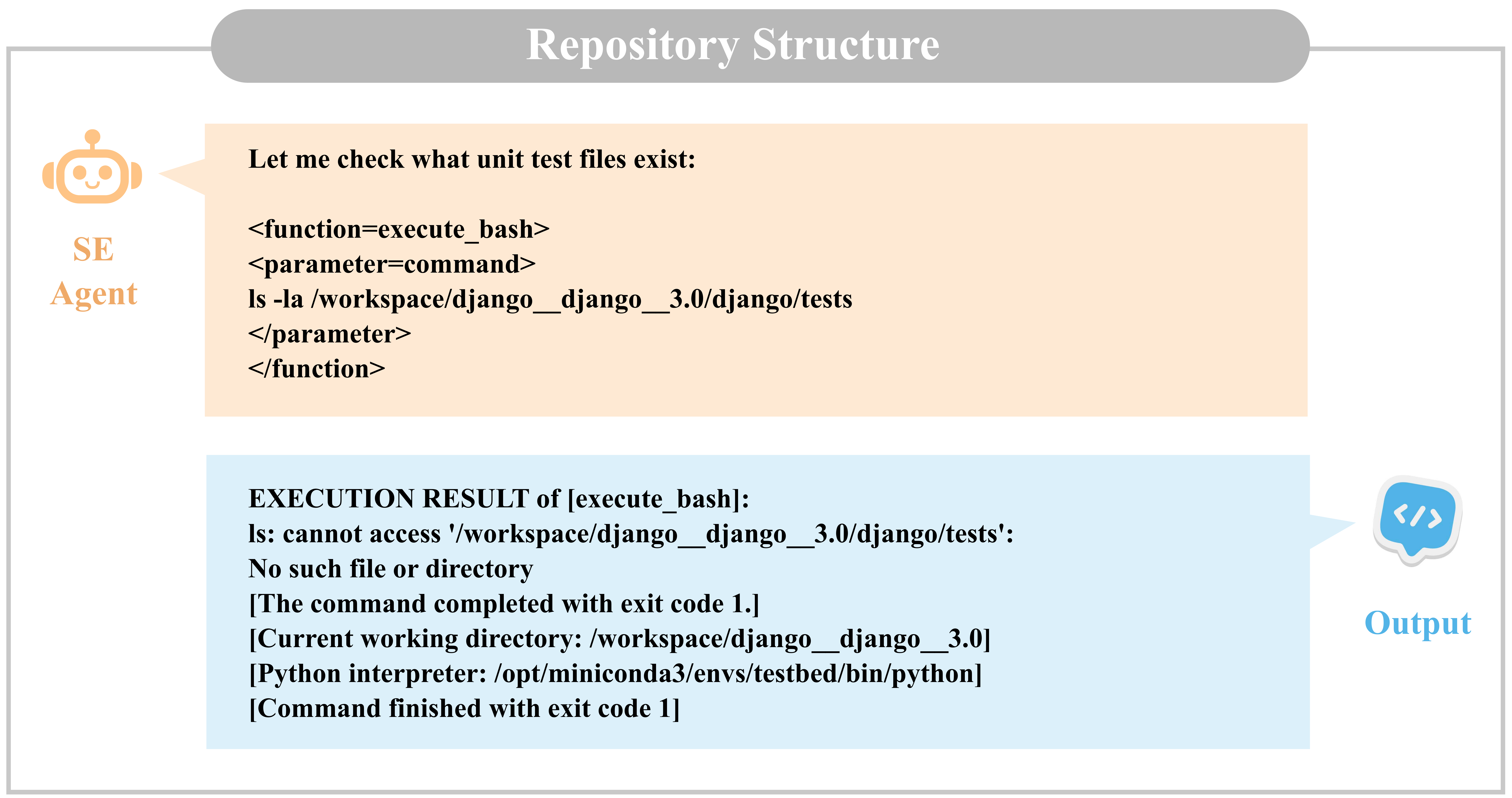}

\caption{\textbf{Failure Case of Repository Structure.} The SE agent fails due to its incorrect understanding of the repository structure.}
\label{fig:preliminary:failure_example1:repo_structure}

\vspace{12pt}

\end{figure}

\begin{figure}[H]

\vspace{-18pt}

\centering
\includegraphics[width=1.0\linewidth]{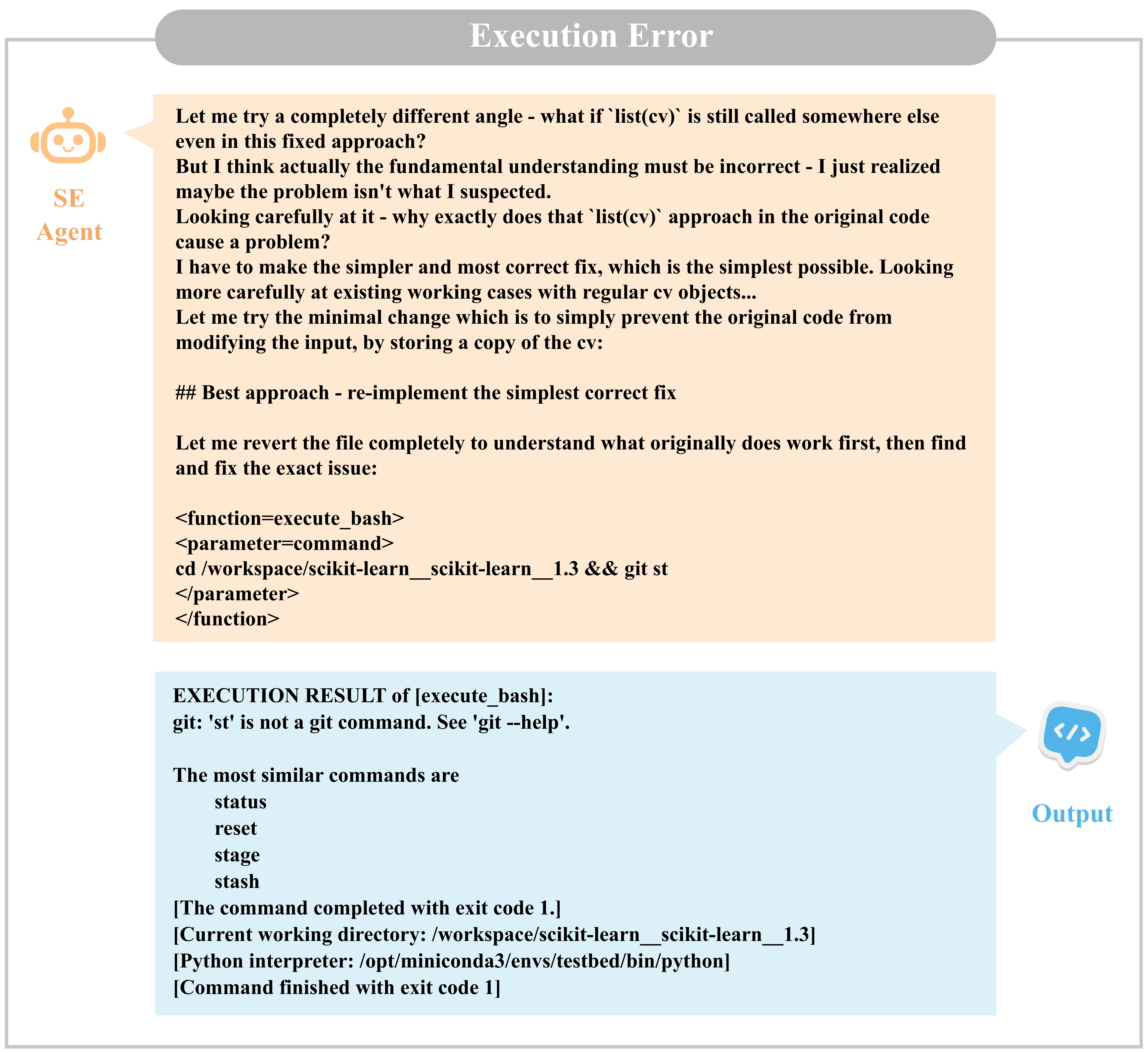}

\caption{\textbf{Failure Case of Execution Error.} The SE agent fails due to execution errors.}
\label{fig:preliminary:failure_example5:execution}

\vspace{-3pt}

\end{figure}

\begin{figure}[H]

\centering
\includegraphics[width=1.0\linewidth]{assets/preliminary_SE_agent_failure_example6_inconsistency.pdf}

\caption{\textbf{Failure Case of Inconsistency.} The SE agent fails due to the inconsistency between its reasoning and action.}
\label{fig:preliminary:failure_example6:inconsistency}

\end{figure}

\begin{figure}[H]

\vspace{-20pt}

\centering
\includegraphics[width=1.0\linewidth]{assets/preliminary_SE_agent_failure_example7_hallucination.pdf}

\caption{\textbf{Failure Case of Hallucination.} The SE agent fails due to hallucinating actions or experiences.}
\label{fig:preliminary:failure_example7:hallucination}

\end{figure}

%% file: figures/preliminary3_memory_generation_prompt.tex
\begin{figure}[H]

\vspace{-28pt}

\centering
\includegraphics[width=1.0\linewidth]{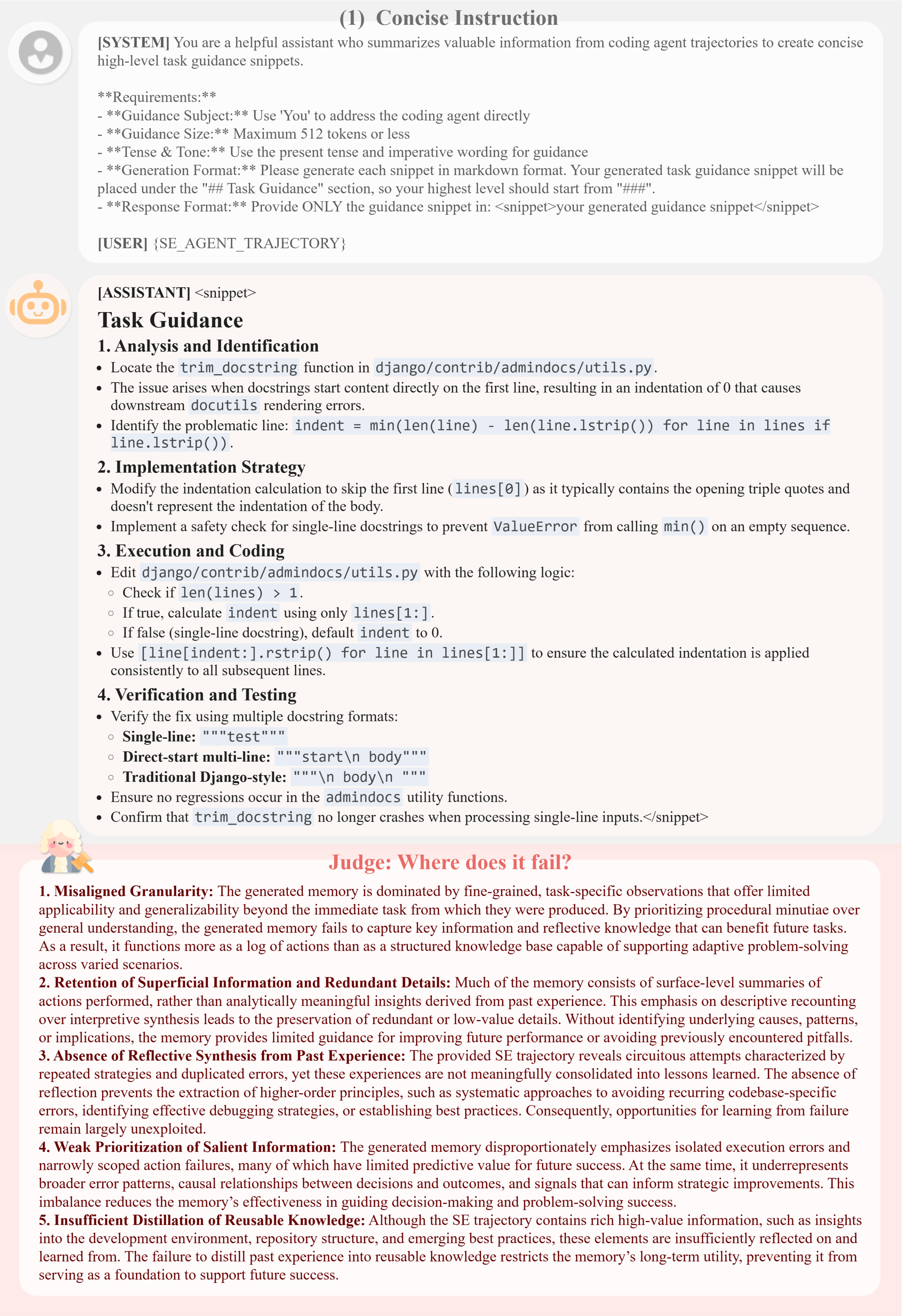}

\end{figure}

\begin{figure}[H]

\vspace{12pt}

\includegraphics[width=1.0\linewidth]{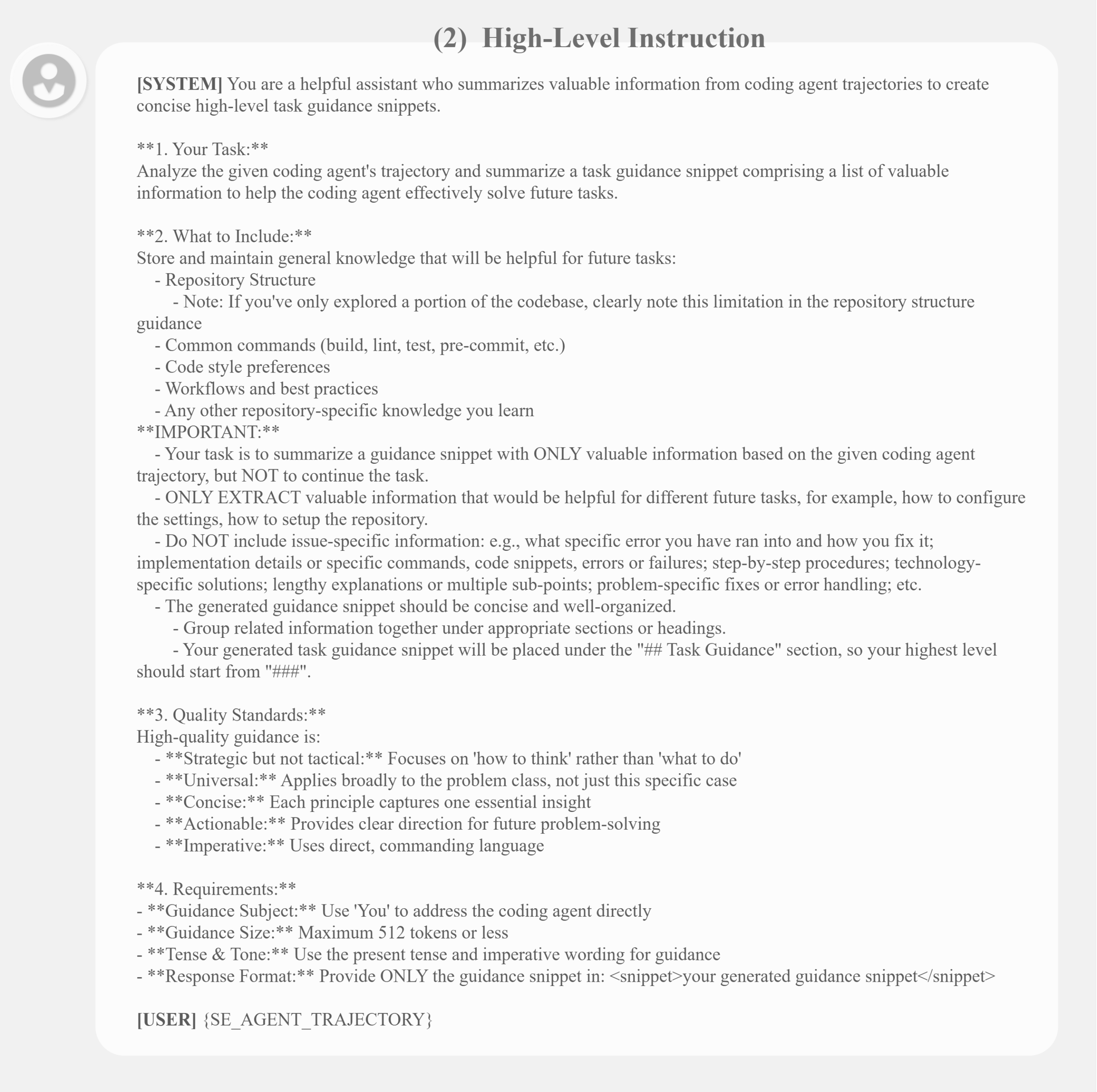}

\end{figure}

\begin{figure}[H]

\vspace{12pt}

\includegraphics[width=1.0\linewidth]{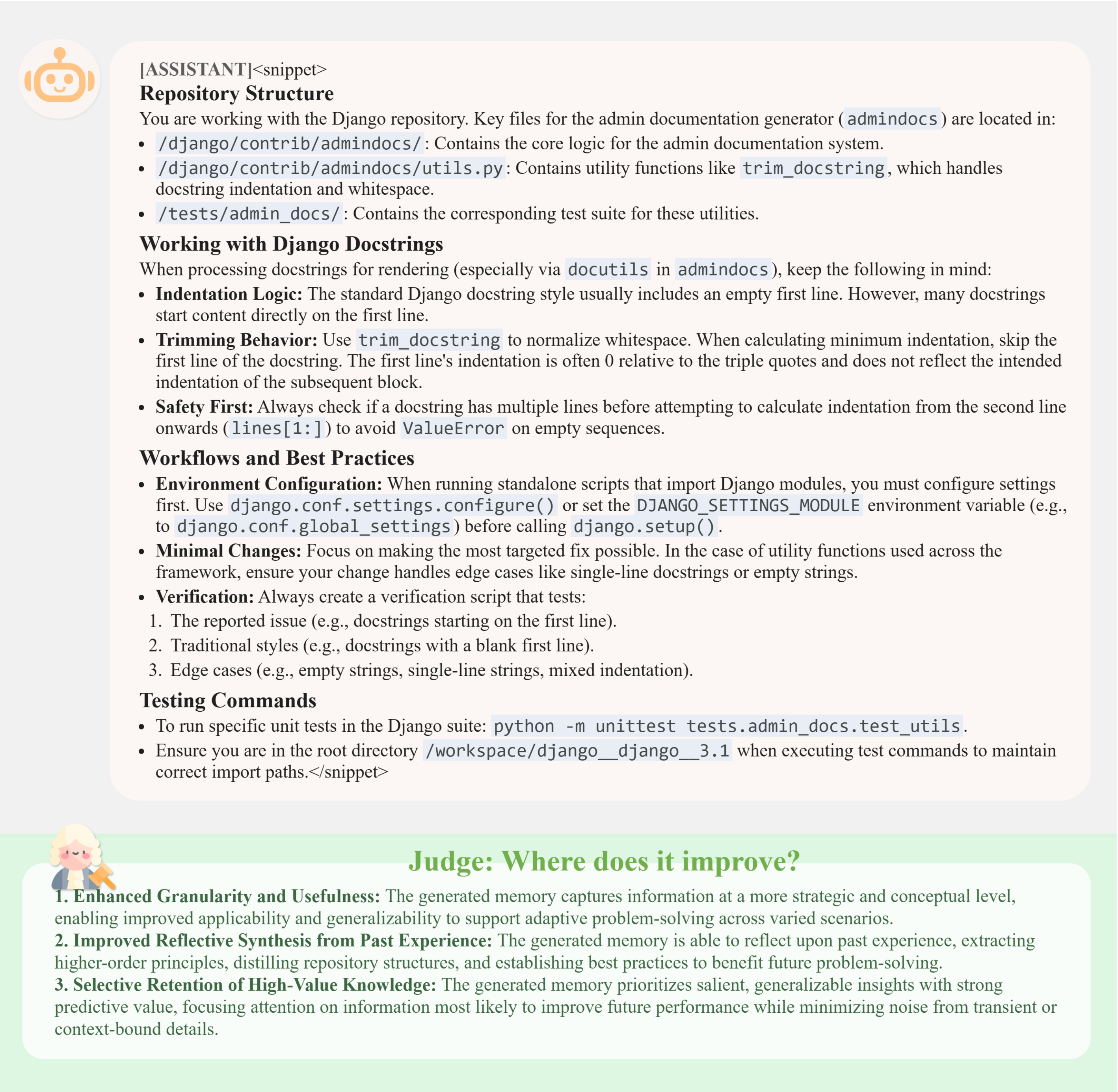}

\end{figure}

\begin{figure}[H]

\includegraphics[width=1.0\linewidth]{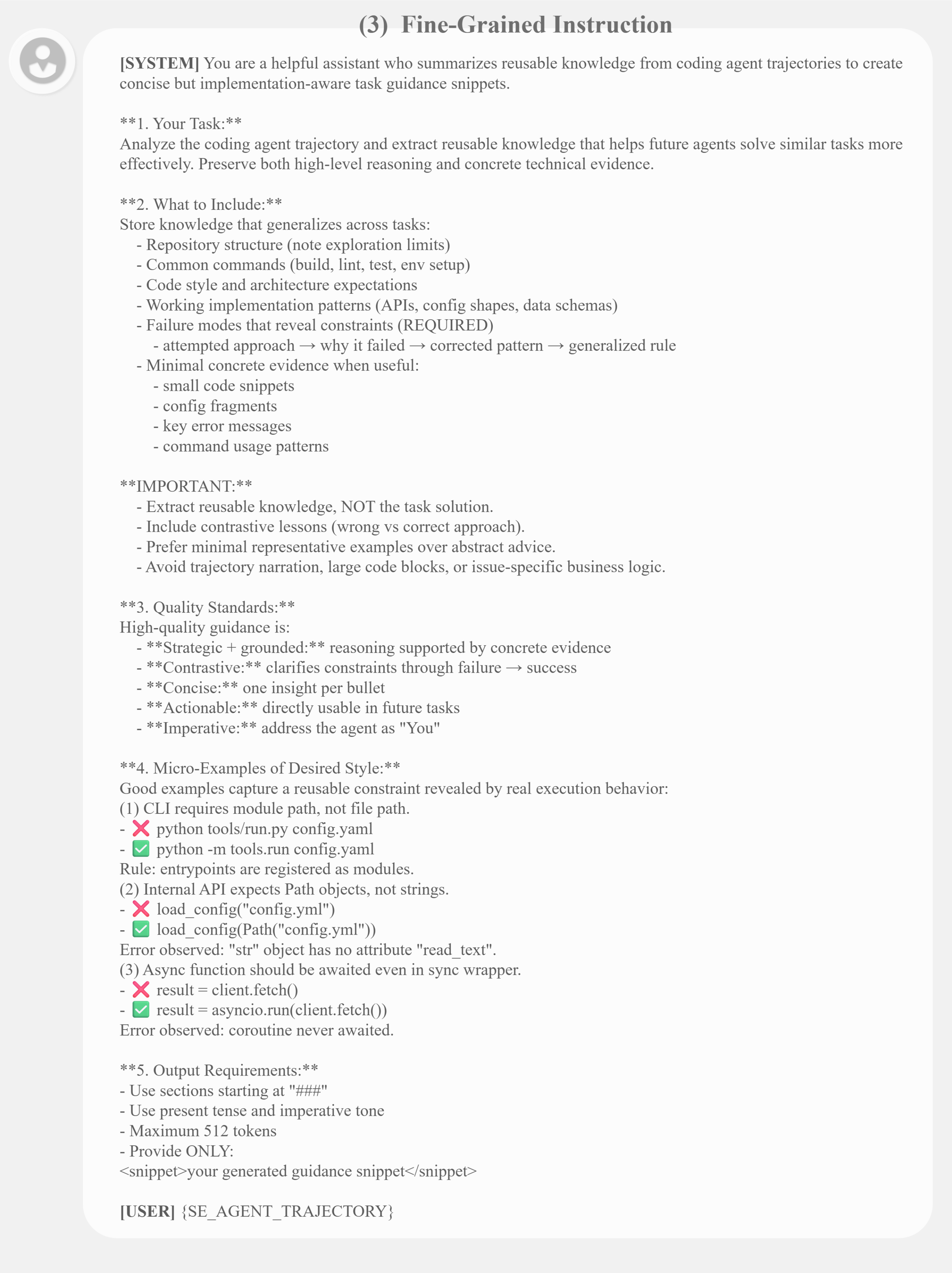}

\end{figure}

\begin{figure}[H]

\includegraphics[width=1.0\linewidth]{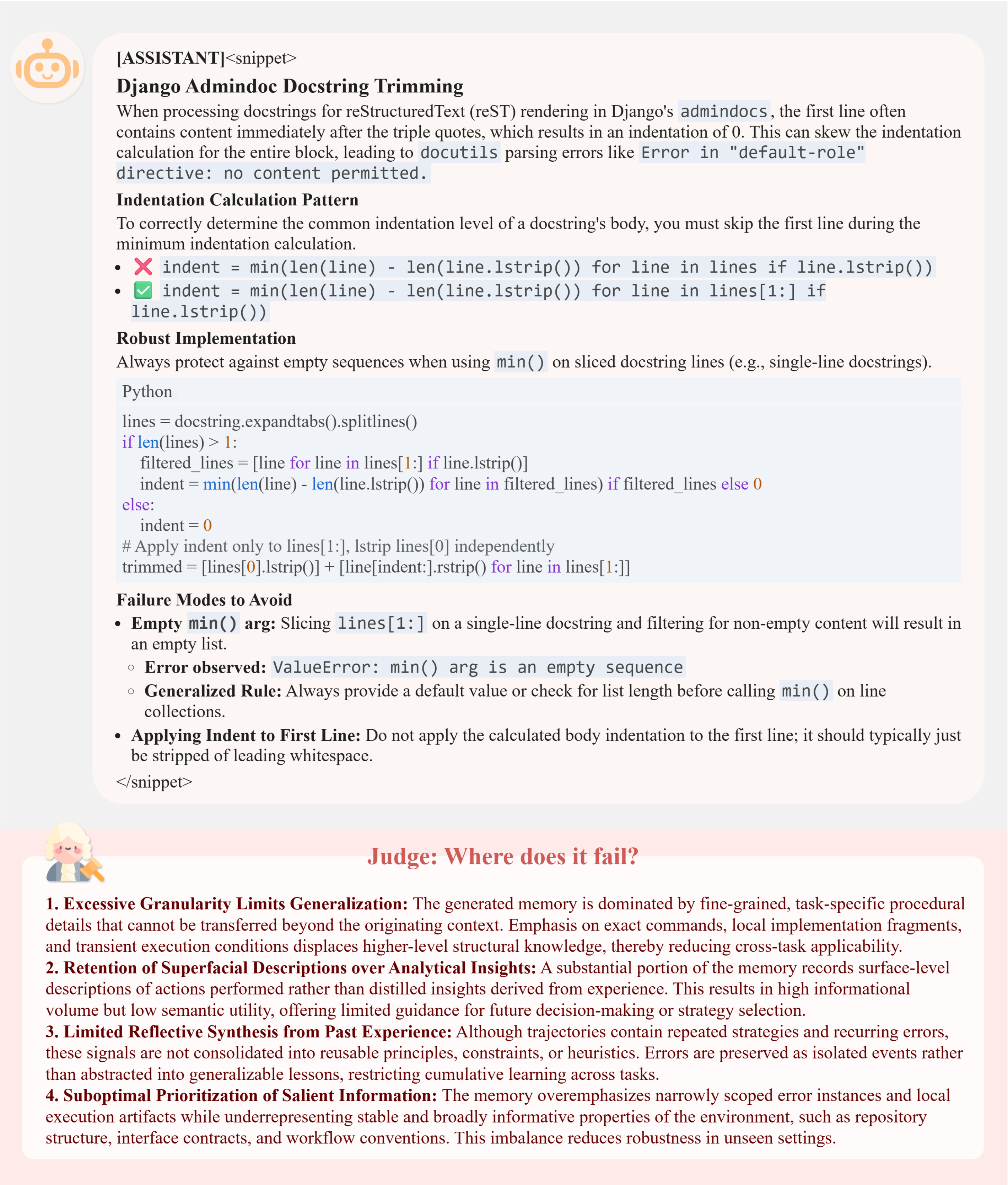}

\caption{\textbf{Memory Instruction.} We conduct the preliminary study on memory reflection through three versions of instructions: (1) \textit{general and concise} instruction, (2) \textit{high-level} instruction, and (3) \textit{fine-grained} instruction.}
\label{fig:preliminary3:memory_generation_prompt}

\vspace{-3pt}

\end{figure}

%% file: figures/exp_repo_wise_delta.tex
\begin{figure}[H]

\centering
\includegraphics[width=1.0\linewidth]{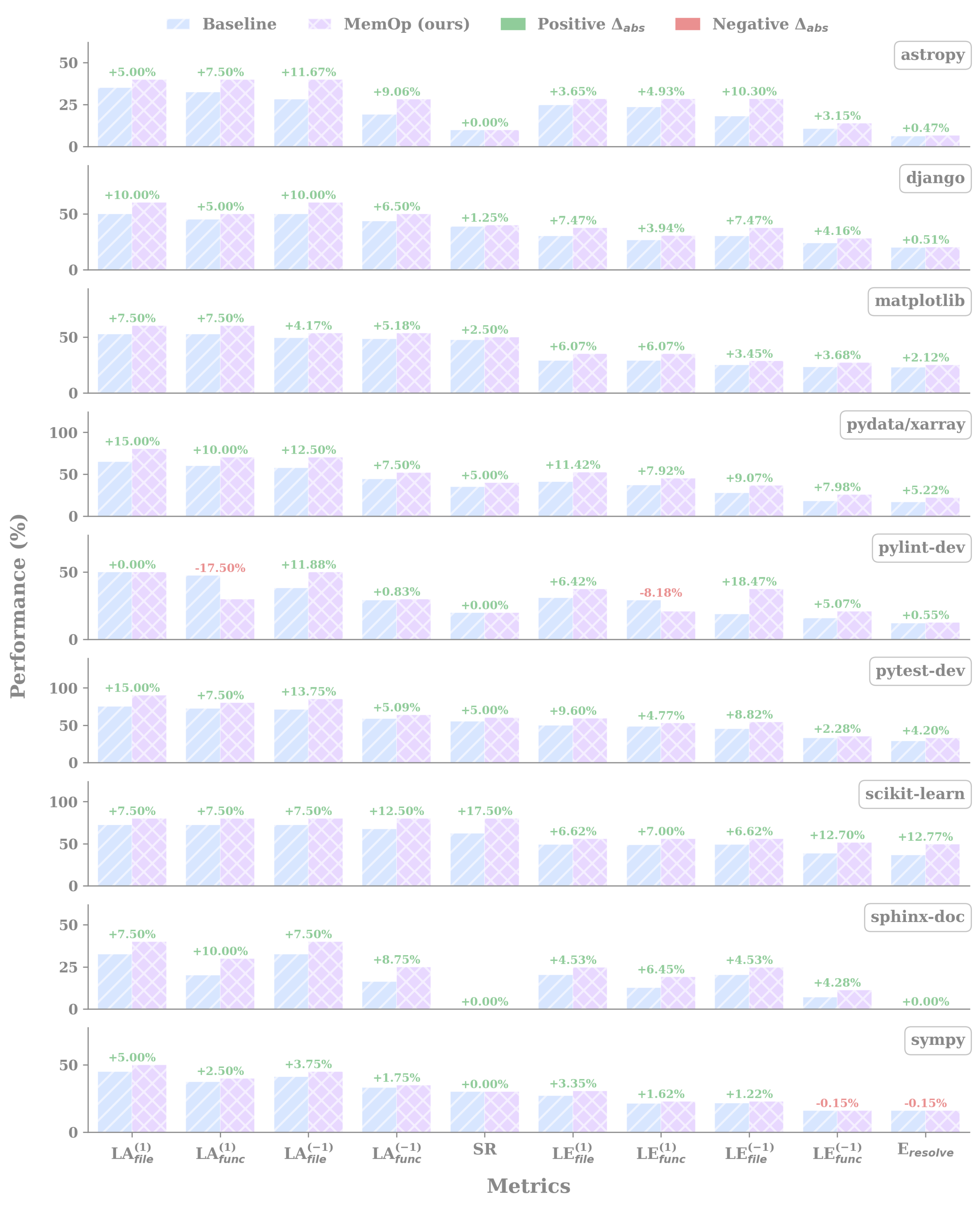}

\caption{\textbf{Repository-Wise Comparison between Baseline and \ours.} Per-repository comparison across our ten metrics (\S\ref{subsec:eval_metrics}) on the SE agent powered by \texttt{Qwen3-Coder-30B-A3B}, augmented with \ours-FT $M_\theta$ (\texttt{Qwen3-4B-T}). \ours consistently outperforms the \textit{\textcolor{gray}{no-$M_\theta$ baseline}} across nine disparate repositories, with gains up to $\Delta_{abs}=\uparrow17.50\%$ in $SR$. This reveals that the benefit is not concentrated in a few favorable codebases.}
\label{fig:repo_wise_delta}

\vspace{-3pt}

\end{figure}

%% file: figures/preliminary3_memory_instruction_result.tex
\begin{figure}[t]

\vspace{0pt}

\centering
\includegraphics[width=1.0\linewidth]{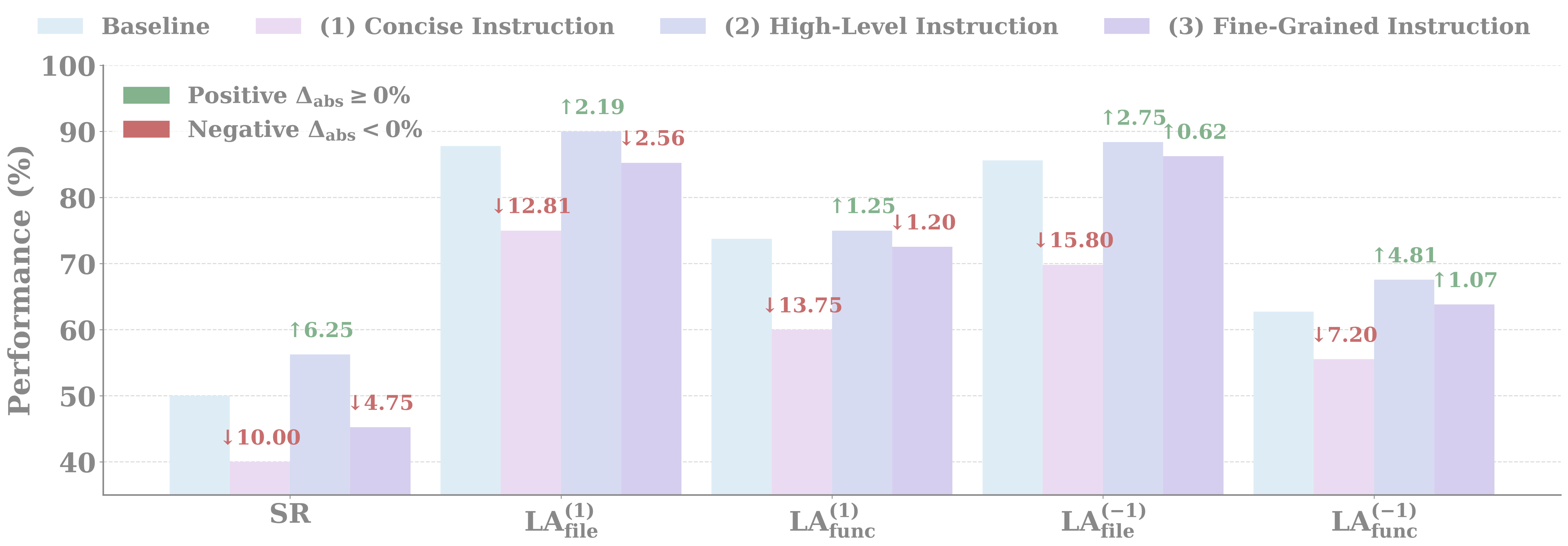}

\vspace{0pt}

\caption{\textbf{Preliminary Study on Memory Instruction.} In addition to qualitative analysis (Fig.~\ref{fig:preliminary3:memory_generation_prompt}), we summarize the quantitative results of three instructions (\S\ref{appendix:subsec:preliminary:memory_prompt}). We use \texttt{Qwen3-Coder-480B} as the coding agent and \texttt{Claude-4-Sonnet} as $M_\theta$, evaluating on $20$ SWE-Bench-\textit{Verified} samples with max-iteration set to $100$ and averaging results over $4$ runs.}
\label{fig:preliminary3:memory_generation_prompt_results}

\vspace{12pt}

\end{figure}

%% file: tables/related_work_taxonomy.tex

\begin{table}[t]

\small
\centering
\renewcommand{\arraystretch}{1.3}
\setlength{\tabcolsep}{4pt}

\caption{\textbf{A Taxonomy of Memory Systems for SE Agents.} We organize prior SE memory systems along five axes, of which the first, \textit{what defines a good memory}, is the axis that separates \ours from all of them. \textit{Agent-agnostic} indicates whether the memory mechanism attaches to a fixed SE agent without adopting that system's own control flow. Systems in the upper block are compared empirically in Tab.~\ref{tab:exp:memory_system_comparison} where a public implementation exists (\S\ref{appendix:subsec:compare_with_other_memory_methods}), while the lower block is compared conceptually.}
\label{tab:related_work_taxonomy}

\begin{tabularx}{\textwidth}{@{}l>{\raggedright\arraybackslash}X>{\raggedright\arraybackslash}X>{\raggedright\arraybackslash}X>{\raggedright\arraybackslash}Xc@{}}

\toprule

\textbf{Memory System} & \textbf{Utility Definition} & \textbf{Supervision} & \textbf{Granularity} & \textbf{Optimization Signal} & \textbf{Agent-Agnostic} \\

\midrule

\makecell[tl]{SWE-Bench-CL\\\citep{joshi2025swebenchcl}} & Not defined (benchmark) & -- & Task sequence & None & -- \\
\addlinespace[4pt]

\makecell[tl]{CTIM-Rover\\\citep{lindenbauer2025knowledge2noise}} & Unvalidated; unfiltered retrieval & None & Episode & Retrieval only & \cmark \\
\addlinespace[4pt]

\makecell[tl]{MemGovern\\\citep{wang2026memgovernenhancingcodeagents}} & Human curation & Human governance & Experience card & Retrieval only & Partial \\
\addlinespace[4pt]

\arrayrulecolor{gray!50}\cmidrule(l){1-6}\arrayrulecolor{black}

\makecell[tl]{Subtask-Level Memory\\\citep{shen2026structurallyalignedsubtasklevelmemory}} & Structural prior over subtasks & Heuristic design & Subtask & Retrieval only & \xmark \\
\addlinespace[4pt]

\makecell[tl]{Repository Memory\\\citep{Wang2025ImprovingCLA}} & Localization gain on one subtask & Task-specific & Repository & Retrieval only & \xmark \\
\addlinespace[4pt]

\makecell[tl]{ExpeRepair\\\citep{mu2026experepair}} & LLM-judged reproduction success & LLM judge & Episodic + semantic & Prompting; no parameter update & \xmark \\
\addlinespace[4pt]

\makecell[tl]{SWE-Exp\\\citep{chen2026sweexpexperience}} & Similarity to gold patches & Gold patches + LLM & Episode & Reranking; no parameter update & \xmark \\
\addlinespace[4pt]

\arrayrulecolor{gray!50}\cmidrule(l){1-6}\arrayrulecolor{black}

\makecell[tl]{\textcolor{ourscolor}{\textbf{\ours (ours)}}} & \textcolor{ourscolor}{\textbf{Measured $\Delta$ over $N_Q{=}10$ metrics (Eq.~\ref{eq:memory_acceptance})}} & \textcolor{ourscolor}{\textbf{None (annotation-free)}} & \textcolor{ourscolor}{\textbf{Episode}} & \textcolor{ourscolor}{\textbf{SFT + RL of $M_\theta$}} & \cmark \\
\addlinespace[4pt]

\bottomrule

\end{tabularx}

\vspace{18pt}

\end{table}

%% file: figures/eval_dataset_distribution.tex
\begin{wrapfigure}{r}{0.5\linewidth}

\vspace{-24pt}

\centering
\includegraphics[width=1.0\linewidth]{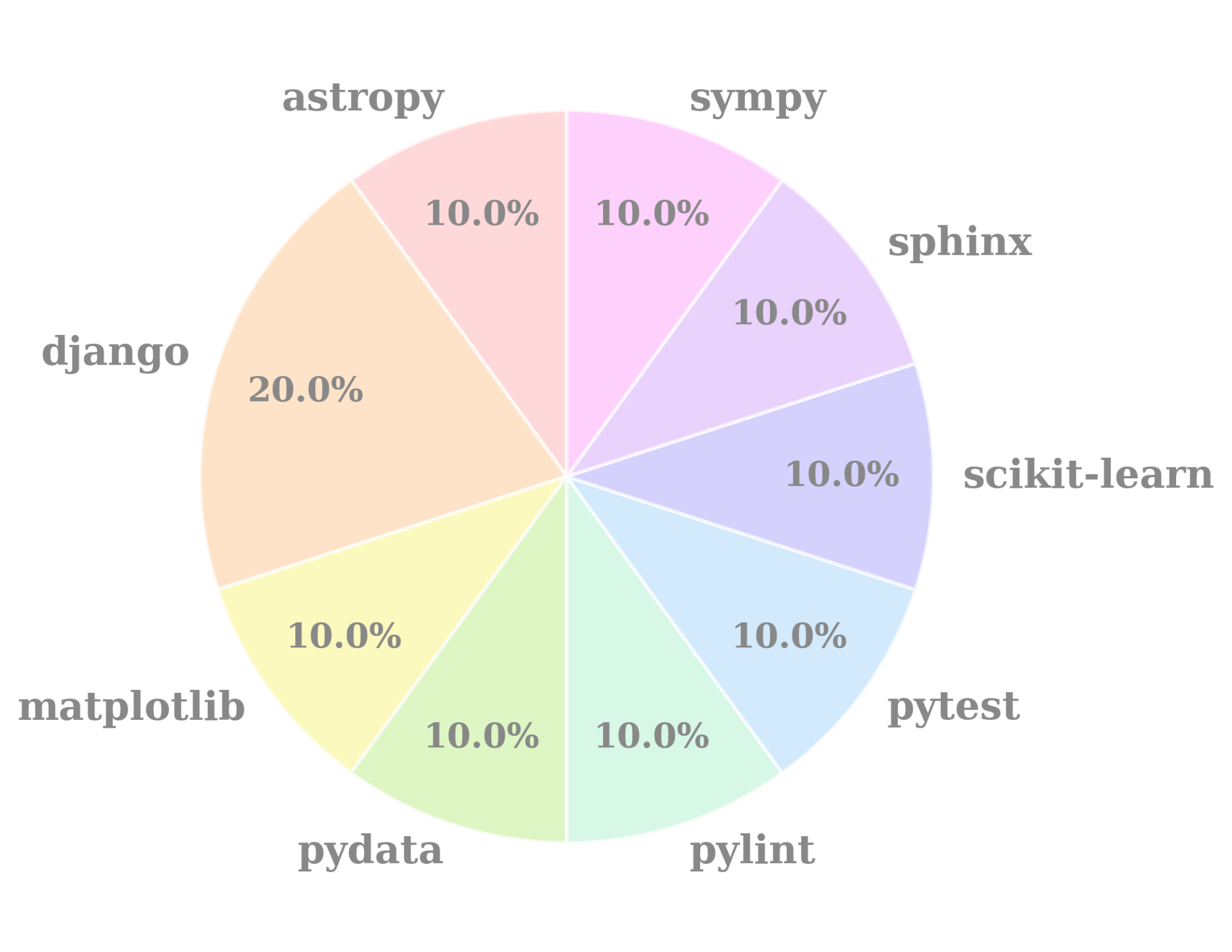}

\vspace{-12pt}

\caption{\textbf{Evaluation Set Distribution}}
\label{fig:eval_dataset_distribution}

\vspace{-20pt}

\end{wrapfigure}

%% file: figures/memory_prompts.tex

\begin{figure}[H]

\vspace{60pt}

\includegraphics[width=1.0\linewidth]{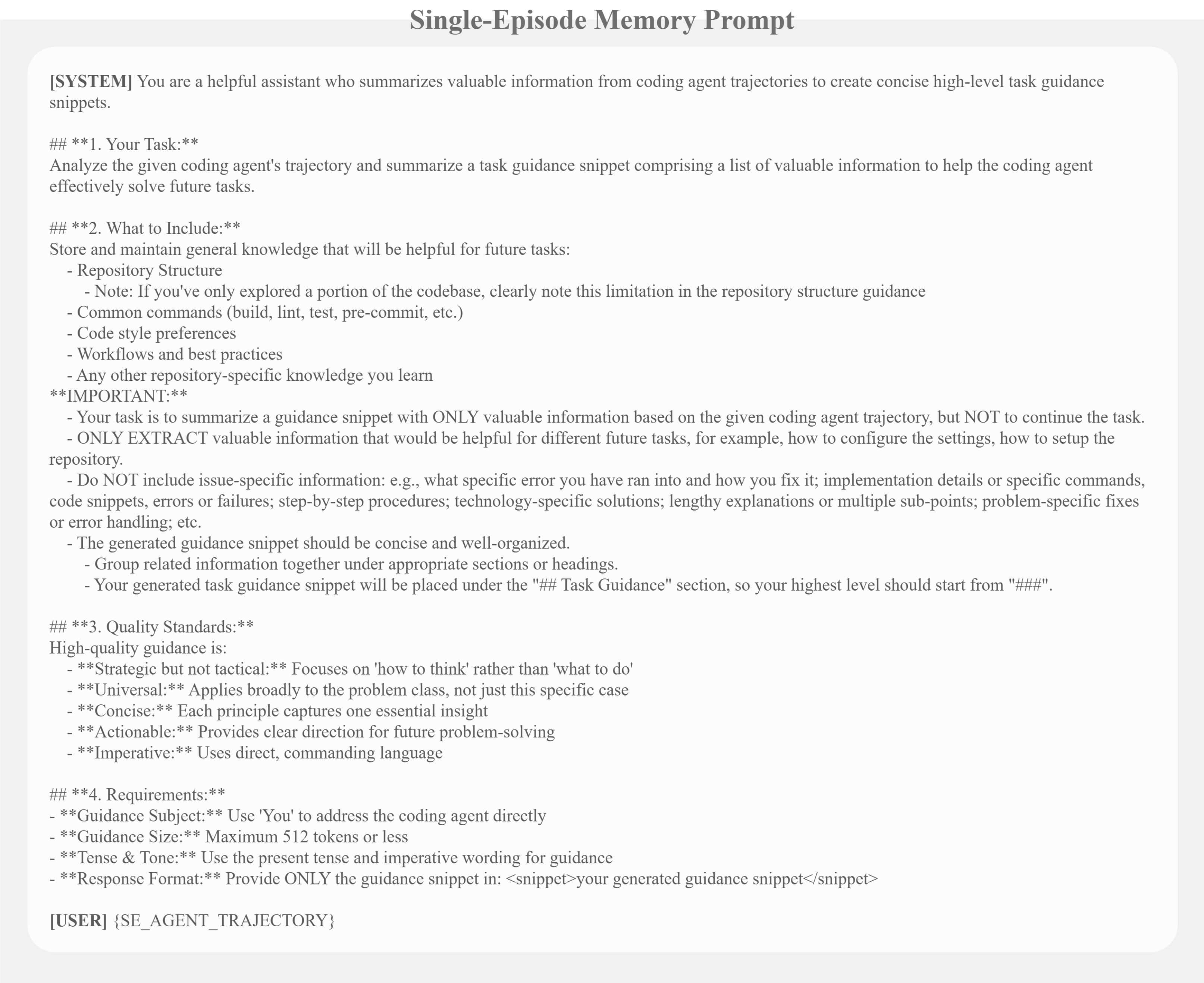}

\vspace{60pt}

\end{figure}

\begin{figure}[H]

\vspace{-28pt}

\includegraphics[width=1.0\linewidth]{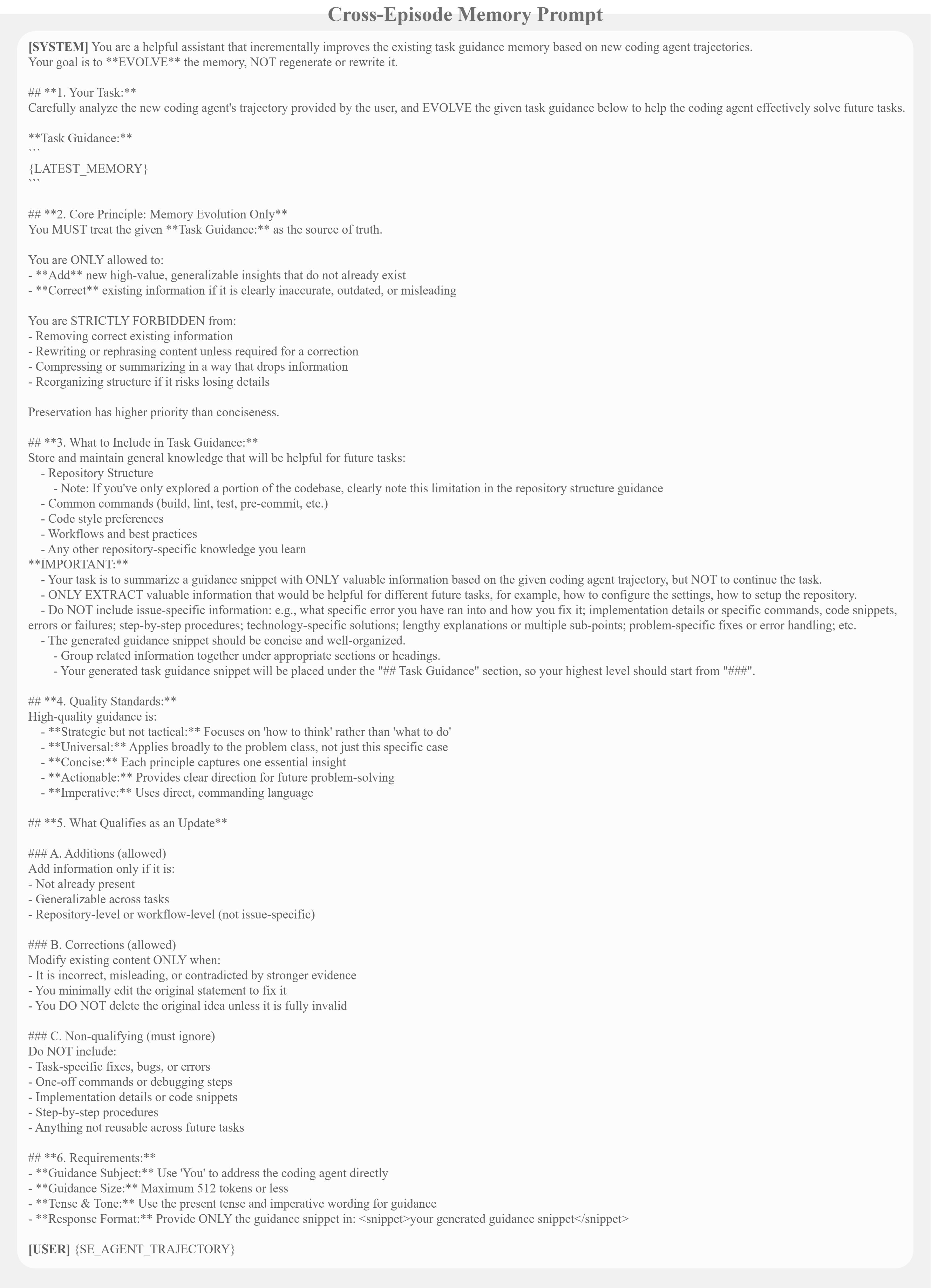}

\caption{\textbf{Instructions for \textit{Single-Episode} and \textit{Cross-Episode} Memory Generation.} Our memory generation instructions for \textit{single-episode} and \textit{cross-episode} memory-augmented software engineering settings.}
\label{fig:two_prompt_for_single_and_cross_episode}

\vspace{-3pt}

\end{figure}

%% file: figures/memory.tex
\begin{figure}[!t]

\vspace{0pt}

\centering
\includegraphics[width=1.0\linewidth]{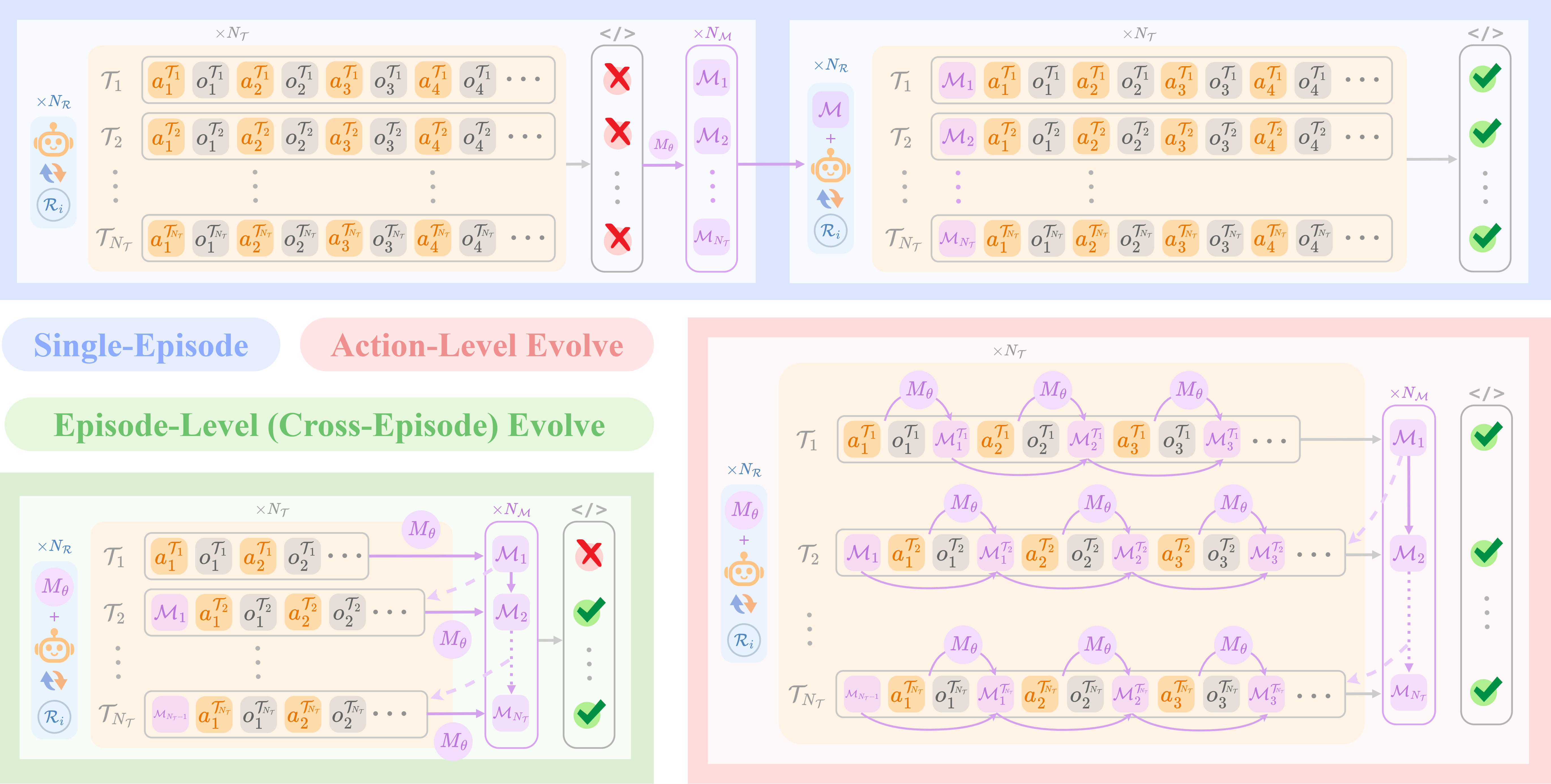}

\caption{\textbf{\ours for Memory-Augmented Software Engineering.} \ours finetunes $M_\theta$ to augment SE agents through adaptive memory generation (\S\ref{subsec:memop_methodology} \& \S\ref{appendix:subsec:exp_evolve_granularity}). Our ablation studies include \textit{single-episode} memory generation, \textit{episode-level} memory evolution (\S\ref{sec:exps}), and \textit{action-level} memory evolution (\S\ref{appendix:subsec:exp_evolve_granularity}).}
\label{fig:ours_memory_method}

\vspace{18pt}

\end{figure}

%% file: tables/training_data.tex
\begin{table}[H]
\small
\centering
\caption{\textbf{Training Data Overview.} Summarization of training datasets in SFT and RL, constructed through trajectory-based rejection sampling (\S\ref{subsec:rejection_sampling}).}
\label{tab:training_data_overview}

\newlength{\coltwo}
\newlength{\colthree}
\newlength{\colfour}
\newlength{\colfive}
\newlength{\colone}

\setlength{\coltwo}{0.18\textwidth}
\setlength{\colthree}{0.18\textwidth}
\setlength{\colfour}{0.18\textwidth}
\setlength{\colfive}{0.18\textwidth}

\setlength{\colone}{\dimexpr\textwidth-\coltwo-\colthree-\colfour-\colfive-8\tabcolsep\relax}

\begin{tabular}{@{}p{\colone}p{\coltwo}p{\colthree}p{\colfour}p{\colfive}@{}}
\toprule
& \centering $N_{\mathcal{R}}$ & \centering Train & \centering Test & \centering\arraybackslash Total \\

\midrule

\multicolumn{5}{c}{\cellcolor{PURPLE_BG}\textbf{SFT}} \\
\noalign{\vskip 0.3em}
$\mathbf{\mathcal{D}_{\text{SFT}}}$ & \centering 10 & \centering 2516 & \centering 280 & \centering\arraybackslash\textbf{2796} \\[0.3em]

\multicolumn{5}{c}{\cellcolor{PURPLE_BG}\textbf{RL}} \\

\noalign{\vskip 0.3em}

$\mathbf{\mathcal{D}_{\text{Pref}} (c=2)}$ & \centering 10 & \centering 1908 & \centering 214 & \centering\arraybackslash\textbf{2122} \\
$\mathbf{\mathcal{D}_{\text{Pref}} (c=4)}$ & \centering 10 & \centering 1310 & \centering 146 & \centering\arraybackslash\textbf{1456} \\

\bottomrule

\end{tabular}
\end{table}

%% file: tables/exp_configuration.tex
\begin{table}[H]
\renewcommand{\arraystretch}{1.3}
\small
\centering

\caption{\textbf{Experiment Configuration.} This table summarizes the key configuration settings during dataset construction, supervised finetuning, reinforcement learning, and SE evaluation. $N_{\mathcal{R}}$, $N_{\mathcal{T}}$, $N_\tau$, and $N_{\mathcal{M}}$ denote the number of repositories, the number of tasks for each repository, the number of trajectories per task, and the number of memory candidates per trajectory, respectively. The number of evaluation metrics used to measure SE agent performance is set to $N_Q=10$ in both dataset construction and performance evaluation, complemented by $\Delta_{abs}$ (Eq.~\ref{eq:eval_metric:deltaabs}) and $\Delta_{rel}$ (Eq.~\ref{eq:eval_metric:deltarel}) to quantify performance differences. For model finetuning, the learning rate is represented by $lr$, $n_{\textit{epoch}}$ denotes the number of training epochs. During RL, $n_{\textit{rollout}}$ denotes the number of rollouts, and $c$ denotes the batch size of preference rollout in $\mathcal{D}_{\textit{RL}}$ (\S\ref{subsec:two_stage_sft_rl_training}). In the \textbf{Construction} row, $N_{\mathcal{R}} \times N_{\mathcal{T}} = 10 \times 10 = 100$ training tasks. In the \textbf{Evaluation} row, $N_{\mathcal{T}}=10$ is the per-repository default: the $100$ evaluation instances span $N_{\mathcal{R}}=9$ repositories, with \texttt{django} contributing $20$ as the largest portion of SWE-Bench-\textit{Verified} (\S\ref{appendix:sec:dataset_construction}, Fig.~\ref{fig:eval_dataset_distribution}).}
\label{tab:exp:configuration}

\begin{tabular*}{\textwidth}{@{\extracolsep{\fill}}ccccccccccc@{}}

\toprule
 
\textbf{Stage} & $\bm{N_{\mathcal{R}}}$ & $\bm{N_{\mathcal{T}}}$ & $\bm{N_\tau}$ & $\bm{N_{\mathcal{M}}}$ & $\bm{N_{\mathcal{A}}}$ & $\bm{N_Q}$ & $\bm{lr}$ & $\bm{n_{\textit{epoch}}}$ & $\bm{n_{\textit{rollout}}}$ & $\bm{c}$ \\

\cmidrule(lr){1-1}
\cmidrule(lr){2-7}
\cmidrule(lr){8-11}

\textbf{Construction} & 10 & 10 & 4 & 4 & 100 & 10 & -- & -- & -- & -- \\

\textbf{SFT} & -- & -- & -- & -- & -- & -- & $1\times10^{-4}$ & 1 & -- & -- \\

\textbf{RL} & -- & -- & -- & -- & -- & -- & $1 \times 10^{-6}$ & 1 & 4 & 2 / 4 \\

\textbf{Evaluation} & 9 & 10 & 1 & 4 & 100 & 10 & -- & -- & -- & -- \\

\bottomrule

\end{tabular*}

\end{table}

%% file: tables/exp_computation_overhead.tex
\begin{table}[H]
\renewcommand{\arraystretch}{1.3}
\small
\centering

\caption{\textbf{Computation Overhead.} This table summarizes the API cost (averaged to \textit{per $100$ instances}) and computation overhead during training and evaluation. $n_{MLLM}$ denotes the number of different backbone MLLMs. During dataset construction, $M_\theta$ is used \emph{exclusively} for memory candidate generation while \emph{does not perform any memory optimization}, and the additional cost of SE agent w/ $M_\theta$ arises solely from $N_{\tau} \times N_{\mathcal{M}}=4 \times 4=16$ rollouts.}
\label{tab:exp:computation_overhead}


\begin{tabular*}{\textwidth}{@{\extracolsep{\fill}}cccccc@{}}
\toprule
\multirow{2}{*}{\textbf{Stage}} 
& \multirow{2}{*}{\textbf{LLM}} 
& \multicolumn{2}{c}{\textbf{API (\$)}} 
& \multicolumn{2}{c}{\textbf{Computation (H100)}} \\

\cmidrule(lr){3-4}
\cmidrule(lr){5-6}

& & \textbf{w/o} $\bm{M_\theta}$ & \textbf{w/} $\bm{M_\theta}$ & \textbf{SE Agent} & $\bm{M_\theta}$ \\

\midrule

\multirow{3}{*}{\textbf{Construction}} & \textbf{Devstral-Small} & 11.73 & 43.52 & -- & -- \\
& \textbf{Qwen3-Coder-480B} & 86.18 & 329.98 & -- & -- \\
& \textbf{Claude-4-Sonnet} & -- & 53.89 & -- & -- \\[1pt]

\cmidrule(l){1-6}

\textbf{SFT} & -- & -- & -- & -- & $80GB \times 2$ \\
\textbf{RL} & -- & -- & -- & -- & $80GB \times 4$ \\

\cmidrule(l){1-6}

\multirow{2}{*}{\textbf{Evaluation}} & \textbf{Devstral-Small} & 11.73 & 10.26 & -- & $80GB \times 1$ \\
& \textbf{Qwen3-Coder-30B} & 9.97 & 8.94 & -- & $80GB \times 1$ \\

\bottomrule

\end{tabular*}

\end{table}

%% file: tables/exp_activity_coverage.tex

\begin{table}[H]

\small
\centering
\renewcommand{\arraystretch}{1.2}
\setlength{\tabcolsep}{4pt}

\caption{\textbf{SE Activity Coverage across Our Four Evaluation Benchmarks.} Every evaluation instance is labeled by the SE activity it involves. Labels are multi-label, so an instance may fall under several activities and column sums exceed the instance count of each benchmark ($100$, $129$, $100$, and $16$, respectively). \textit{Scored under} names the signal by which each activity is evaluated. Labels are generated by \texttt{Claude-4.5-Sonnet} and serve to characterize evaluation coverage rather than to support a quantitative claim, and we make no significance claim for activities with fewer than ten instances.}
\label{tab:exp:activity_coverage}

\begin{tabularx}{\textwidth}{@{}>{\raggedright\arraybackslash}X>{\raggedright\arraybackslash}Xccccc@{}}

\toprule

\multirow{2}{*}{\textbf{SE Activity}} & \multirow{2}{*}{\textbf{Scored Under}} & \multicolumn{4}{c}{\textbf{Labeled Instances}} & \multirow{2}{*}{\textbf{Total}} \\
\cmidrule(lr){3-6}
& & \textbf{SWE-Bench-\textit{V}} & \textbf{Multi-SWE} & \textbf{TestGen} & \textbf{Commit0} & \\

\midrule

Program comprehension & $LA$ (Eq.~\ref{eq:eval_metric:LA}), $LE$ (Eq.~\ref{eq:eval_metric:LE}) & 100 & 129 & 100 & 16 & \textbf{345} \\
Bug fixing & fail-to-pass tests & 75 & 76 & 0 & 0 & \textbf{151} \\
Testing / test-related & fail-to-pass tests, coverage & 1 & 3 & 100 & 16 & \textbf{120} \\
Error handling \& validation & fail-to-pass tests & 17 & 34 & 50 & 15 & \textbf{116} \\
Documentation change & $LA$, $LE$, fail-to-pass tests & 7 & 41 & 40 & 6 & \textbf{94} \\
API design \& deprecation & fail-to-pass, regression tests & 4 & 17 & 35 & 16 & \textbf{72} \\
Feature implementation & fail-to-pass tests & 12 & 29 & 2 & 16 & \textbf{59} \\
Refactoring & fail-to-pass, regression tests & 13 & 12 & 0 & 0 & \textbf{25} \\

\arrayrulecolor{gray!50}\cmidrule(l){1-7}\arrayrulecolor{black}

Performance optimization & fail-to-pass, regression tests & 5 & 8 & 0 & 4 & 17 \\
Security fix & fail-to-pass, regression tests & 1 & 5 & 3 & 2 & 11 \\
Build / dependency / config & fail-to-pass, build success & 0 & 7 & 1 & 0 & 8 \\

\bottomrule

\end{tabularx}

\vspace{12pt}

\end{table}

%% file: figures/exp_evolve_granularity.tex
\begin{wrapfigure}{r}{0.5\linewidth}

\vspace{-12pt}

\centering
\includegraphics[width=1.0\linewidth]{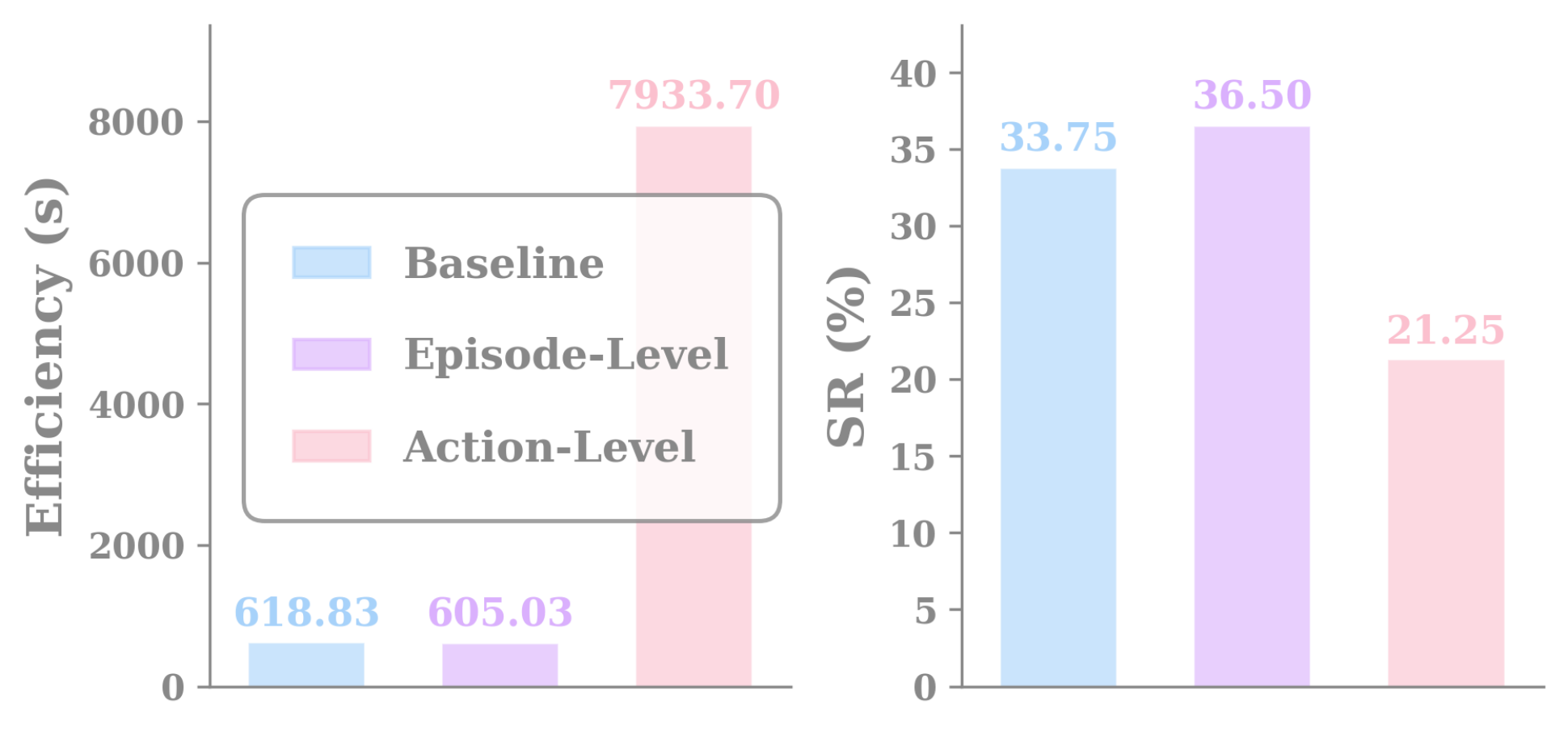}

\caption{\textbf{Memory Evolution Granularity.} We compare \textit{action-level} and \textit{episode-level} memory evolution against the \textit{\textcolor{gray}{no-$M_\theta$ baseline}}, using \ours-FT $M_\theta$ (\texttt{Qwen3-4B-T}) on the SE agent powered by \texttt{Qwen3-Coder-30B-A3B}. \textit{Action-level} evolution requires up to $\times11.82$ more time per task and lowers $SR$ by $\Delta_{abs}=\downarrow12.50\%$, whereas \textit{episode-level} evolution raises $SR$ by $\Delta_{abs}=\uparrow2.75\%$ at reduced overhead, motivating our episode-level design.}
\label{fig:exp_evolution_granularity}

\vspace{0pt}

\end{wrapfigure}

%% file: figures/exp_ablation_rl_option.tex
\begin{figure}[H]

\vspace{0pt}

\centering
\includegraphics[width=1.0\linewidth]{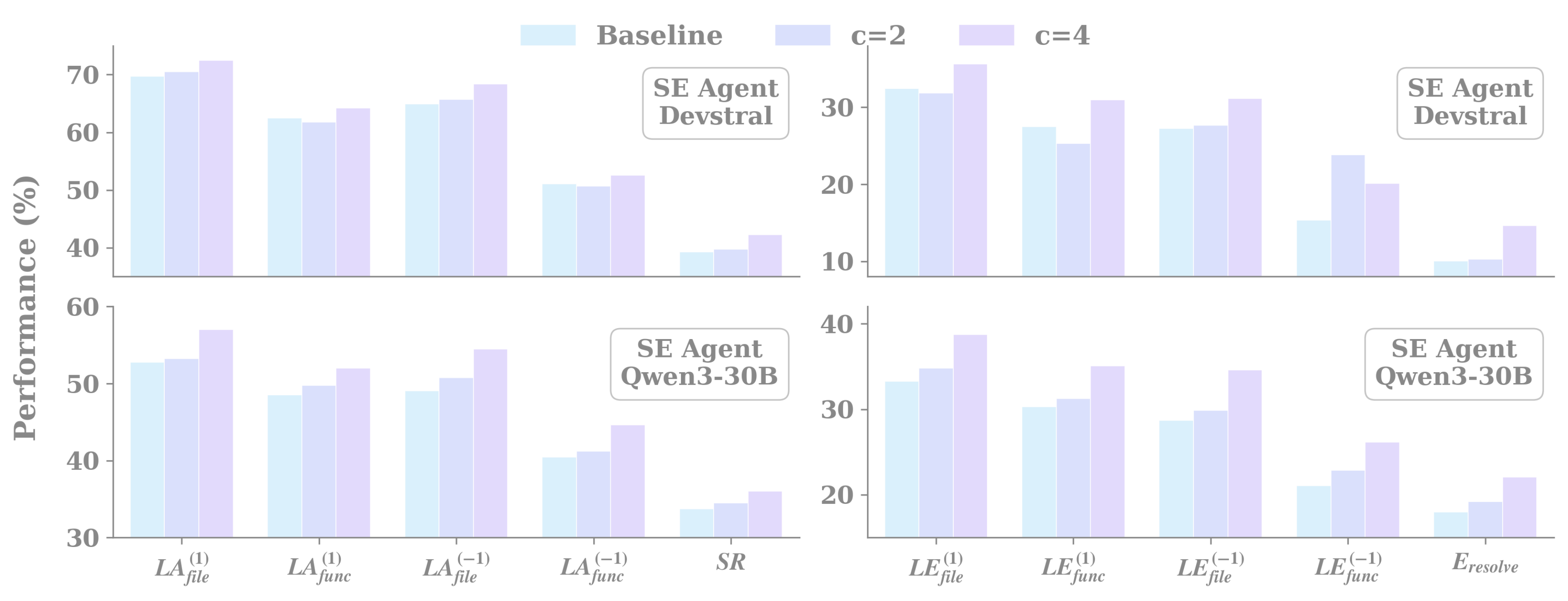}

\caption{\textbf{Effects of Preference Rollout Batch Size Configuration on $M_\theta$ Optimization.} With the $M_\theta$ backbone held fixed at \texttt{Qwen3-4B-Thinking}, we compare Stage~II finetuning on $\mathcal{D}_{\textit{RL}}$ constructed with $c=2$ and $c=4$ contrastive candidates (\S\ref{subsec:rejection_sampling}), evaluated across our ten metrics (\S\ref{subsec:eval_metrics}). Both settings improve over the \textit{\textcolor{gray}{no-$M_\theta$ baseline}}, and $c=4$ gives broader gains ($\Delta_{abs} \geq 2.25\%$ in $SR$, against $\Delta_{abs} \leq 0.75\%$ for $c=2$), indicating that a larger contrastive group sharpens the group-relative advantage in Eq.~\ref{eq:rl_objective}.}
\label{fig:exp_ablation_rl_option}

\vspace{0pt}

\end{figure}

%% file: tables/exp_memory_system_comparison.tex

\begin{table}[H]

\small
\centering
\renewcommand{\arraystretch}{1.3}

\caption{\textbf{Comparison against Existing SE Memory Systems.} All systems are evaluated under an identical setting: the same SE agent (\texttt{Devstral-Small-2507}), the same evaluation set (\S\ref{appendix:sec:dataset_construction}), and the same metric suite (\S\ref{subsec:eval_metrics}), so that only the memory mechanism differs. We compare systems that attach to a fixed SE agent without replacing the agent stack and provide a usable public implementation. The remaining systems have no public implementation and are compared conceptually in Tab.~\ref{tab:related_work_taxonomy} (\S\ref{appendix:subsec:related_work_extension:memory4SE}). \textbf{\textit{Absolute performance differences}} (Eq.~\ref{eq:eval_metric:deltaabs}) against the \textit{\textcolor{gray}{no-$M_\theta$ baseline}} are shown as \textit{positive} \textcolor{deltagreen}{$\uparrow \Delta_{\textit{abs}}\%$} or \textit{negative} \textcolor{deltared}{$\downarrow \Delta_{\textit{abs}}\%$}.}
\label{tab:exp:memory_system_comparison}

\begin{tabular*}{\textwidth}{@{\extracolsep{\fill}}lcccc@{}}

\toprule

\textbf{Memory System} & \textbf{Public Implementation} & $\bm{SR}$ & $\bm{\textbf{\textit{LA}}_{\textit{file}}^{(-1)}}$ & $\bm{E_{\textit{resolve}}}$ \\

\midrule

\textcolor{gray}{\textbf{(Baseline)}} & \textcolor{gray}{--} & \textcolor{gray}{39.25} & \textcolor{gray}{64.89} & \textcolor{gray}{10.01} \\

\arrayrulecolor{gray!50}\cmidrule(l){1-5}\arrayrulecolor{black}

CTIM-Rover~\citep{lindenbauer2025knowledge2noise} & \cmark & \absdeltapct{39.25}{32.25} & \absdeltapct{64.89}{59.32} & \absdeltapct{10.01}{8.12} \\
MemGovern~\citep{wang2026memgovernenhancingcodeagents} & \cmark & \absdeltapct{39.25}{40.50} & \absdeltapct{64.89}{65.31} & \absdeltapct{10.01}{8.87} \\

\arrayrulecolor{gray!50}\cmidrule(l){1-5}\arrayrulecolor{black}

Subtask-Level Memory~\citep{shen2026structurallyalignedsubtasklevelmemory} & \xmark & -- & -- & -- \\
Repository Memory~\citep{Wang2025ImprovingCLA} & \xmark & -- & -- & -- \\

\arrayrulecolor{gray!50}\cmidrule(l){1-5}\arrayrulecolor{black}

\textcolor{ourscolor}{\textbf{\ours (ours)}} & \cmark & \absdeltapctbf{39.25}{41.75} & \absdeltapctbf{64.89}{68.37} & \absdeltapctbf{10.01}{14.64} \\

\bottomrule

\end{tabular*}

\end{table}

%% file: tables/exp_matched_budget.tex

\begin{table}[H]

\small
\centering
\renewcommand{\arraystretch}{1.3}

\caption{\textbf{Deployed versus Matched-Budget Memory Gain.} \ours generates $N_{\mathcal{M}}=4$ memory candidates per instance (Alg.~\ref{alg:rejection_sampling}) and deploys a single unselected one, so the deployed gain is an \textit{avg@4} quantity. Under \textit{max@4} the selection budget is matched on both sides: the four memory candidates for \ours, and four independent rollouts for the \textit{\textcolor{gray}{no-$M_\theta$ baseline}}. Both settings use \ours-FT $M_\theta$ (\texttt{Qwen3-4B-T}), and $SR$ (Eq.~\ref{eq:eval_metric:SR}) is shown with \textbf{\textit{absolute performance differences}} (Eq.~\ref{eq:eval_metric:deltaabs}) against the baseline of the same column.}
\label{tab:exp:matched_budget}

\begin{tabular*}{\textwidth}{@{\extracolsep{\fill}}llcc@{}}

\toprule

\textbf{SE Agent} & \textbf{Condition} & $\bm{SR}$ \textbf{(\%, \textit{avg@4})} & $\bm{SR}$ \textbf{(\%, \textit{max@4})} \\

\midrule

\multirow{2}{*}{\textbf{Devstral-Small-2507}}
  & \textcolor{gray}{no memory} & \textcolor{gray}{39.25} & \textcolor{gray}{43.00} \\
  & \textcolor{ourscolor}{+ \ours} & \absdeltapctbf{39.25}{41.75} & \absdeltapctbf{43.00}{48.50} \\

\arrayrulecolor{gray!50}\cmidrule(l){1-4}\arrayrulecolor{black}

\multirow{2}{*}{\textbf{Qwen3-Coder-30B-A3B}}
  & \textcolor{gray}{no memory} & \textcolor{gray}{33.75} & \textcolor{gray}{38.25} \\
  & \textcolor{ourscolor}{+ \ours} & \absdeltapctbf{33.75}{36.00} & \absdeltapctbf{38.25}{44.25} \\

\bottomrule

\end{tabular*}

\end{table}

%% file: figures/exp_improved_efficiency.tex
\begin{figure}[H]

\vspace{0pt}

\centering
\includegraphics[width=1.0\linewidth]{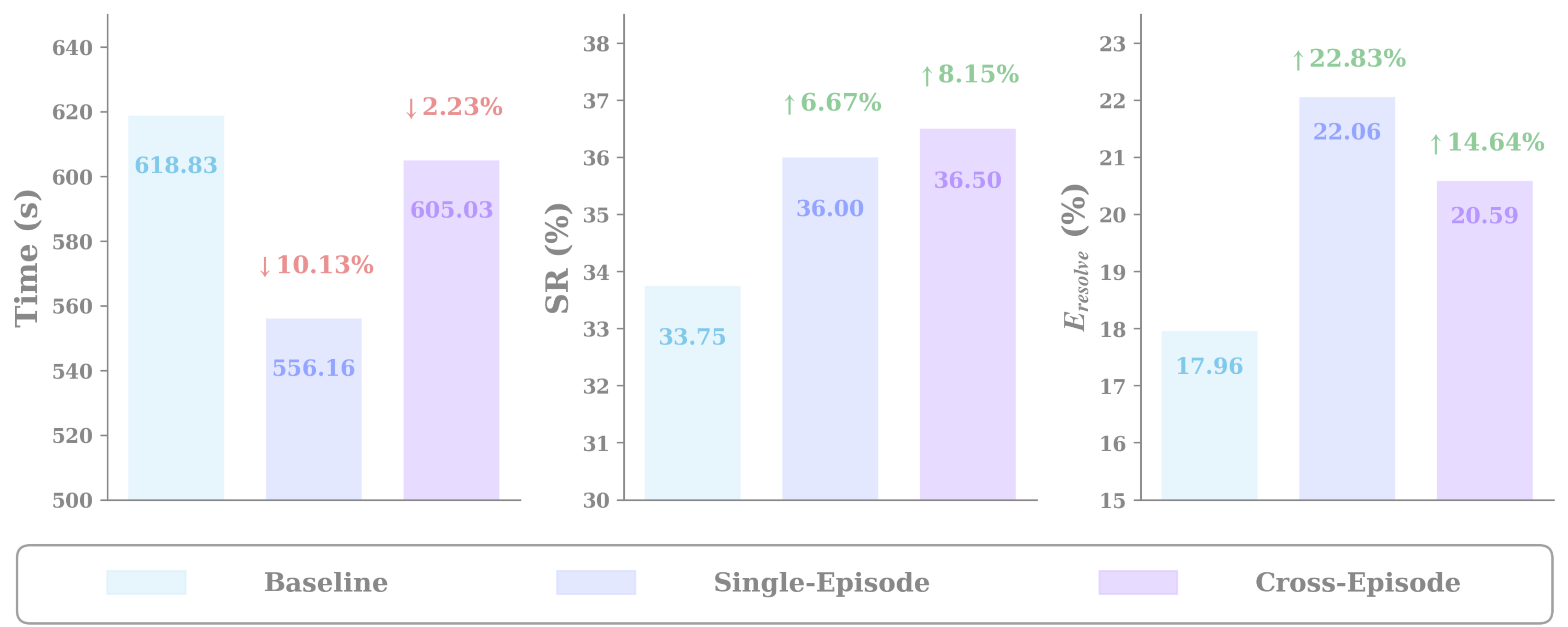}

\caption{\textbf{\ours Improves SE Performance at Reduced Deployment Cost.} Average computation time per task, $SR$, and $E_{\textit{resolve}}$ for the \textit{\textcolor{gray}{no-$M_\theta$ baseline}}, \textit{single-episode} \ours, and \textit{cross-episode} \ours, using \ours-FT $M_\theta$ (\texttt{Qwen3-4B-T}) on the SE agent powered by \texttt{Qwen3-Coder-30B-A3B}. Both regimes raise success rate and resolve efficiency while spending less time per task, so accuracy and cost improve together rather than trading against each other.}
\label{fig:improved_SE_efficiency}

\vspace{0pt}

\end{figure}

%% file: figures/exp_error_bar.tex
\begin{figure}[H]

\vspace{0pt}

\centering
\includegraphics[width=1.0\linewidth]{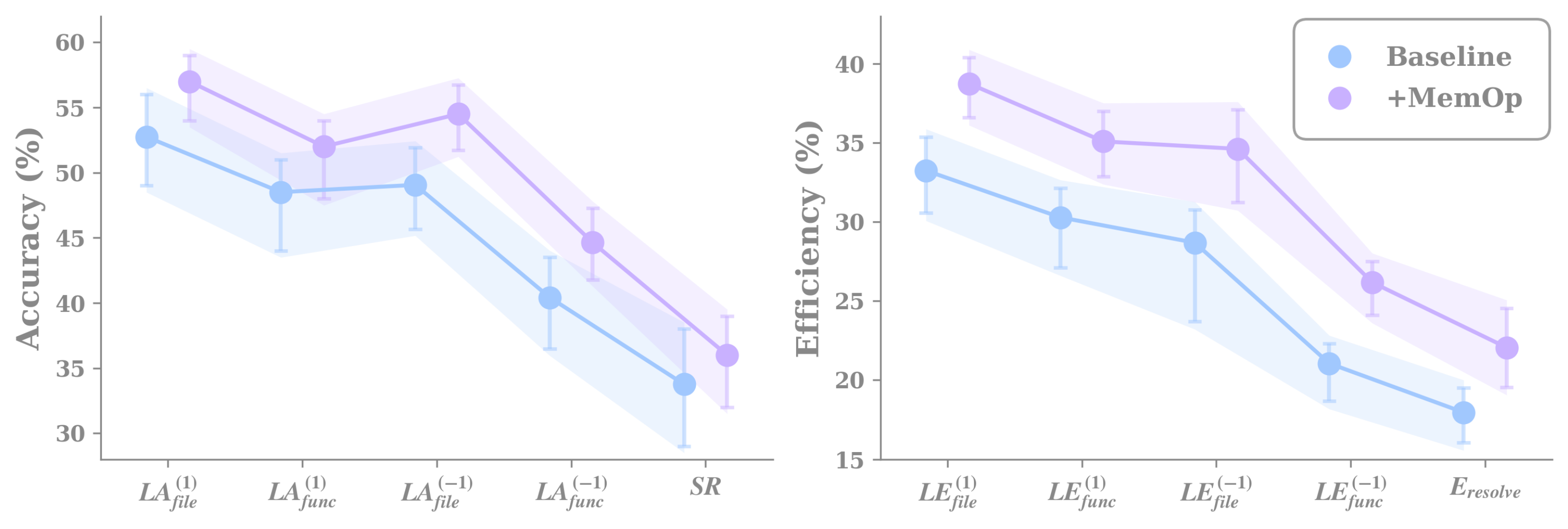}

\caption{\textbf{\ours Enhances SE Agent Robustness.} Mean and variance across rollouts for all ten metrics (\S\ref{subsec:eval_metrics}), comparing the \textit{\textcolor{gray}{no-$M_\theta$ baseline}} against \ours-FT $M_\theta$ (\texttt{Qwen3-4B-T}) on the SE agent powered by \texttt{Qwen3-Coder-30B-A3B}. \ours attains higher mean performance on every metric while exhibiting tighter variance bands, particularly on localization accuracy (\textit{left}). This indicates that utility-grounded memory stabilizes problem-solving against the stochasticity inherent in long-horizon agentic SE.}
\label{fig:error_bar_robustness}

\vspace{0pt}

\end{figure}

%% file: figures/qualitative_analysis_success.tex
\begin{figure}[H]

\vspace{-28pt}

\centering
\includegraphics[width=1.0\linewidth]{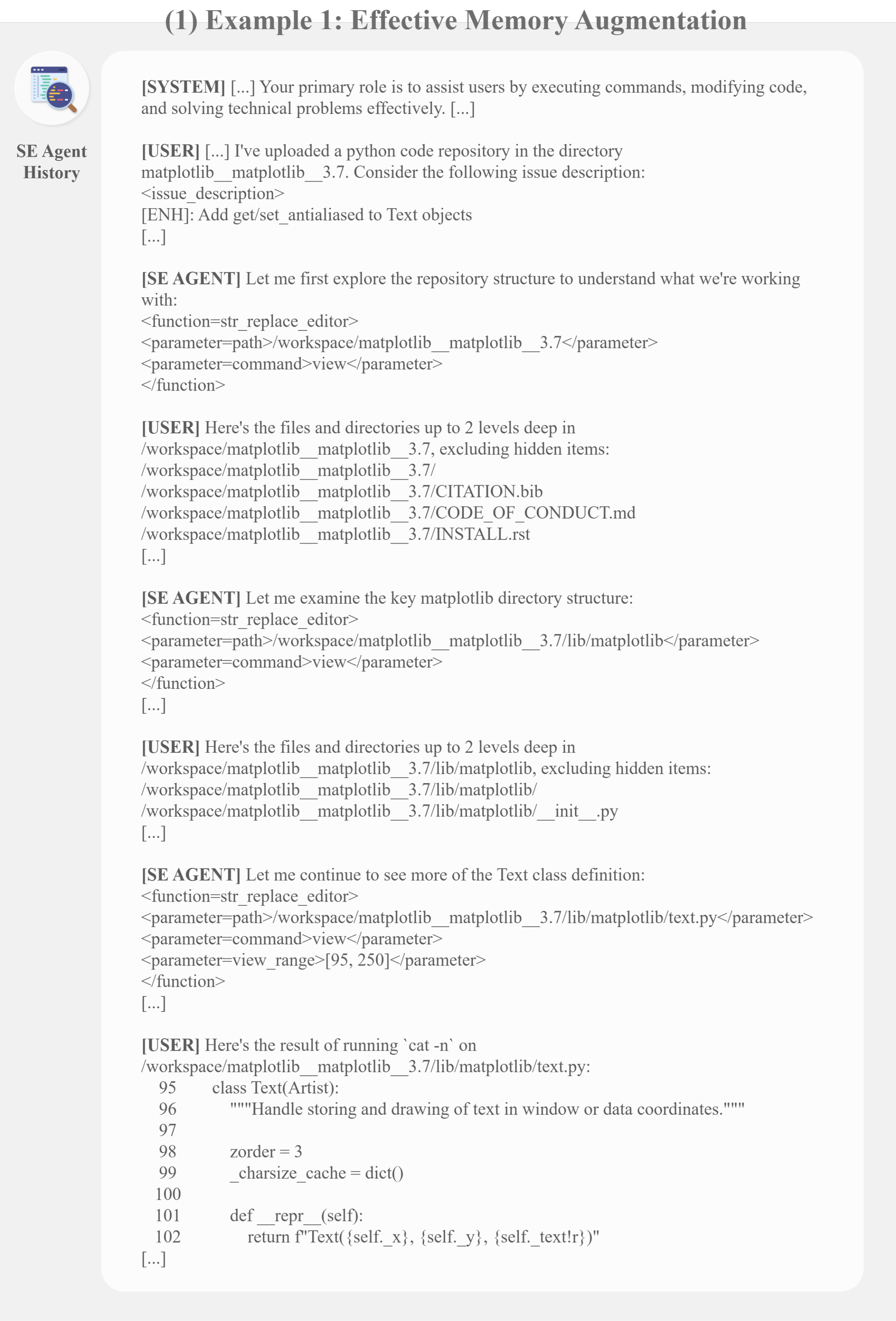}

\vspace{0pt}

\end{figure}

\begin{figure}[H]

\vspace{12pt}

\centering
\includegraphics[width=1.0\linewidth]{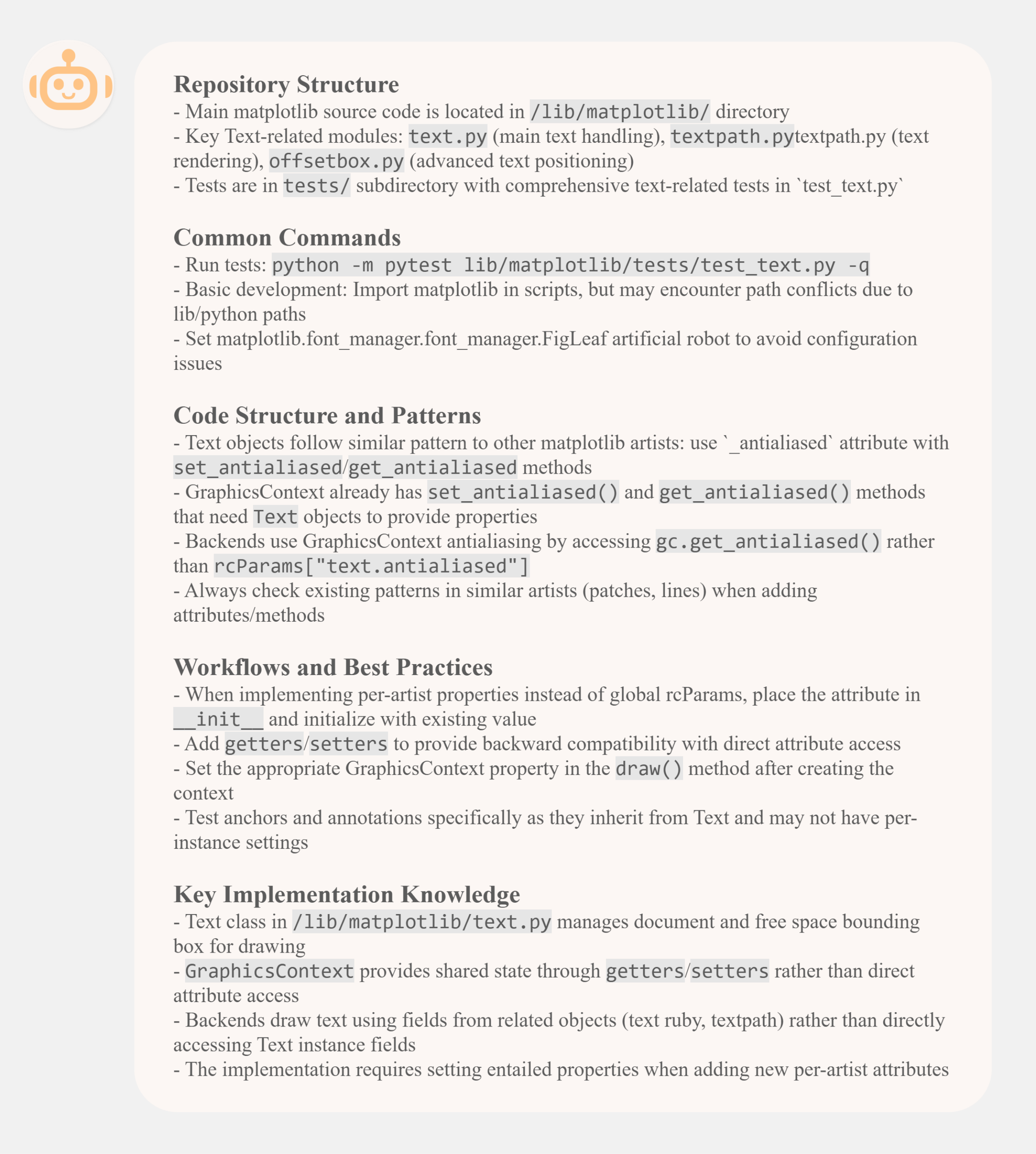}

\vspace{0pt}

\end{figure}

\begin{figure}[H]

\vspace{-28pt}

\centering
\includegraphics[width=1.0\linewidth]{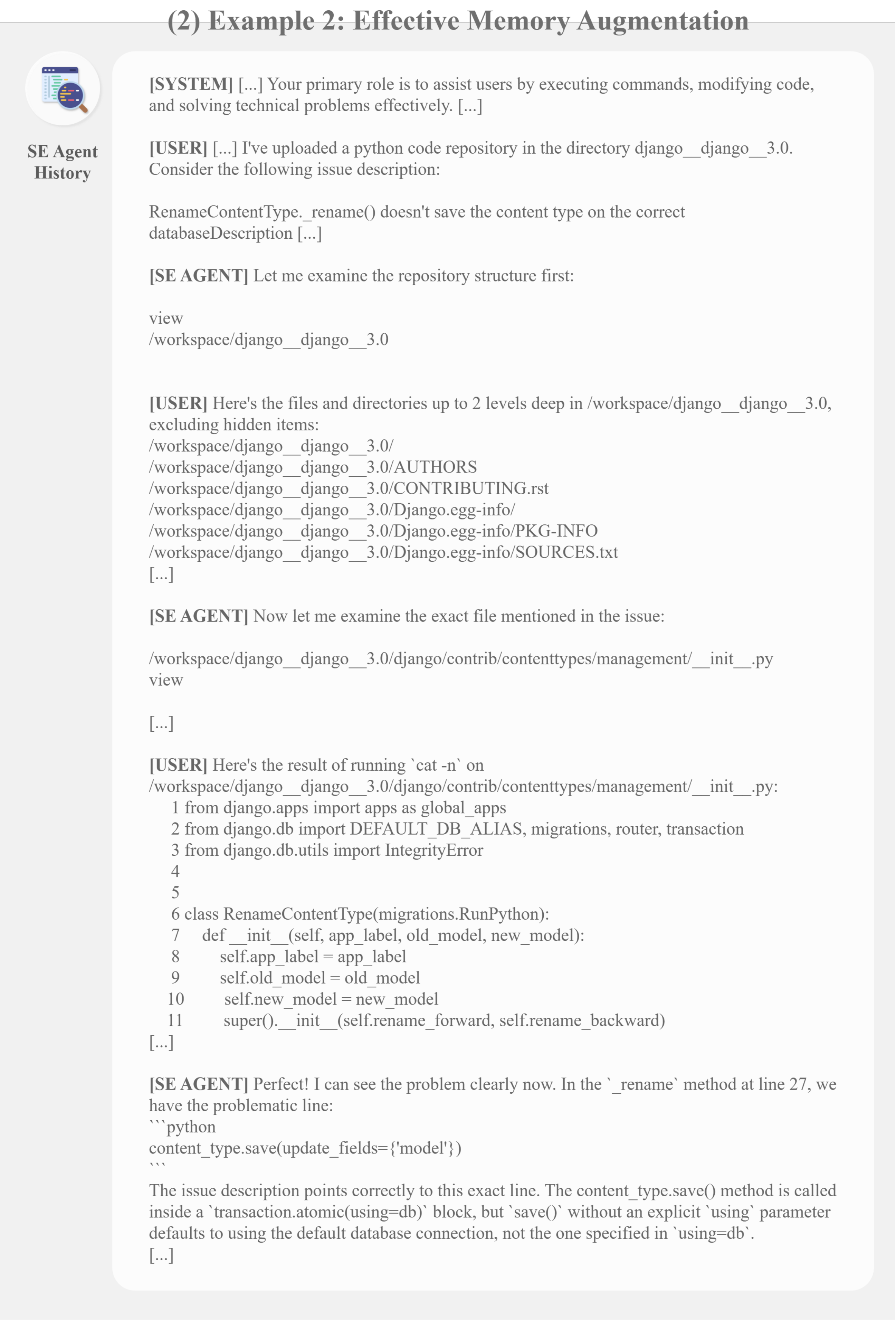}

\vspace{0pt}

\end{figure}

\begin{figure}[H]

\vspace{12pt}

\centering
\includegraphics[width=1.0\linewidth]{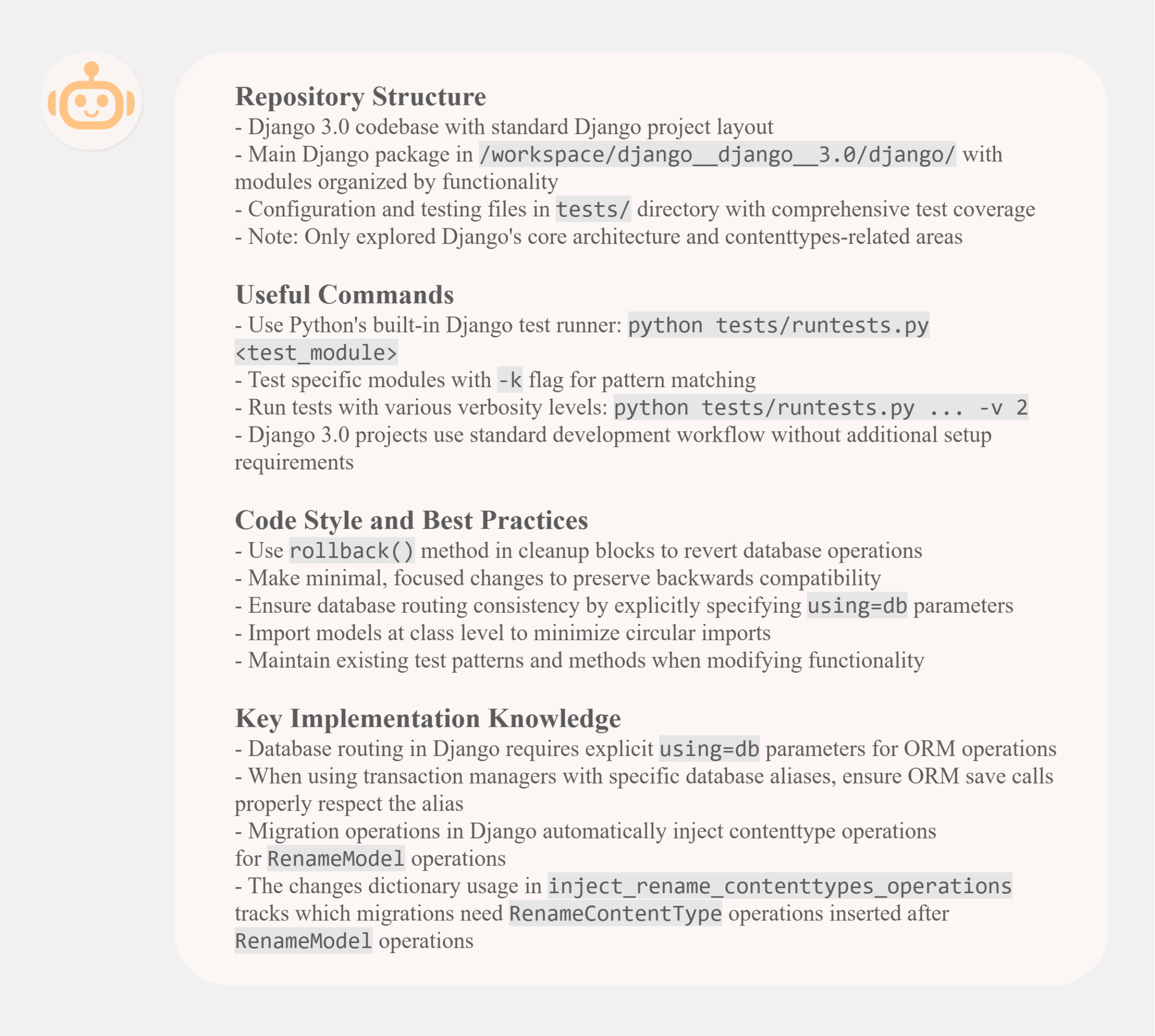}

\vspace{0pt}

\end{figure}

\begin{figure}[H]

\vspace{28pt}

\centering
\includegraphics[width=1.0\linewidth]{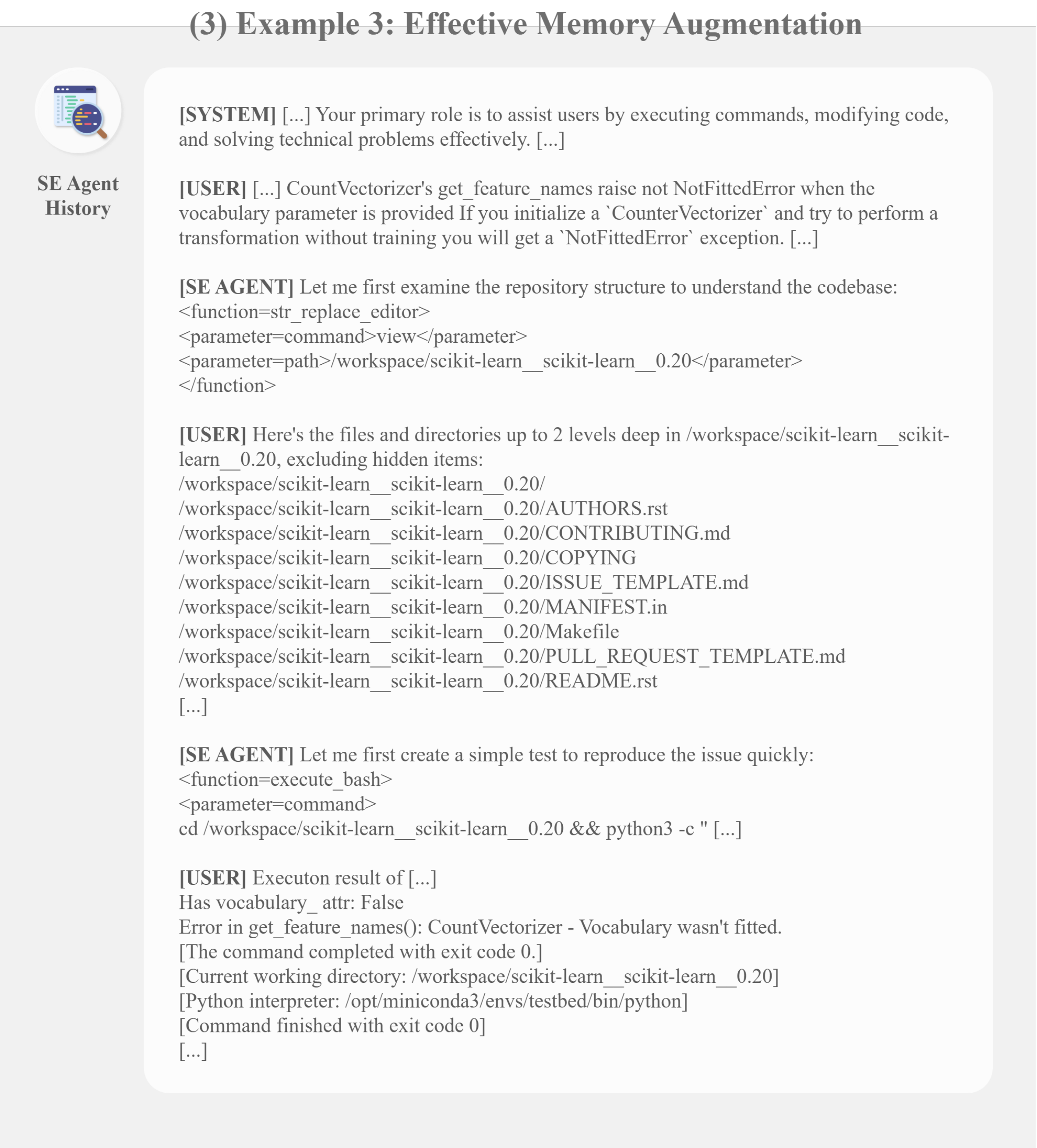}

\vspace{0pt}

\end{figure}

\begin{figure}[H]

\vspace{-12pt}

\centering
\includegraphics[width=1.0\linewidth]{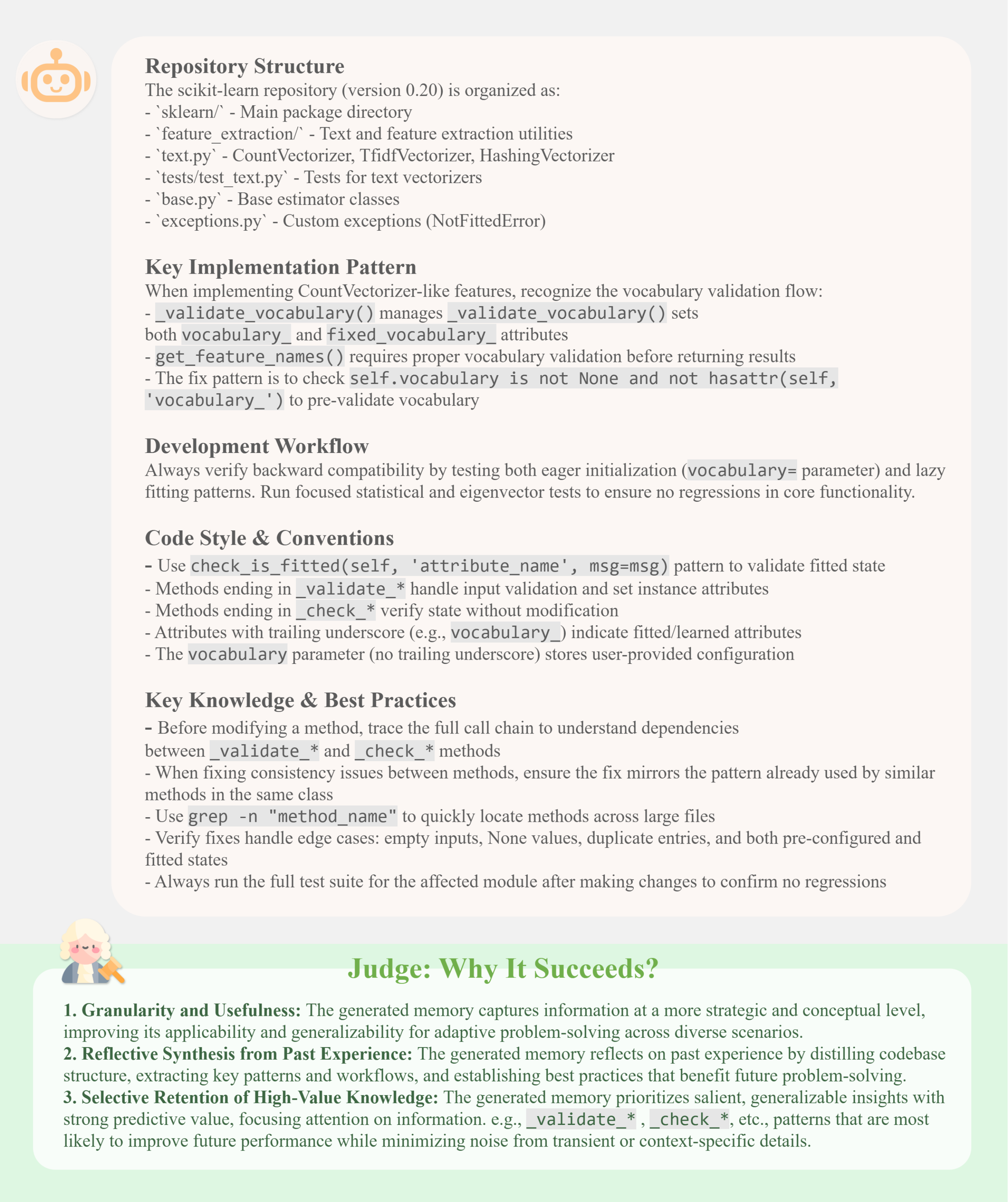}

\caption{\textbf{Examples of Effective Memory Reflection.} The generated memories effectively support SE agent to successfully resolve the task with enhanced problem-solving efficiency. Examples extracted from SE agent \textit{powered by} \texttt{Qwen3-Coder-30B-A3B} and \ours with $M_\theta$ \textit{powered by} \texttt{Qwen3-4B-T} (FT). Trajectory history details are omitted as \texttt{[...]} for clarity.}
\label{fig:qualitative_analysis:success_examples}

\vspace{0pt}

\end{figure}

%% file: figures/qualitative_analysis_failure.tex
\begin{figure}[H]

\vspace{-28pt}

\centering
\includegraphics[width=1.0\linewidth]{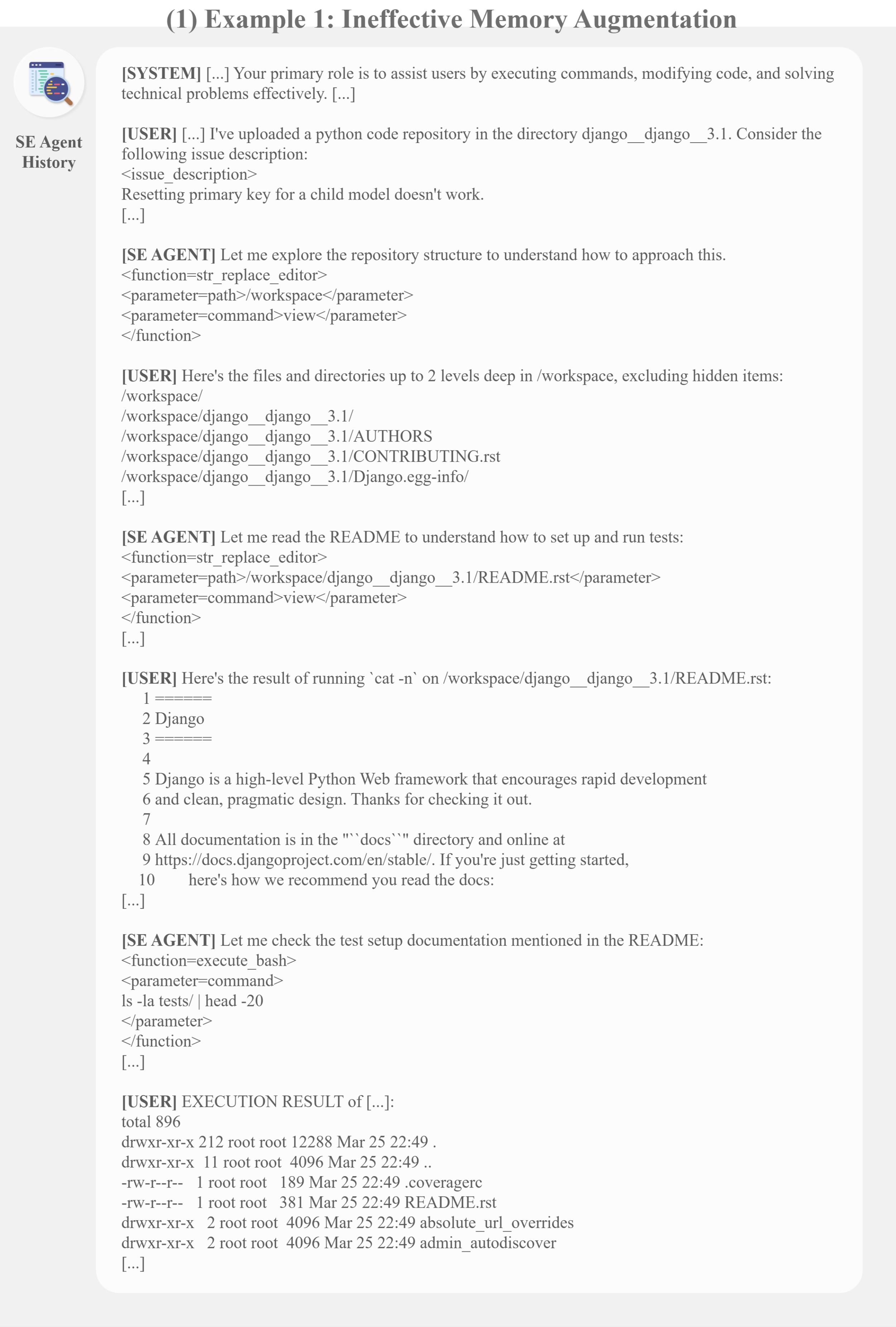}

\vspace{0pt}

\end{figure}

\begin{figure}[H]

\vspace{28pt}

\centering
\includegraphics[width=1.0\linewidth]{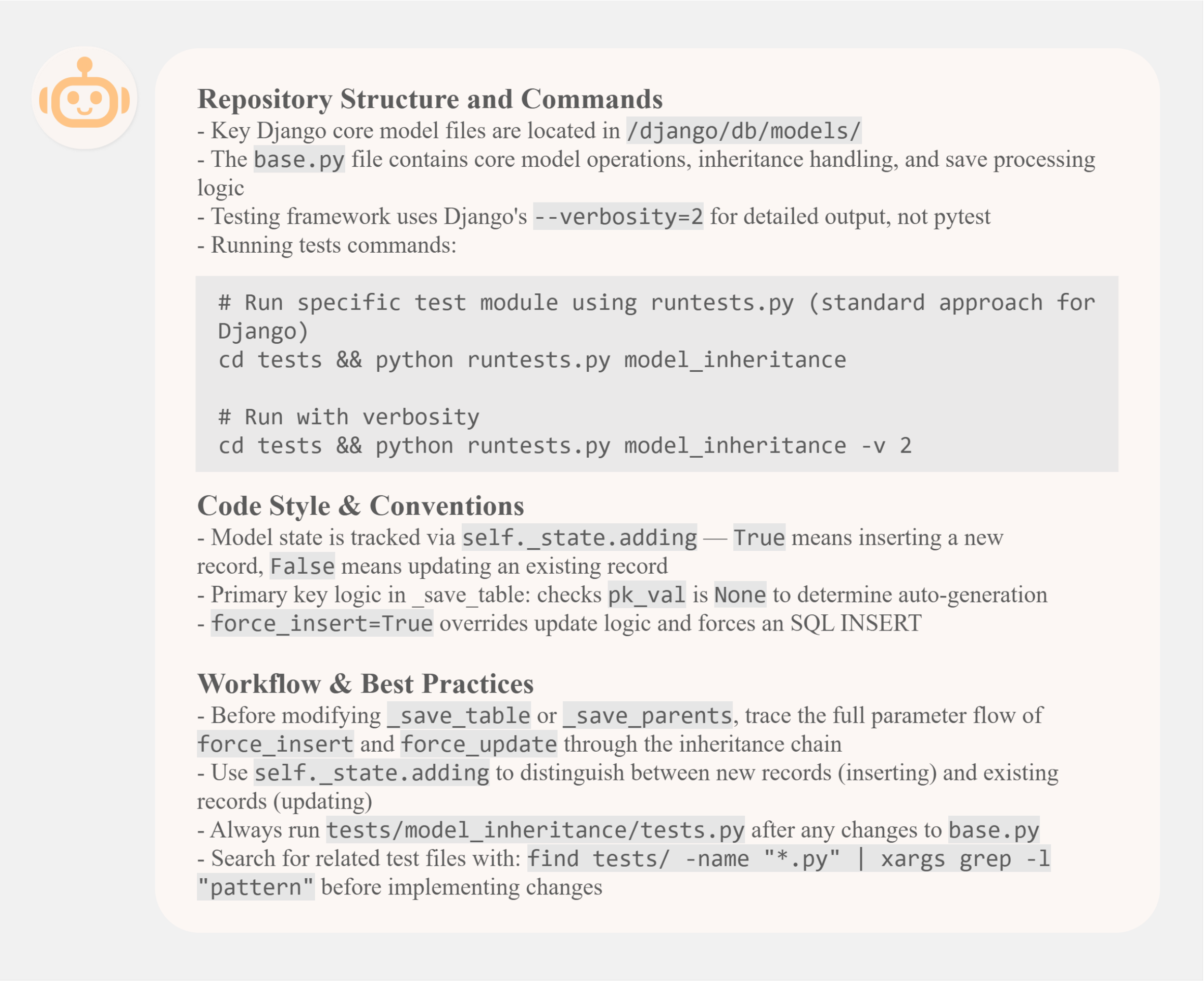}

\vspace{0pt}

\end{figure}

\begin{figure}[H]

\vspace{-12pt}

\centering
\includegraphics[width=1.0\linewidth]{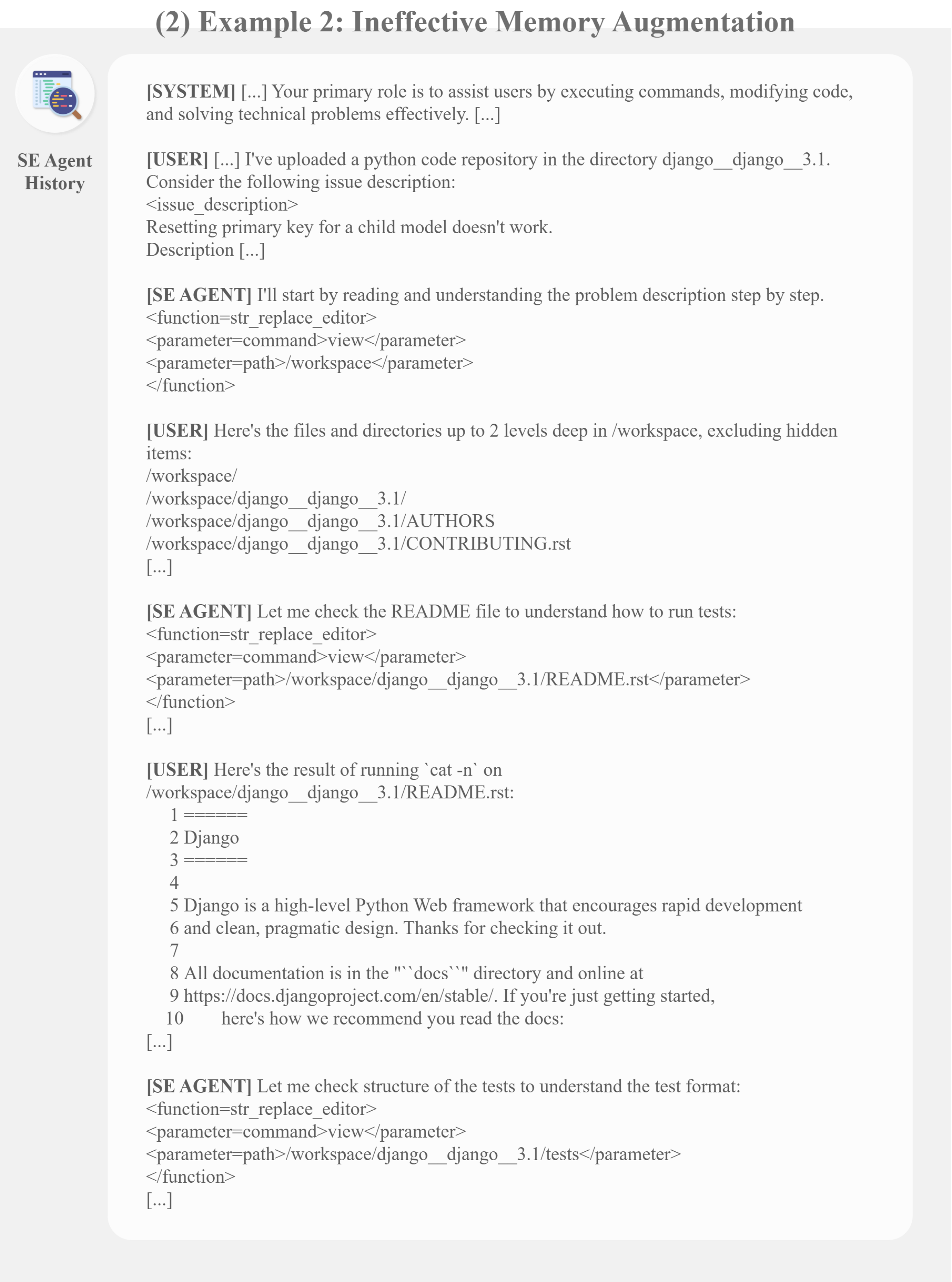}

\vspace{0pt}

\end{figure}

\begin{figure}[H]

\vspace{0pt}

\centering
\includegraphics[width=1.0\linewidth]{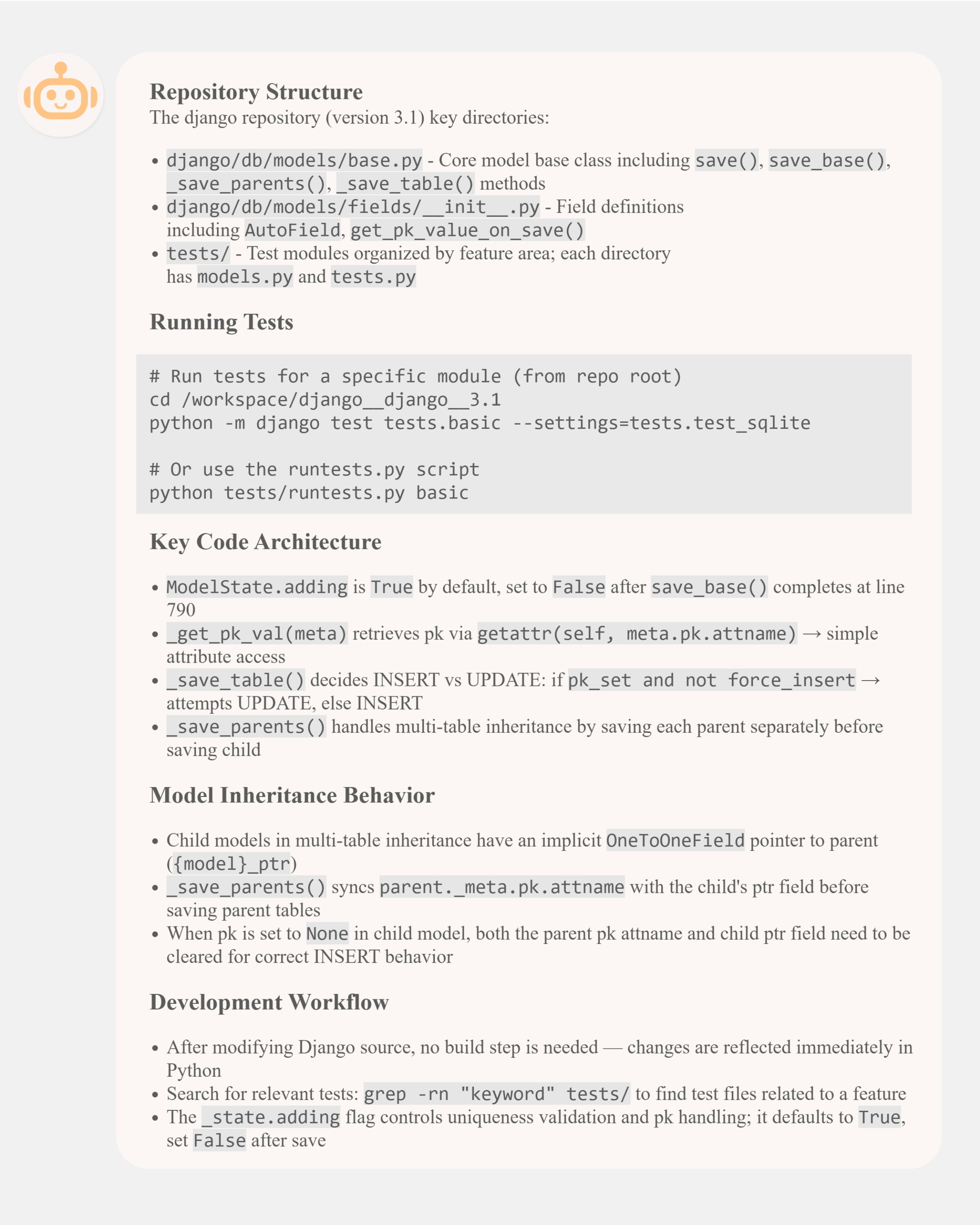}

\vspace{0pt}

\end{figure}

\begin{figure}[H]

\vspace{-0pt}

\centering
\includegraphics[width=1.0\linewidth]{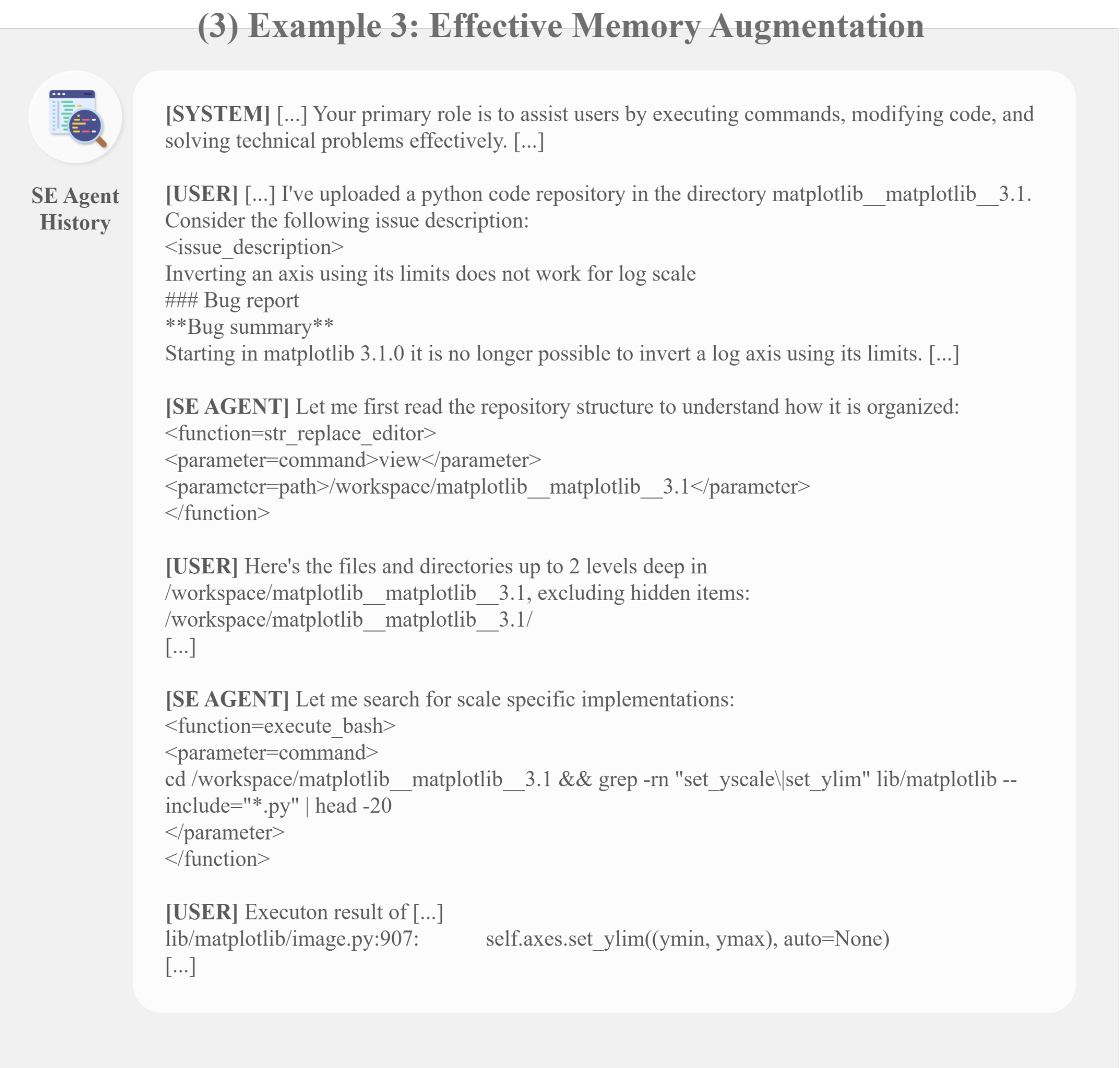}

\vspace{0pt}

\end{figure}

\begin{figure}[H]

\vspace{-18pt}

\centering
\includegraphics[width=1.0\linewidth]{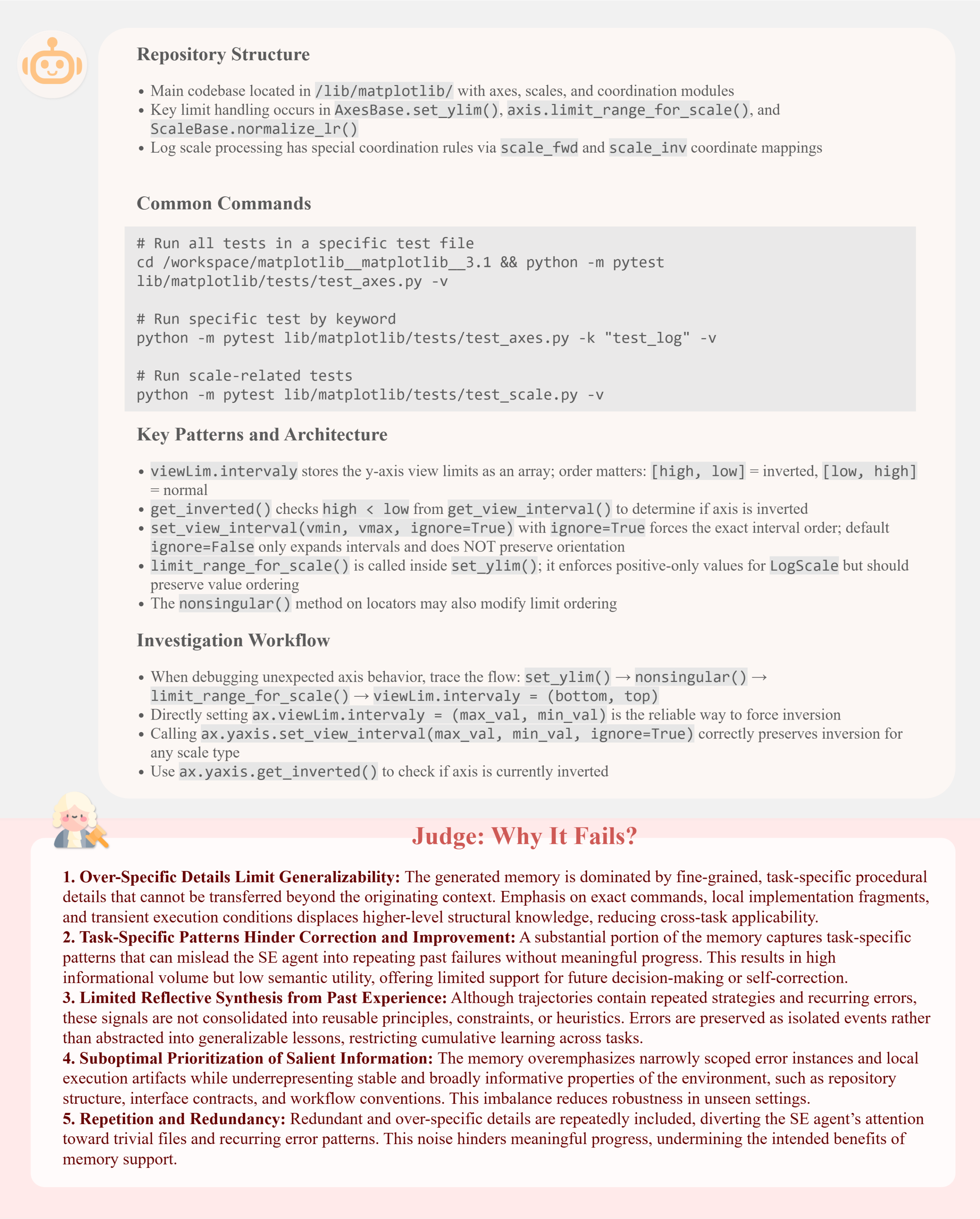}

\caption{\textbf{Examples of Ineffective Memory Reflection.} The generated memories fail to effectively support SE agent. Examples extracted from SE agent \textit{powered by} \texttt{Qwen3-Coder-30B-A3B} and \ours with $M_\theta$ \textit{powered by} \texttt{Claude-4-Sonnet} (NFT). Trajectory history details are omitted as \texttt{[...]} for clarity.}
\label{fig:qualitative_analysis:failure_examples}

\vspace{0pt}

\end{figure}